\documentclass[a4paper,11pt]{article}
\pdfoutput=1 

\usepackage{jheppub} 

\usepackage[T1]{fontenc} 
\usepackage{graphicx}
\usepackage{epstopdf}
\usepackage{bm}
\usepackage{epsfig}
\usepackage{graphics}
\usepackage{xspace}
\usepackage{hyperref}
\usepackage{slashed}
\usepackage{xcolor}
\usepackage{longtable}
\usepackage[small]{subfigure}
\usepackage[normalem]{ulem}

\newcommand{\nb}[0]{{\overline{n}}}

\newcommand{\mbar}{\overline{m}}

\newcommand{\MSb}{\overline{\mathrm{MS}}}
\newcommand{\Ord}{\mathcal{O}}

\newcommand{\dd}{\mathrm{d}}

\newcommand{\bra}[1]{\left\langle #1 \right|}
\newcommand{\ket}[1]{\left| #1 \right\rangle}

\title{\boldmath Massive Event-Shape Distributions at N$^2$LL}

\preprint{\begin{flushright} IFT-UAM/CSIC-20-79, UWThPh 2020-17\end{flushright}\vspace*{-2cm}}

\author[a,b]{Alejandro Bris}
\author[b,c]{, Vicent Mateu}
\author[d]{and Moritz Preisser}

\affiliation[a]{Departamento de F\'isica Te\'orica, Universidad Aut\'onoma de Madrid,\\Cantoblanco, 28049, Madrid, Spain}
\affiliation[b]{Instituto de F\'isica Te\'orica UAM-CSIC, E-28049 Madrid, Spain}
\affiliation[c]{Departamento de F\'isica Fundamental e IUFFyM,\\Universidad de Salamanca, E-37008 Salamanca, Spain}
\affiliation[d]{University of Vienna, Faculty of Physics, Boltzmanngasse 5, A-1090 Wien, Austria}

\emailAdd{alejandro.bris@uam.es}
\emailAdd{vmateu@usal.es}
\emailAdd{moritz.preisser@univie.ac.at}

\abstract{
In a recent paper we have shown how to optimally compute the differential and cumulative cross sections for massive event-shapes at $\mathcal{O}(\alpha_s)$ in full QCD. In the present
article we complete our study by obtaining resummed expressions for non-recoil-sensitive observables to N$^2$LL + $\mathcal{O}(\alpha_s)$ precision. Our results can be used for thrust,
heavy jet mass and C-parameter distributions in any massive scheme, and are easily generalized to angularities and other event shapes. We show that the so-called E- and P-schemes
coincide in the collinear limit, and compute the missing pieces to achieve this level of accuracy: the \mbox{P-scheme} massive jet function in Soft-Collinear Effective Theory (SCET)
and boosted Heavy Quark Effective Theory (bHQET). The resummed expression is subsequently matched into fixed-order QCD to extend its validity towards the tail and far-tail
of the distribution. The computation of the jet function cannot be cast as the discontinuity of a forward-scattering matrix element, and involves phase space integrals in $d=4-2\varepsilon$
dimensions. We show how to analytically solve the renormalization group equation for the P-scheme SCET jet function, which is significantly more complicated than its 2-jettiness counterpart,
and derive rapidly-convergent expansions in various kinematic regimes. Finally, we perform a numerical study to pin down when mass effects become more relevant.}

\begin{document}
\maketitle
\flushbottom

\section{Introduction}\label{sec:intro}
Since the late 70s, a class of observables called \emph{event shapes} has been used to test and determine fundamental properties of QCD (for a review see ~\cite{Dasgupta:2003iq,Kluth:2006bw}),
most notably to measure the strong coupling. As the name suggests, these observables contain information about the geometric momentum distribution of the final-state particle momenta. The
(historic) main fields of application are $e^+e^-$ collisions and deep inelastic scattering (DIS), but today there also exist adaptations developed specifically for $pp$ colliders.
In high-energy experiments most of the time it is sufficient to use the approximation that all particles in the final state are massless. If one is interested in high-precision calculations or in
cases where the quark mass is a dominant effect, this approximation is no longer valid.

While the theoretical computation of event-shape distributions for massless quarks at $e^+e^-$ colliders has been pushed to unprecedented precision in recent years, including fixed-order
results to $\mathcal{O}(\alpha_s^3)$~\cite{GehrmannDeRidder:2007bj, GehrmannDeRidder:2007hr,Weinzierl:2008iv, Ridder:2014wza, Weinzierl:2009ms,DelDuca:2016ily} and resummation
at next-to-next-to-leading-logarithm (N$^2$LL)~\cite{Hornig:2009vb,Becher:2012qc,Bell:2018gce} and next-to-next-to-next-to-leading-logarithm
(N$^3$LL)~\cite{Becher:2008cf,Chien:2010kc,Hoang:2014wka,Moult:2018jzp,HJM-theo}, computations for massive quarks remain at a lower precision. Such high level of resummation for massless
quarks has been achieved in most cases using Soft-Collinear Effective Theory (SCET)~\cite{Bauer:2000ew, Bauer:2000yr, Bauer:2001ct, Bauer:2001yt, Bauer:2002nz} (or in the equivalent formalism
by Collins, Soper and Sterman~\cite{Collins:1981uk,Korchemsky:1998ev, Korchemsky:1999kt,Korchemsky:2000kp, Berger:2003iw}), in which the most singular terms of the distribution are written
in a factorized form, and summation of large logarithms is carried out with standard renormalization-group evolution. On the other hand, using the coherent branching formalism~\cite{Catani:1992ua},
the resummation can be automatized up to N$^2$LL with numeric codes~\cite{Banfi:2004yd,Banfi:2014sua}. The results have been extended to the case of oriented event shapes in
Ref.~\cite{Mateu:2013gya}, and can be used to extract the strong coupling with high precision by comparing the theoretical expressions to LEP data, see
e.g.~\cite{Becher:2012qc,Chien:2010kc,Abbate:2010xh,Abbate:2012jh,Gehrmann:2012sc,Hoang:2015hka,HJM-fit}.

A first step towards having a firmer theoretical control over massive event shapes was taken in Ref.~\cite{Lepenik:2019jjk}. In that article, the fixed-order differential and cumulative cross sections
were computed at $\mathcal{O}(\alpha_s)$ for any event shape. Since quark masses screen collinear divergences, there are only soft singularities that translate at threshold into
two types of singular terms:\footnote{The threshold of an event-shape distribution is located at its minimal value, referred collectively to as $e_{\rm min}$ in this article.} a Dirac delta and a plus function (at higher orders other singular functions might appear). In Ref.~\cite{Lepenik:2019jjk} the coefficients of the delta and plus functions
were computed analytically, providing a closed $1$-dimensional integral form for the former, and showing that the latter is universal. Furthermore, the delta function coefficient was provided
in an analytic form for most event shapes. An algorithm was devised to compute the non-singular terms through a single integral (which can be carried out analytically in a
few cases), a method much faster and accurate than binning the distribution in conjunction with a Monte Carlo integrator. In the present paper we complement those developments by adding
resummation of large Sudakov logarithms at N$^2$LL. This kind of resummation for boosted quarks has been worked out only for the hemisphere-mass (doubly differential) distribution in
Refs.~\cite{Fleming:2007qr,Fleming:2007xt}, which can be easily marginalized into heavy jet mass and 2-jettiness (and with extra little work, also into a variable called C-Jettiness in
Ref.~\cite{Lepenik:2019jjk}). In both cases, and only through recent computations that will be reviewed in the next paragraph, N$^3$LL precision has been achieved.
In this article we take the necessary steps to bring the accuracy to N$^2$LL~+~$\mathcal{O}(\alpha_s)$ for SCET-I type observables in the massive E- and P-schemes. These massive schemes were defined to better understand soft hadronization and affect the way in which the energy and momentum of each
massive particle enters the computation of the event shape, while they have no effect on massless particles. In the P-scheme only three-momentum information
is considered, whereas in the E-scheme the energy and particle direction are used.

Although SCET was first developed as an effective theory for massless particles, it was quickly generalized to include massive quarks in Ref.~\cite{Leibovich:2003jd}. Resummation is achieved
in SCET by writing the most singular terms of the cross section as the product or convolution of various pieces: the hard matrix element, which is the squared modulus of the matching
coefficient between SCET and QCD; the jet function, describing radiation collinear to the initiating partons; and the soft function, accounting for wide-angle soft radiation. Quark masses are
infrared modes of the effective theory and therefore do not show up in the hard matching coefficient, but they do contribute to the jet (starting at one-loop) and soft (starting at two loops) functions.
In the limit in which the difference between the jet and the heavy-quark massess is much smaller than either of those, a new physical scale emerges, requiring additional resummation of
large logarithms. This is achieved by matching SCET onto boosted heavy-quark effective theory (bHQET). For jettiness all the ingredients necessary to build an N$^3$LL-accurate cross section are known:
bHQET jet function~\cite{Jain:2008gb}, SCET massive jet function~\cite{Hoang:2019fze}, and bHQET matching coefficient~\cite{Hoang:2015vua}, as well as the secondary production
contributions~\cite{Gritschacher:2013pha,Pietrulewicz:2014qza} (all up to two-loop level). The computations carried out in this article make possible N$^2$LL resummation both in SCET and bHQET
for other classes of event shapes. Masses can also modify the endpoint of the distribution, which is their only effect at LL and NLL. This threshold modification depends solely on the
scheme used to define the massive event shape.

In the approach followed in \cite{Butenschoen:2016lpz} to build the differential cross section, kinematic power corrections are taken into account in fixed-order perturbation theory by matching the
SCET cross section onto full QCD. This setup was successfully applied to phenomenological
analyses of massless event shapes, and is crucial to obtain reliable predictions in the tail and far-tail of the distribution. In the case of massive event shapes the situation is a
bit more complex for two reasons: a)~the partonic threshold is different in full QCD and the EFT, and b)~the EFT prediction does not completely reproduce the singular terms at threshold.
The reason is simple to understand: since the mass is an infrared mode, it is power-counted together with other infrared scales such as the quark virtuality, in such a way that
power-suppressed terms include kinematic as well as mass corrections. Since the main focus of this article is on event shapes with mass-independent thresholds, issue a) is not relevant. At
leading power, SCET and bHQET only predict the leading $\hat m$ behavior of the singular terms in QCD.\footnote{Here and in what follows,
$Q$ denotes the $e^+e^-$ center of mass energy and $\hat m=m/Q$ the reduced mass.} It is however possible to modify the hard and jet functions to fully account for the singular
terms in the factorization theorems, such that power corrections are not distributions and behave well close to threshold. We will show how to apply this prescription to
event shapes in the E- and P-schemes.

Non-vanishing quark masses imply that the mass sensitivity can be tuned using different schemes for the event-shape definition. To the best of our knowledge, this possibility has only
been studied in the context of fixed-order perturbation theory in Ref.~\cite{Lepenik:2019jjk}, and in the present work this analysis is continued by considering resummation of large logarithms.
Furthermore, we complete the list of ingredients necessary for a full N$^2$LL + $\mathcal{O}(\alpha_s)$ computation, which are applied to the case study of P-scheme thrust, making our
work a useful reference for upcoming studies. Our result is essential to consistently include bottom-quark mass effects in analyses that aim to extract $\alpha_s$ from
fits to LEP data, and to clarify the top quark mass interpretation problem.

This article is organized as follows: in Sec.~\ref{sec:dijet} we discuss kinematics in the dijet limit, which is the core for factorizing cross sections; in Sec.~\ref{sec:schemes} we review the
concept of massive schemes, explore how they affect the mass sensitivity of cross sections, and derive the implications they have for the SCET power counting; the factorization theorems
in SCET and bHQET are presented in Sec.~\ref{sec:factorization}, in which also large-logarithm resummation is introduced; we carry out the computation of the SCET massive
jet function in Sec.~\ref{sec:computation}, and use this result in Sec.~\ref{sec:FO-SCET} to write down the fixed-order expression of the cross section in the dijet limit; in
Sec.~\ref{sec:jet-bHQET} we compute the bHQET jet function, and derive analytic results for the running of the non-distributional pieces of the SCET jet function
in Sec.~\ref{eq:evolution}, presenting some useful expansions that can be implemented in different regions of the spectrum; kinematic, massive and non-perturbative power corrections are discussed
in Sec.~\ref{sec:power}, while some numerical investigations are carried out in Sec.~\ref{eq:numerics}; finally, our conclusions are contained in Sec.~\ref{eq:conclusions}. Some technical aspects
of the computations are relegated to Appendices~\ref{sec:secDec} and \ref{sec:yReal}.

\section{Dijet Kinematics}\label{sec:dijet}
In this section we study the kinematics of dijet events: two narrow, nearly back-to-back jets, plus additional soft radiation. In this situation the value of the event-shape is not far from its
minimal value 
$e_{\min}$, and the SCET power-counting rules apply. Particles in such events can be either soft, $n$-collinear or $\bar n$-collinear, with
momenta whose light-cone coordinates $p^\mu = (p^+,p^-,p^\perp)$ scale like $p_s^\mu\sim Q(\lambda^2,\lambda^2,\lambda^2)$, $p_n^\mu \sim Q(\lambda^2,1,\lambda)$ and
$p_{\bar n}^\mu \sim Q(1,\lambda^2,\lambda)$ respectively, with $\lambda$ the SCET power-counting parameter. Since we are interested in the primary production of heavy quarks and
$p^2=m^2$, for consistency one has that $\hat m\sim \lambda$ and soft particles are either massless or have $\hat m_s\lesssim\lambda^2$. The SCET scaling holds for momenta defined
in the P- and E-schemes as well. In this limit it is possible to write the event-shape measurement $e$ as the sum of contributions from collinear (in both directions) and soft
particles~\cite{Bauer:2008dt,Mateu:2012nk}:
\begin{equation}\label{eq:es_sum}
\overline{e} = e_n + e_{\bar n} + e_s\,,
\end{equation}
where $\overline{e}$ denotes the event-shape measurement in the dijet limit at leading power, and we consider only soft perturbative particles. For SCET-I type observables, the three terms
scale like $\lambda^2$ and are equally important. Moreover, $e_n$, $e_{\bar n}$ and $e_s$ can be written as a sum over single-particle contributions. If the event shape is already defined
by a single sum of final-state particle momenta (as it is the case for thrust or angularities) this statement is trivial. When the event shape correlates momenta of final-state particles
[\,e.g.\ the definition involves a double sum (C-parameter) or there is a single sum squared (jet masses)\,] the situation is more complicated. In the latter case one has to show explicitly that in
this limit the leading contribution to the event shape can be written in the form of Eq.~\eqref{eq:es_sum}, as was done e.g.\ in Ref.~\cite{Hoang:2014wka} for C-parameter.
Let us work out this decomposition for hemisphere masses, which are defined as the square of the total four-momentum flowing into one of the hemispheres, being those delimited by the
plane normal to the thrust axis. Assuming the $z$ axis in the thrust direction, the mass of the plus hemisphere takes the following form in light-cone coordinates:
\begin{equation}\label{eq:rho+-}
Q^2 \rho_+ = \sum_{i \in +} p_i^+ \sum_{j \in +} p_j^- - \biggl( \sum_{i \in +} p^{\perp}_i\biggr)^{\!\!2}
= \sum_{i \in +} p_i^+ \sum_{j \in +} p_j^-,
\end{equation}
where we have used that the component of the total hemisphere momenta normal to the thrust axis is identically zero. 
Let us assume the negative direction of the $z$ axis pointing towards the plus hemisphere such that it does not contain $\bar n$-collinear particles (with $p_z>0$). In the dijet
limit, particles $i$ and $j$ can be either soft or collinear, but if both are soft the corresponding contribution is $\rho^{s,s} \propto\mathcal{O} (\lambda^4)$ and therefore power suppressed.
Next we consider that $i$ is soft and $j$ is collinear, such that the leading contribution comes from the $p_j^-$ term:
\begin{equation}
Q^2 \rho^s_+ = \biggl( \sum_{j \in n} p_j^- \biggr)\! \biggl(\sum_{i \in s_+} p_i^+ \biggr) = 2\biggl( \sum_{j \in n} E_j
\biggr)\! \biggl( \sum_{i \in s_+} p_i^+ \biggr) = Q\! \sum_{i \in s_+} p_i^+ = Q P_{s_+}^+ ,
\end{equation}
where we have used that up to power corrections in the dijet limit $p_i^-=2E_i$ and the total available energy $Q$ is carried by collinear particles only and equally divided into each hemisphere,
such that the total minus momentum flowing into the plus hemisphere is $Q$ up to power corrections. The set of soft particles that belong to the plus hemisphere is denoted by $s_+$. With an identical
computation we get for the collinear-collinear contribution
\begin{equation}\label{eq:rho-col}
Q \rho^n_+ = \sum_{i \in c_+} p_i^+ = P_n^+\,,
\end{equation}
with $c_+$ the set of $n$-collinear particles. In the dijet limit we have $Q\rho_+ = P_{s_+}^+ + P_n^+$, which is equal to the total plus momentum entering the minus hemisphere. An identical reasoning
leads to $Q\rho_- = P_{s_-}^- + P_{\bar n}^-$. Since we have not made any assumption on the mass of the particles, our result is valid for massless and massive quarks.

\section{Massive Schemes }\label{sec:schemes}
In this section we review the generalization of event-shape measurements for massive particles that go under the name of ``massive schemes'' (which should not be mistaken with the quantum field theory
mass schemes such as pole, PS, 1S, MSR, $\MSb$, \ldots). These schemes were introduced in an article by Salam and Wicke~\cite{Salam:2001bd} to study the effects of hadron masses on hadronization
power corrections, which were further studied in \cite{Mateu:2012nk}. Both studies consider light quarks only, such that massive schemes have no effect on partonic computations, but change the size
of non-perturbative power corrections. For massive quarks, however, switching schemes can dramatically change the cross section, in particular its sensitivity to the quark mass, which is obviously of
high interest either in cases where very accurate computations demand including quark mass corrections, or when they are the leading effect. A propaedeutic study of massive schemes for heavy quarks in
event-shape cross sections was carried out in Ref.~\cite{Lepenik:2019jjk}, where fixed-order results were computed and subsequently analyzed. A more complete study demands resummation of large
logarithms in the peak and tail of the distribution, and we aim to fill in this gap here.

Massless particles travel at the speed of light and their four-momenta satisfy $p^2=0$, which when translated into energy and momentum in a given frame implies $E_p=|\vec{p}\,|$. Therefore on can
interchange $E_p \leftrightarrow |\vec{p}\,|$ in the event-shape definition with no visible effect. If particles become massive, $p^2=m^2$ and one has $E_p>|\vec{p}\,|$, meaning in turn that replacing
the momentum magnitude by the energy (or vice versa) changes the value of the event shape. In this context, various massive schemes were defined to quantify this possibility, which reproduce the
original ``massless'' definition in the limit $m\to 0$:
\begin{enumerate}
\item \textbf{E-scheme}: As indicated by its name, one replaces momenta by energies with the substitution $(p_i^0,\vec{p}_i)\to p_i^0\,(1, \vec{p}_i/|\vec{p}_i|)$. The scalar product takes the
following form:
\begin{equation}
p_E \!\cdot\! q_E = E_p E_q \biggl( 1 - \frac{\vec{p}\! \cdot \vec{q}}{| \vec{p}\,|| \vec{q}\, |} \biggr)\,,
\end{equation}
such that $p_E^2=0$ even for massive particles. One advantage of this prescription is that hadronization corrections become universal. The variables called angularities~\cite{Berger:2003pk}
were originally defined in this scheme.
\item \textbf{P-scheme}: Again the name suggests that energy gets replaced by momenta as $(p_i^0,\vec{p}_i)\to |\vec{p}_i|\,(1, \vec{p}_i/|\vec{p}_i|)$, and it happens that most of the
classical event shapes were originally defined in this scheme: thrust~\cite{Farhi:1977sg}, C-parameter~\cite{Parisi:1978eg,Donoghue:1979vi} and broadening~\cite{Rakow:1981qn}. The scalar
product now reads $p_P \!\cdot\! q_P = | \vec{p} \,| | \vec{q} \,| - \vec{p} \cdot \!\vec{q}$, which again implies $p_P^2=0$ for massless and massive particles.
\item \textbf{M-scheme}: The name ``massive scheme'' is used for event shapes that in their original definition were neither in the P- nor in the E-scheme, such as heavy jet
mass~\cite{Clavelli:1979md, Chandramohan:1980ry, Clavelli:1981yh}. Their definition contains both energy and momentum and are the most sensitive to quark masses, in particular because
in this scheme the usual relation $p^2=m^2$ is satisfied.
\end{enumerate}
It is important to realize that four-momenta as defined in the E- and P-schemes are frame-dependent, and that event shapes are usually defined in terms of magnitudes measured in the center-of-mass
frame. The usual light-cone decomposition applies in either scheme $S$
\begin{equation}\label{eq:light-cone}
p_S\!\cdot\! q_S=\frac{1}{2}\,p_S^+ q_S^- + \frac{1}{2}\,q_S^+ p_S^- - \vec{p}_{S,\perp}\! \cdot \vec{q}_{S,\perp}\,,
\end{equation}
with $S=E,P$. The specific definition of the event shapes just introduced can be found e.g.\ in the original papers and will not be repeated here unless necessary. 
They are also summarized, including a discussion on massive schemes, in Ref.~\cite{Lepenik:2019jjk}.

\subsection{Mass Sensitivity}
When studying the sensitivity of event shapes at parton level\footnote{We consider for now partonic final states, assuming stable massive quarks.} the leading order contribution for $e^+e^-$
annihilation comes from the production of a heavy quark-antiquark pair without additional radiation. In this case, the thrust axis is parallel to the three-momenta of the quarks, which makes trivial to
calculate the threshold for two particles in the final state with equal mass $m$. Moreover, this simple computation sets the lower threshold even if additional gluons and (massless) quarks are
radiated.
\begin{table}[t!]
\centering
\begin{tabular}{l|cccc}
\hline
\hspace{1cm} &$\tau$ & $\tau_a$ & $C$ & $\rho$\\
\hline
M-scheme & $1-\beta$ & $(1-\beta)^{\frac{2-a}{2}}(1+\beta)^{\frac{a}{2}}$ & $12\hat m^2(1-\hat m^2)$ & $\hat m^2$\\
P- and E- schemes & $0$ & $0$ & $0$ & $0$\\
\hline
\end{tabular}\label{tab:eventshapes}
\caption{\label{tab:ESthres}Threshold position for various event shapes in the case of primary production of a stable quark-antiquark pair in different massive schemes.
We use \mbox{$\beta=\sqrt{1-4\hat m^2}$}, the velocity of the quarks at threshold in natural units.}
\end{table}
The results in Tab.~\ref{tab:ESthres} show that for events in which a massive stable quark-antiquark pair is produced (\emph{primary production}) only the \mbox{M-scheme} is sensitive to
the quark mass while P- and E-schemes are not. In most of the events there will be some extra radiation present which will modify the former dijet into two fatter jets or an
even more isotropic momentum distribution. For the observables we study, such processes will mainly contribute for event-shape values away from threshold adding subleading
mass sensitivity (i.e. suppressed by a factor of $\alpha_s$) even in the P- and E-schemes, but will not substantially change the leading sensitivity of the M-scheme definition since it
comes from the tree-level peak position. From this we can conclude that the M-scheme is preferred if the aim is a mass-sensitive observable (e.g.\ for quark mass determinations), but
in case that one seeks a mass-insensitive observable, the P- and \mbox{E-schemes} are a better choice.\footnote{If the massive partons enter the final state via gluon splitting in a massive
quark-antiquark pair (that is, through secondary production) the sensitivity to the quark mass will again be subleading (now suppressed by a factor of $\alpha_s^2$).}

\subsection{Massive Schemes in the Collinear Limit}
Collinear particles in the $n$ direction satisfy $E_p = (p^+ + p^-)/2= p^-/2 + \mathcal{O}(\lambda^2)$ and also $|\vec{p}\,| = \sqrt{E_p^2 -m^2}= p^-/2 + \mathcal{O}(\lambda^2)$ such that the E-scheme
($Q=\sum_i E_i$) and P-scheme \mbox{($Q_p\equiv\sum_i |\vec{p}_i|$)} normalizations are the same at leading power. Let us compute the four-momenta of massive $n$-collinear particles in the
E- and P-schemes. Since we have seen that $E_p/|\vec{p}\,|=1+ \mathcal{O}(\lambda^2)$, the ``large'' (or label) components $p^-$ and $p^\perp$ are the same in any scheme. Let us then focus
in the small $p^+$ momenta, which in the massive scheme takes the following form (for $n$-collinear particles the $z$ component of momenta is negative):
\begin{equation}\label{eq:m-pplus}
p^+ = E_p + p_z = E_p - \sqrt{| \vec{p} \,|^2 - | \vec{p}_{\perp} |^2} = E_p - \sqrt{E^2_p - E_{\perp}^2}
\simeq \frac{E_{\perp}^2}{2E_p} +\mathcal{O} (\lambda^4)\, ,
\end{equation}
where we have used the so-called perpendicular energy $E_\perp\equiv \sqrt{| \vec{p}_{\perp}|^2 + m^2}$. The computation of $p^+$ in the P-scheme is very similar since one only has to
use $|\vec{p}\, |$ instead of $E_p$
\begin{equation}\label{eq:ppplus}
p_P^+ = |\vec{p}\, | + p_z = | \vec{p}\, | - \sqrt{| \vec{p}\,|^2 - | \vec{p}_{\perp} |^2} \simeq \frac{| \vec{p}_{\perp} |^2}{2 E_p}+\mathcal{O} (\lambda^4) =
p^+ - \frac{m^2}{p^-} +\mathcal{O} (\lambda^4)\,,
\end{equation}
where we notice that in the second step $p_P^+$ takes the same analytic form as in Eq.~\eqref{eq:m-pplus} with the replacements $E_p\to| \vec{p}\,|$ and $E_{\perp}\to | \vec{p}_{\perp} |$,
in the third step we use $| \vec{p}\,| = E_p + \mathcal{O} (\lambda^2)$, and in the last step we replace $|\vec{p}_{\perp}|^2=p^+p^- - m^2$ and $E_p=p^-/2 + \mathcal{O} (\lambda^2)$. It is
important to notice that in general $p_P^+\neq p^+$ because $m^2/(p^+p^-)\simeq\mathcal{O}(1)$. Moreover, the mass that appears in Eq.~\eqref{eq:ppplus} (and in any other event-shape
measurement function) comes from kinematic \mbox{on-shell} considerations and therefore corresponds to the pole scheme. Finally, let us compute the E-scheme $p^+$ component, and for that
we only need to make the replacement \mbox{$p_z\to p_z E_p/| \vec{p}\,|$} in Eq.~\eqref{eq:m-pplus}:
\begin{equation}
p_E^+ =E_p + \frac{E_p}{| \vec{p} \,|} \,p_z = \frac{E_p}{| \vec{p}\,|}\, p_P^+ = p_P^+ +\mathcal{O} (\lambda^4) \,,
\end{equation}
where in the last step we again have used $E_p/|\vec{p}\,|=1+ \mathcal{O}(\lambda^2)$. Since for collinear particles (in any direction) $p_P^\mu=p_E^\mu$ at leading power, we can safely
conclude that at this order the collinear measurements for all event shapes take the same form in the P- and \mbox{E-schemes}, but is in general different from the M-scheme. In what follows we will
work out the collinear measurement for a few event shapes.

\subsubsection*{Thrust}
The original thrust definition is already in the P-scheme and reads
\begin{equation}
\tau^P = \frac{1}{Q_P}\min_{\hat t}\sum_i(|\vec p_i| - |\hat t\cdot \vec p_i|)\,,
\end{equation}
with $\hat t$ the thrust axis. For $n$-collinear particles one has $|\hat t\cdot \vec p_i|=-p_z$ and therefore up to power corrections we have
\begin{equation}
Q \tau^P_c=Q \tau^E_c=\sum_{i\in +} p_{P,i}^+ = \sum_{i\in +} \biggl(p_i^+ - \frac{m_i^2}{p_i^-} \biggr) \,,
\end{equation}
where we already indicate that the collinear measurement is the same in the E-scheme. In Ref.~\cite{Stewart:2009yx} an M-scheme generalization of thrust, dubbed 2-jettiness,
was introduced
\begin{equation}
\tau^J = \frac{1}{Q}\min_{\hat t}\sum_i(E_i - |\hat t\cdot \vec p_i|)\,,
\end{equation}
such that its collinear limit is the total plus momentum flowing into the plus hemisphere \mbox{$Q \tau^J_c=\sum_{i\in +} p_i^+$}. Since the measurement is completely inclusive, the computation of the
jet function can be carried out as the imaginary part of a forward-scattering matrix element. This is not the case for the E- and P-schemes if quark masses are non-vanishing.

\subsubsection*{Hemisphere Jet Masses}
We already worked out the collinear measurement for heavy jet mass in Eq.~\eqref{eq:rho-col}, and getting the P-scheme measurement is equally simple since Eq.~\eqref{eq:rho+-} still applies
with minimal modifications:
\begin{equation}\label{eq:rhoP}
Q^2 \rho_+^P = \biggl(\sum_{i \in +} |\vec p_i|\biggr)^{\!\!2} - \biggl(\sum_{i \in +} p^z_i\biggr)^{\!\!2} =
\biggl[\,\sum_{i \in +}( |\vec p_i| + p^z_i )\biggr]\! \biggl[\,\sum_{j \in +}( |\vec p_j| - p^z_j )\biggr] =
\sum_{i \in +} p_{P,i}^+ \sum_{j \in +} p_{P,j}^-\,,
\end{equation}
where we again use that the total perpendicular momentum vanishes and use the identity $a^2-b^2=(a+b)(a-b)$. With this result we can trivially obtain the collinear measurement using that
$p_{P,j}^- = 2 E_j + \mathcal{O}(\lambda^2)$:
\begin{equation}\label{eq:rhoE}
Q\rho_{c,+}^P = Q\rho_{c,+}^E = \sum_{i\in +} \biggl(p_i^+ - \frac{m_i^2}{p_i^-} \biggr) \,,
\end{equation}
that matches the P-scheme thrust result. The total perpendicular momentum does not vanish in the E-scheme, since there is not such thing as E-scheme three-momentum conservation
(in the same way, P-scheme energy is not conserved either). However, in the dijet limit the perpendicular components are already $\mathcal{O}(\lambda)$ and therefore
$p^{\perp}_E = p^{\perp}+ \mathcal{O}(\lambda^3)$, making $\sum_{i \in +} p^{\perp}_{E,i} \propto \mathcal{O}(\lambda^3)$, thence power suppressed, such that the result in
Eq.~\eqref{eq:rhoE} is also valid for the E-scheme.

\subsubsection*{C-parameter}
In Ref.~\cite{Hoang:2014wka} it was shown how the C-parameter measurement splits into the sum of soft and collinear contributions in the dijet limit. The proof relied on the particles being massless,
so it cannot be taken for granted that it will work when quarks have a non-zero mass. Here we carry out a similar proof valid for massive particles as well. C-parameter is defined already in the P-scheme as
\begin{align}
C^P = &\,\frac{3}{2Q_P^2}\sum_{i,j}|\vec p_i||\vec p_j|\sin^2(\theta_{ij}) =
\frac{3}{2Q_P^2}\sum_{i,j} |\vec{p}_i | | \vec{p}_j | [1 + \cos (\theta_{ij})] [1 - \cos(\theta_{ij})]\\
= &\, \frac{3}{2Q_P^2}\sum_{i,j} \frac{(| \vec{p}_i || \vec{p}_j | + \vec{p}_i \!\cdot\! \vec{p}_j)
(| \vec{p}_i	|| \vec{p}_j | - \vec{p}_i\! \cdot\! \vec{p}_j)}{| \vec{p}_i | | \vec{p}_j |} \,,\nonumber
\end{align}
where we simply use $\sin^2(\theta_{ij}) = 1 - \cos^2(\theta_{ij})=[1 + \cos (\theta_{ij})] [1 - \cos(\theta_{ij})]$ and the definition of the euclidean scalar
product to get to the final form. Next one can express the result in terms of P-scheme light-cone coordinates using Eq.~\eqref{eq:light-cone} as follows
\begin{align}
C^P = &\,\frac{3}{2Q_P^2}\sum_{i,j}\frac{(2| \vec{p}_i || \vec{p}_j | - p_{P, i} \!\cdot\! p_{P, j})\, p_{P, i} \!\cdot \!p_{P, j}}{| \vec{p}_i | | \vec{p}_j |} \\
= &\,\frac{3}{2Q_P^2}\sum_{i, j} \frac{(p_{P, i}^+\, p_{P, j}^+ + p_{P, i}^-\, p_{P, j}^- + 2
\vec{p}_{\perp, i} \!\cdot \vec{p}_{\perp, j}) (p_{P, i}^+ \,p_{P, j}^- + p_{P,i}^- \,p_{P, j}^+ - 2 \vec{p}_{\perp, i} \!\cdot \vec{p}_{\perp, j})}{(p_{P, i}^+
+ p_{P, i}^-) (p_{P, j}^+ + p_{P, j}^-)}\,,\nonumber
\end{align}
with a similar result in the E-scheme. Arguments analogous to those used in Ref.~\cite{Hoang:2014wka} apply, and we focus on the collinear measurement only. First consider that both
$i$ and $j$ are $n$-collinear such that the SCET scaling implies
\begin{equation}\label{eq:Cnn}
C^P_{nn}= \frac{3}{Q^2}\biggl[\biggl(\sum_{i\in n} p^+_{P, i}\biggr)\!\biggl(\sum_{j\in n} p^-_{P, j}\biggr)\!-
\biggl(\sum_{i\in n} \vec{p}_{\perp, i}\biggr)\!\cdot\!\biggl(\sum_{j\in n} \vec{p}_{\perp, j}\biggr)\! \biggr] 
= \frac{3}{Q}\sum_{i\in n} p^+_{P, i} + \mathcal{O}(\lambda^3)\,,
\end{equation}
where once again we use that the total collinear perpendicular momenta flowing into the plus hemisphere is zero up to power corrections. One gets an analogous result for $C^P_{\bar n\bar n}$, while
if $i$ is $n$-collinear and $j$ is $\bar n$-collinear we get (we already include a factor of $2$ to account for the case in which $i$ is $\bar n$-collinear and $j$ is $n$-collinear)
\begin{align}\label{eq:Cnnbar}
C^P_{n\bar n}\,=\,& \frac{3}{Q^2}\biggl[\!\biggl(\sum_{i\in n} p^+_{P, i}\biggr)\!\biggl(\sum_{j\in \bar n} p^+_{P, i}\biggr)\!
+\! \biggl(\sum_{i\in n} p^-_{P, i}\biggr)\!\biggl(\sum_{j\in \bar n} p^-_{P, i}\biggr) \!+
2\biggl(\sum_{i\in n} \vec{p}_{\perp, i}\biggr)\!\cdot\!\biggl(\sum_{j\in \bar n} \vec{p}_{\perp, j}\biggr) \!\biggr] \nonumber\\
\,=\, & \frac{3}{Q}\biggl(\sum_{i\in n} p^+_{P, i}+\sum_{i\in \bar n} p^-_{P, i}\biggr) \!+ \mathcal{O}(\lambda^3)\,.
\end{align}
Summing up the contributions in Eqs.~\eqref{eq:Cnn} and \eqref{eq:Cnnbar} we obtain the collinear measurement in the P- and E-schemes:
\begin{equation}
Q C^P_c=Q C^E_c=6\sum_{i\in +} p_{P,i}^+ = 6\sum_{i\in +} \biggl(p_i^+ - \frac{m_i^2}{p_i^-} \biggr) \,,
\end{equation}
identical to that of thrust or the hemisphere masses up to a factor of $6$. The computation for the E-scheme is identical, relies on arguments already exposed, and therefore will not be repeated.
For the M-scheme C-jettiness variable introduced in Ref.~\cite{Gardi:2003iv} it was shown in Ref.~\cite{Preisser:mthesis} that the collinear measurement takes the simple form
$Q C^J_c=6\sum_{i\in +} p_i^+$. Therefore, for the reduced \mbox{C-parameter} variable $\widetilde{C}\equiv C/6$ the collinear measurement for the three event shapes we consider coincide in
every massive scheme.

\section{Factorization Theorems}\label{sec:factorization}
For simplicity we consider the well-known case of thrust (in either massive scheme), which can be easily modified to obtain the corresponding factorized results for C-parameter (changing the soft function and taking into account factors of $6$ in the measurement function) or heavy jet mass (using the hemisphere jet and
soft functions, and including two convolutions, one per hemisphere).
After having shown that in the three schemes considered Eq.~\eqref{eq:es_sum} holds, the derivation of the factorization theorem is obtained following the steps outlined in Ref.~\cite{Bauer:2008dt}.
The leading hadronization corrections (which are soft) can also be factorized as an extra convolution with the so-called shape function, and even though they are included in our numerical analysis, for
the sake of conciseness we ignore them in this section. Likewise, we include kinematic and mass power corrections in our final analysis, but postpone their discussion until Sec.~\ref{sec:power}

\subsection{SCET}
The value of the quark mass can have different hierarchies with respect to the (EFT) hard, jet, and soft scales. These are the natural scales of the various matrix
elements appearing in the factorization theorem shown in Eq.~\eqref{eq:FacMom}, that is, the scales that produce no large logarithms. They can be identified with physical
scales: $\mu_H$ is similar to the center-of-mass energy $Q$, $\mu_J$ can be associated to the perpendicular momentum of a jet with respect to its axis, and
$\mu_S$ is of the order of the soft-particles' energies.
In Refs.~\cite{Gritschacher:2013pha,Pietrulewicz:2014qza} it was extensively
discussed how to setup a consistent variable-flavor number scheme for final-state jets accounting for primarily and secondarily produced massive quarks. Four scenarios can be defined for the cases
in which $m>\mu_H$ (scenario I, which is of no interest for primary quarks since there is no energy to produce them), $\mu_H> m > \mu_J$ (scenario II, relevant for very boosted heavy quarks, and
better described in bHQET), $\mu_J> m > \mu_S$ (scenario III) and $m < \mu_S$ (scenario IV). Each of them has a different factorization theorem and renormalization group evolution setup. Even
though the heavy quark mass is a fixed parameter, the jet and soft scales depend on the event-shape value and therefore they change along the spectrum, such that several scenarios might occur
in a given distribution. For simplicity, we assume the quark mass is always smaller than the soft scale, such that we stay in scenario IV even in the peak of the distribution.\footnote{Strictly speaking, the SCET counting $m\sim Q\lambda$ only applies in Scenarios II and III: in Scenario I no massive collinear particle can be produced while in IV the mass can be power-counted away. In practice the $m\to 0$ limit is smooth and using the counting $m\lesssim Q\lambda$ in
Scenario IV captures some power corrections that make the transition between Scenarios III and IV smooth.} In this way, we avoid having
to deal with integrating out the heavy quark mass and the partonic factorization formula reads\footnote{Extending our results to take into account different scenarios poses no difficulty.}
\begin{equation}\label{eq:FacMom}
\frac{1}{\sigma_0} \frac{\text{d} \hat\sigma_{\rm SCET}}{\text{d} \tau} = Q^2 H (Q, \mu)\!
\!\int_0^{Q(\tau-\tau_{\rm min})}\!\! \text{d} \ell\, J_{\tau} (Q^2 \tau - Q \ell, \mu) S_{\tau} (\ell, \mu)\,,
\end{equation}
with $\sigma_0$ the Born or point-like (massless) cross section, $H$ and $S_{\tau}$ the hard and soft functions, respectively, and
\begin{equation}\label{eq:2-hem-jet}
J_{\tau} (s, \mu) \equiv\! \int_{s_{\rm min}}^{s-s_{\rm min}}\! \text{d} s' J_n (s - s', \mu) J_n (s',\mu)\,,
\end{equation}
the thrust jet function, which is the convolution of two single-hemisphere jet functions. The M-scheme hemisphere jet function has support for $s>s_{\rm min}=m^2$, what sets
the integration limits in Eq.~\eqref{eq:2-hem-jet}. Accordingly, the thrust jet function has support for $s>2s_{\min}$, implying that the minimal value for 2-jettiness is
$\tau^J_{\rm min} = 2\hat m^2$. We shall present the computation of the M- and P-scheme SCET jet function in Sec.~\ref{sec:computation}. The definition of the soft function in terms of Wilson lines can be found
e.g.\ in Ref.~\cite{Bauer:2008dt} and the corresponding expression for the jet function will be given in Sec.~\ref{sec:computation}. The factorization
formula takes a simpler form in Fourier space
\begin{equation}\label{eq:FacPos}
\frac{1}{\sigma_0} \frac{\text{d} \hat\sigma_{\rm SCET}}{\text{d} \tau} = \frac{Q}{2 \pi} H (Q, \mu)\!\! \int\! \text{d} x\, e^{i x p} \tilde{J}_{\tau}\!\biggl( \frac{x}{Q}, \mu \biggr) \tilde{S}_{\tau} (x, \mu) \,,
\end{equation}
with $p=Q (\tau - \tau_{\rm min})$ and $\tilde{J}_{\tau}$ and $ \tilde{S}$ the Fourier transforms of the jet and soft functions, respectively. The thrust jet function in position space is the square of its
hemisphere counterpart, and can be computed as follows
\begin{equation}\label{eq:jet-fourier}
\tilde{J}_\tau (y, \mu) = \!\int_{0}^{\infty} \!\!\text{d} \bar s\, e^{- i \bar s y}
J_\tau (\bar s + s_{\min}, \mu) = \tilde{J}_n (y, \mu)^2\,.
\end{equation}
In Eqs.~\eqref{eq:FacMom} and~\eqref{eq:FacPos} all matrix elements are evaluated at the same renormalization scale $\mu$. In order to minimize large logarithms that appear in each of
them one should use RGE equations to evaluate them at their respective natural scales, denoted by $\mu_H\sim Q$, $\mu_J\sim Q\sqrt{\tau}$ and $\mu_S\sim Q\tau$, such that for small $\tau$
there is a strict hierarchy among those: $\mu_H> \mu_J > \mu_S$ and the SCET scaling parameter takes the value $\lambda\sim \sqrt{\tau}$. The form and solution of the renormalization group
equations is also simpler in position space. 
Using those and changing variables to $y=x/p$ one arrives at
\begin{equation}\label{eq:FacPosSum}
\frac{1}{\sigma_0} \frac{\text{d} \hat\sigma_{\rm SCET}}{\text{d} \tau} = \frac{H (Q,\mu_H)}{p} \biggl( \frac{e^{\gamma_E} \mu_S}{p} \biggr)^{\!\!\tilde{\omega}} R
(Q, \mu_i) \!\int\! \frac{\text{d} y}{2 \pi} \,e^{i y} (i\,y)^{\tilde{\omega}} \tilde{J}_{\tau}\! \biggl( \frac{y}{Q p}, \mu_J \biggr)
\tilde{S}_{\tau}\! \biggl( \frac{y}{p}, \mu_S \biggr) \,,
\end{equation}
where $\mu_i$ denotes collectively all renormalization scales (including the common $\mu$) and we use the following compact notation
\begin{align}
R (Q, \mu_i) =\,& Q \biggl( \frac{\mu_H}{Q} \biggr)^{\!\!-2\tilde{\omega}_H} e^{\tilde{k}} \biggl( \frac{\mu^2_J}{Q \mu_S}\biggr)^{\!\!\tilde{\omega}_J}\,, & &\nonumber\\
\tilde{k} =\, & \tilde{k}_H + \tilde{k}_J + \tilde{k}_S\,, & \tilde{\omega}& = \tilde{\omega}_J - 2 \tilde{\omega}_S\,,\nonumber\\
\tilde{\omega}_S =\, & \tilde{\omega}_{\Gamma_c} (\mu_S, \mu),
&\tilde{\omega}_J &= \tilde{\omega}_{\Gamma_c} (\mu_J, \mu),
\\
\tilde{\omega}_H =\, & \tilde{\omega}_{\Gamma_c} (\mu_H, \mu)\,,
& \tilde{k}_S& = \tilde{\omega}_{\gamma_S} (\mu_S, \mu) -
2 \tilde{k}_{\Gamma_c} (\mu_S, \mu)\,, \nonumber\\
\tilde{k}_H =\, & \tilde \omega_{\gamma_H} (\mu_H, \mu) - 2
\tilde{k}_{\Gamma_c} (\mu_H, \mu)\,, & \tilde{k}_J &=
\tilde{\omega}_J (\mu_J, \mu) + 4 \tilde{k}_{\Gamma_c} (\mu_J, \mu) \,,
\nonumber
\end{align}
with $\tilde{\omega}$ and $\tilde{k}$ the exponential running kernels defined in terms of integrals over the SCET and QCD anomalous dimensions as follows
\begin{align}\label{eq:kernel}
\tilde{\omega}_{\gamma} (\mu_0, \mu) =\, & 2 \!\int^{\alpha_\mu}_{\alpha_0} \!\text{d} \alpha\, \frac{\gamma
(\alpha)}{\beta_{\rm QCD} (\alpha)}\,, \\
\tilde{k}_{\gamma} (\mu_0, \mu) =\, & 2 \!\int^{\alpha_{\mu}}_{\alpha_0}\!
\text{d} \alpha \,\frac{\gamma (\alpha)}{\beta_{\rm QCD} (\alpha)}
\int^{\alpha}_{\alpha_0} \text{d} \alpha' \frac{1}{\beta_{\rm QCD}
(\alpha')} \,. \nonumber
\end{align}
Here $\gamma$ can refer to cusp or non-cusp anomalous dimensions, and their dependence on $\alpha$ is in the form of perturbative series that define their
respective coefficients
\begin{equation}
\!\!\!\beta_{\rm QCD} (\alpha) = - 2\, \alpha_s\! \sum_{n = 1} \beta_{n - 1} \Bigl( \frac{\alpha}{4 \pi}\Bigr)^{\!\!n} ,\quad 
\Gamma_{\rm cusp}(\alpha) =\!\sum_{n = 1} \Gamma_n \Bigl( \frac{\alpha}{4\pi} \Bigr)^{\!\!n} , \quad
\gamma(\alpha) =\! \sum_{n = 1} \gamma_n \Bigl(\frac{\alpha}{4 \pi} \Bigr)^{\!\!n} .
\end{equation}
The integrals in Eq.~\eqref{eq:kernel} can be solved analytically in terms of the anomalous-dimension coefficients if an expansion in $\alpha_s$ is carried out.
Their explicit form up to N$^3$LL can be found e.g.\ in Ref.~\cite{Abbate:2010xh}. General expressions valid for arbitrarily high order can also be derived and will
be given elsewhere.

The jet function of a massive quark contains terms which are distributions, and hence easy to Fourier transform, plus others which are regular functions, and
to the best of our knowledge it seems impossible to find an analytic expression in position space for them. Up to one loop, the momentum-space hemisphere jet function
can be decomposed in the following form:
\begin{align}\label{eq:jetDecomposition}
J_n(\bar s+s_{\rm min},\mu) =\,& \delta(\bar s) + \frac{\alpha_s(\mu)}{4\pi} C_F \Bigl[J_{\rm dist}(\bar s,\mu) +
\frac{1}{m^2}J_{\rm nd}\Bigl(\frac{\bar s}{m^2}\Bigr)\Bigr] + \mathcal{O}(\alpha_s^2) \,,\\
J_{\rm dist}(\bar s,\mu) =\,& \frac{1}{\mu^2}J_{m=0}\Bigl(\frac{\bar s}{\mu^2}\Bigr) + \frac{1}{m^2} J_m\Bigl(\frac{\bar s}{m^2}\Bigr) \,,\nonumber
\end{align}
where the massive corrections, either with distributions $J_m$ or fully non-distributional $J_{\rm nd}$, are $\mu$-independent dimensionless functions with support for positive values of
their (dimensionless) arguments. The $\mu$ dependence of $J_{\rm dist}$ is entirely determined from the jet and QCD anomalous dimensions, does not depend on the quark mass, and
therefore can be fully accounted for in the massless jet function of Eq.~\eqref{eq:jetDecomposition}. The only piece that needs an explicit computation in the E- and P-schemes is $J_{\rm nd}$,
since the rest can be obtained using consistency conditions and results obtained in Refs.~\cite{Lepenik:2019jjk} and \cite{Fleming:2007xt}.

The integral in Eq.~\eqref{eq:FacPosSum} can be easily solved for all terms involving only distributions, and generic formulas can be found for instance in Ref.~\cite{Becher:2008cf}. For the
non-distributional piece of the jet function we carry out resummation in momentum space, and at one loop it is multiplied by the hard and soft functions at tree-level only. Therefore,
using Eq.~\eqref{eq:jet-fourier} in \eqref{eq:FacPosSum} and carrying out the $y$ integration, the non-distributional part of the 1-loop partonic cross section reads
\begin{align}\label{eq:ndRun}
\frac{1}{\sigma_0} \frac{\text{d} \hat\sigma_{\rm nd}}{\text{d} \tau} =\, & \frac{C_F\alpha_s(\mu_J)}{2\pi}\frac{R(Q, \mu_i)}{\hat m^2\Gamma (- \tilde{\omega})} (Q e^{\gamma_E}
\mu_S)^{\tilde{\omega}}\! \int_{2s_{\rm min}}^{Q^2 \tau}\!\! \text{d} s\, J_{\rm nd} \Bigl(\frac{s-2s_{\rm min}}{m^2}\Bigr) (Q^2 \tau - s)^{- 1 - \tilde{\omega}}\nonumber\\
=\, & \frac{C_F\alpha_s(\mu_J)}{2\pi}\frac{R(Q, \mu_i)}{\hat m^2} \biggl[ \frac{\mu_S e^{\gamma_E}}{Q(\tau-\tau_{\rm min})} \biggr]^{\!\tilde{\omega}}
I_{\rm nd}\Bigl(\tilde \omega, \frac{\tau - \tau_{\rm min}}{\hat m^2}\Bigr)\, ,\\
I_{\rm nd}(\tilde \omega, y)=\, & \frac{y^{\tilde \omega}}{\Gamma (- \tilde{\omega})}\int_0^y \text{d} x\, (y - x)^{- 1 - \tilde{\omega}} J_{\rm nd}(x)
= \frac{1}{\Gamma (- \tilde{\omega})}\int_0^1 \text{d} z \,(1 - z)^{- 1 - \tilde{\omega}} J_{\rm nd} (z y)\,.\nonumber
\end{align}
The lower limit of integration in the first line has been moved to $2s_{\rm min}$ since below that value the jet function has no support. In the E- and P-schemes
$s_{\min} = 0$ so we have not lost any generality. To get to the second line we have switched variables in the integral to $s=x m^2 + 2s_{\rm min}$, and 
to obtain the second expression for $I_{\rm nd}(\tilde \omega, y)$ we switch variables to $x = zy$. For the partonic cumulative distribution one gets instead
\begin{align}
\!\hat\Sigma_{\rm nd}(\tau_c) \,&\equiv \frac{1}{\sigma_0}\!\int_0^{\tau_c}\!\! {\rm d}\tau\, \frac{\text{d} \sigma_{\rm nd}}{\text{d} \tau}\\
\, &=
\frac{C_F\alpha_s(\mu_J)}{2\pi}\frac{R(Q, \mu_i)}{\hat m^2} \frac{\mu_S e^{\gamma_E}}{Q}\!\biggl[ \frac{\mu_S e^{\gamma_E}}{Q(\tau_c - \tau_{\rm min})} \biggr]^{\tilde{\omega}-1}
I_{\rm nd}\Bigl(\tilde \omega - 1, \frac{\tau_c - \tau_{\rm min}}{\hat m^2}\Bigr).\nonumber
\end{align}
To make the function $I(\tilde \omega, y)$ smooth in the no resummation limit, achieved when $\tilde \omega\to 0$, one can integrate by parts to obtain
\begin{equation}
I_{\rm nd}(\tilde \omega, y) = \frac{1}{\Gamma (1 - \tilde{\omega})}\biggl[ \,y\!\! \int_0^1 {\rm d} z (1 - z)^{- \tilde{\omega}} J_{\rm nd}^\prime(z y) + J_{\rm nd} (0)\biggr]\,,
\end{equation}
with $J_{\rm nd}^\prime$ the derivative of the $J_{\rm nd}$ function. This form is particularly useful if the integration has to be carried out numerically, making it more convergent
and defining its analytic continuation to values $0 < \tilde \omega < 1$. Further integration by parts can be implemented to define the integral for even larger values of $\tilde \omega$.
If a closed analytical form is found, this procedure is unnecessary.

Although the discussion in this section has been carried out assuming the pole mass for the heavy quark, it is straightforward to convert the result to a short-distance scheme.
In the scenarios in which SCET applies, the $\MSb$ scheme is perfectly adequate. In scenario II it is more convenient to employ low-scale short-distance schemes such as
the MSR mass~\cite{Hoang:2008yj,Hoang:2017suc}.

\subsection{bHQET}\label{sec:bHQET}
If the heavy quark mass is large enough or if the jet is very narrow one enters scenario~II, in which the jet and heavy quark masses are close to each other, corresponding to very
boosted quarks. In this kinematic situation, a new physical scale emerges $\mu_B\sim Q\tau/{\hat m}\sim \mu_S/{\hat m}$, such that there is a new hierarchy between scales:
$\mu_H > m > \mu_B > \mu_S$.\footnote{In practice one can still identify $\mu_B$ with the jet scale ($\mu_B=\mu_J$), and we will do so in what follows.} A practical way
to see how this becomes manifest is looking at the structure of the one-loop jet function in Eq.~\eqref{eq:jetDecomposition}. Since the non-distributional terms are power
suppressed when $s\to s_{\rm min}$, it is enough to focus on the terms with distributions, which generically read
\begin{align}\label{eq:jet-generic}
\frac{1}{\mu^2}J_{m=0}\Bigl(\frac{\bar s}{\mu^2}\Bigr)=\,& A\,\delta(\bar s) + \frac{B}{\mu^2} \biggl[\frac{\mu^2}{\bar s}\biggr]_+
+ \frac{C}{\mu^2} \biggl[\frac{\mu^2\log(\bar s/\mu^2)}{\bar s}\biggr]_+\,,\\
\frac{1}{m^2}J_m\Bigl(\frac{\bar s}{m^2}\Bigr) =\,& A_m\,\delta(\bar s) + \frac{B_m}{m^2} \biggl[\frac{m^2}{\bar s}\biggr]_+
+ \frac{C_m}{m^2} \biggl[\frac{m^2\log(\bar s/m^2)}{\bar s}\biggr]_+\,,\nonumber
\end{align}
where the coefficients $A,\,B$ and $C$ (with or without subindex $m$) depend neither on $\mu$ nor on $m$. In scenarios III and IV one has $\bar s \lesssim m^2$ and therefore the choice
$\mu^2\sim \bar s$ makes sure there are no large logarithms in neither term (the massless limit is smooth since $J_m+J_{\rm nd}\to 0$ when $m\to 0$ and no new class of large logarithms emerges).
On the other hand, if $\bar s \ll m^2$ the choice $\mu^2\sim \bar s$ cannot prevent the logarithms in $J_m$ from becoming large.\footnote{The bHQET limit should not be confused with the threshold
limit, yet another interesting physical situation in which radiation other than the heavy quark is soft as compared with the quark mass, while there is no hierarchy between $m$ and $Q$.}

The massive shell of the heavy quark carrying momentum $p=m v + k$ with $v^2=1$ gets integrated out as a dynamical degree of freedom giving raise to 
heavy-quark effective theory~\cite{Eichten:1989zv,Isgur:1989vq,Isgur:1989ed,Georgi:1990um}. The remaining degrees of freedom are referred to as ultracollinear and carry residual momentum
$k$. In the heavy quark rest frame [\,in which $v^\mu=(1,1,\vec{0}_{\perp})$\,] they are soft $k^\mu = \Delta (1,1,1)$, with $\Delta \ll m$ a low-energy scale,\footnote{For an unstable top quark,
this scale is of the order of its width $\Delta\sim\Gamma$, but for a stable bottom quark it can be identified with $\Delta\sim Q^2\tau/m\sim (Q^2\tau_J - 2m^2)/m$ for thrust and 2-jettiness,
respectively.} being able to interact with each other and with color sources representing the integrated-out heavy quarks. In the center-of-mass frame these momenta get boosted
and a hierarchy is generated among their light-cone components\footnote{If only two back-to-back quarks are produced, their velocity equals
$\beta = \sqrt{1-4\hat m^2}\simeq 1-2\hat m^2$, and therefore the boost factor reads $\gamma=1/(2\hat m)$. When boosting momenta in light-cone coordinates the plus/minus
components get multiplied/divided by $\gamma(1-\beta)\simeq 1/\hat m$.}
\begin{align}\label{eq:bhqet-count}
v_{+}^{\mu}=\,&\biggl(\frac{m}{Q}, \frac{Q}{m}, \vec{0}_{\perp}\!\biggr)\,, &k_{+}^{\mu}& \sim \Delta\biggl(\frac{m}{Q}, \frac{Q}{m}, 1\!\biggr)\,,
&q_s^\mu& = \frac{m\Delta}{Q}(1,1,1)\,,\\
v_{-}^{\mu}=\,&\biggl(\frac{Q}{m}, \frac{m}{Q}, \vec{0}_{\perp}\!\biggr)\,, &k_{-}^{\mu}& \sim \Delta\biggl(\frac{Q}{m}, \frac{m}{Q}, 1\!\biggr)\,,&&\nonumber
\end{align}
where we have included momenta $q_s$ which is soft in the center-of-mass frame. The two boosted copies of HQET are matched onto SCET in order to account for global soft radiation, such that the
heavy quark and ultracollinear particles can interact with soft degrees of freedom. The typical off-shellness of ultracollinear particles is softer than for collinear degrees of freedom which are part
of SCET.

Using this framework, it is possible to derive a factorization theorem for the partonic cross section~\cite{Fleming:2007qr} which effectively separates physics at the different involved
scales.
\begin{equation}
\!\!\!\frac{1}{\sigma_0} \frac{\text{d} \hat\sigma_{\rm bHQET}}{\text{d} \tau} \!=\! Q^2 H (Q,\mu_m) H_m\! \biggl( \!m, \frac{Q}{m}, \mu_m, \mu \!\biggr) \!\!\!\int \!\!\text{d} \ell\,
B_{\tau}\! \biggl( \!\frac{Q^2 (\tau - \tau_{\rm min}) - Q \ell}{m}, \mu \!\biggr)\! S_{\tau}(\ell, \mu),
\end{equation}
where the hard and soft functions are the same as in the SCET factorization theorem, but there is an additional matching coefficient $H_m$ between SCET and bHQET. The jet
function $B_{\tau}(\hat s)$ is different from $J_\tau$ in SCET, has support for $\hat s > 0$, its mass dependence is only through a global $1/m$ factor and contains only
distributions. It is also the convolution of two hemisphere bHQET jet functions $B_n$:
\begin{equation}
B_{\tau} (\hat{s}, \mu) = m\! \int_0^{\hat{s}}\! \text{d} \hat{s}' B_n (\hat{s}- \hat{s}', \mu) B_n (\hat{s}', \mu) \,,
\end{equation}
whose operator definition shall be given in Sec.~\ref{sec:jet-bHQET}. Since $H$ and $H_m$ are the same for all event shapes and $S_\tau$ does not depend on the quark mass (it sees only
light degrees of freedom), the anomalous dimension of the $B_n$ function is the same in any massive scheme. In turn this implies that all terms in the jet function except for the Dirac delta
are fixed by consistency and hence are the same in the M-, P- and E-schemes. Knowing $H$, $H_m$ and $S_\tau$ at one loop, the delta function coefficient in $B_n$ can be
obtained taking the $\hat m\to0$ limit of the result quoted in Ref.~\cite{Lepenik:2019jjk} for the full QCD prediction of the threshold delta function coefficient. In this sense, our
computation in Sec.~\ref{sec:jet-bHQET} will be just a sanity check. The $H_m$ matching coefficient and bHQET jet function satisfy
$J_\tau(s+s_{\rm min}) = H_m B_\tau(s/m) [1 + \mathcal{O}(s/m^2)]$, such that both factorization theorems smoothly join. For the various arguments exposed in this paragraph,
this also implies that the coefficients $B_m$ and $C_m$ in Eq.~\eqref{eq:jet-generic} do not depend on the massive scheme, and they are known from the 2-jettiness computation
of Ref.~\cite{Fleming:2007xt}. Therefore, different massive schemes can differ only in $A_m$ and the $J_{\rm nd}$, but knowing the one-loop hard and soft functions, $A_m$ can
be obtained again from the massless limit of the full-QCD threshold result.

Carrying out resummation in bHQET is identical to massless SCET. Since we will limit our numerical analysis to situations in which it is sufficient to use SCET with masses,
we will not give further details on how to solve the corresponding RGE equations, which can be found elsewhere. Moreover, we will not provide a detailed discussion on how to
switch to a short-distance mass scheme in this setup.

\section{SCET Jet Function Computation}\label{sec:computation}
The jet function accounts for the dynamics of collinear particles within the hemisphere. Since the collinear measurement function in the P- and E-schemes is not the total plus
momentum, it cannot be computed as the discontinuity of a forward-scattering amplitude, as was done in~\cite{Fleming:2007qr,Fleming:2007xt}. Instead, one has to use the
definition given in Ref.~\cite{Bauer:2008dt}, which after a small modification to match the form of our factorization theorem and minimal manipulations can be cast into the form:
\begin{align}\label{eq:jet-def}
J_{n}(s,\mu) &= \!\int \!\frac{{\rm d}\ell^+\!}{2\pi}\, \mathcal{J}_n(s,\ell^+)\,, \\
\mathcal{J}_n(s,\ell^+) & = \frac{1}{4N_c} {\rm Tr}\!\!\int\! {\rm d}^dx\, e^{i\ell x} \bra{0}\slashed{\nb}\chi_{n}(x)\, \delta(s-Q^2 \hat{e}_n)\, \overline{\chi}_{n,Q}(0)\!\ket{0}\,,\nonumber
\end{align}
with $\chi_{n,Q}$ the jet field with total minus momentum equal to $Q$, $d=4-2\varepsilon$ the space-time dimension in dimensional regularization, $\ell^-=Q$ and
$\vec{\ell}_\perp=\vec{0}$ due to label momentum conservation, and the trace is taken over spin and color indices. The $n$-collinear event-shape operator $\hat{e}_n$ acting on
some final state $\ket{X}$ pulls out $e_n(X)$, the contribution from $n$-collinear particles to the value of the event shape:
$\hat{e}_n\ket{X}=e_n(X)\ket{X}$. To simplify the expression in the second line of Eq.~\eqref{eq:jet-def} we insert the identity $I = \sum_X \ket{X}\!\bra{X}$ after the delta function and shift
the field $\chi_{n,Q}(x)$ to $x=0$ employing the momentum operator. Using the label operators for the large components of the momenta, the sum over $X$ can be carried out
and we obtain the following convenient expression:
\begin{equation}\label{eq:Jet-cut}
J_n(s,\mu) = \frac{(2 \pi)^{d - 1}}{N_C} \mathrm{Tr} \biggl[ \frac{\overline{n} \! \! \! /}{2} \langle 0 | \chi_n (0) \delta(s - Q\hat{e}_n)
\delta^{(d - 2)} (\vec{\mathcal{P}}_X^{\perp}) \delta (\mathcal{\bar P} - Q) \bar{\chi}_n (0) | 0 \rangle \biggr]\,.
\end{equation}
For practical computations one inserts a complete set of states after $\delta (\mathcal{\bar P} - Q)$
\begin{equation}
\sum_X \ket{X}\!\bra{X} \equiv \sum_{n = 1}\,\sum_{\rm spin} \,\int\prod_{i = 1}^n \frac{{\rm d}^{d - 1} \vec{p}_i}{(2 \pi)^{d -1} (2 E_i)} | X_n \rangle \langle X_n |\,,
\end{equation}
where we exclude the vacuum from the sum because it does not contribute to the jet function. Each term in the sum over $n$ can include several contributions, accounting for various
particle species (heavy or light quarks and gluons), and the sum over polarizations affects all particles in the final state. The perturbative expansion of the jet function in powers of
$\alpha_s$ is obtained by adding more particles to the sum as well as more virtual (loop) contributions to the matrix elements that appear after inserting the identity, which in
compact form can be written as
\begin{equation}
J_n(s,\mu) =\frac{(2 \pi)^{d - 1}}{N_C} \sum_X \delta^{(d - 2)} (\vec{p}_X^{\perp}) \delta
(p_X^- - Q) \delta [s - Q^2e_n (X)] {\rm Tr}\! \biggl[ \gamma^0
\frac{\overline{n} \! \! \! /}{2} \left| \left\langle \: 0| \chi_n (0) | X \: \right\rangle \right|^2 \biggr]\,.
\end{equation}
For the computation of the P-scheme hemisphere jet function one does not need any regularization beyond taking the space-time dimension from $4$ to $d=4-2\varepsilon$.
In the following we carry out the computation of the jet function using Eq.~\eqref{eq:Jet-cut} for both 2-jettiness and P-scheme thrust. Although the result is already known for the
former, it is instructive to repeat its computation to highlight the differences between the two approaches. In a way, the computation that uses Eq.~\eqref{eq:Jet-cut} can be obtained
applying Cutkosky rules to directly obtain the imaginary part of the forward-scattering matrix element.

\begin{figure}[t!]\centering
\includegraphics[width=0.6\textwidth]{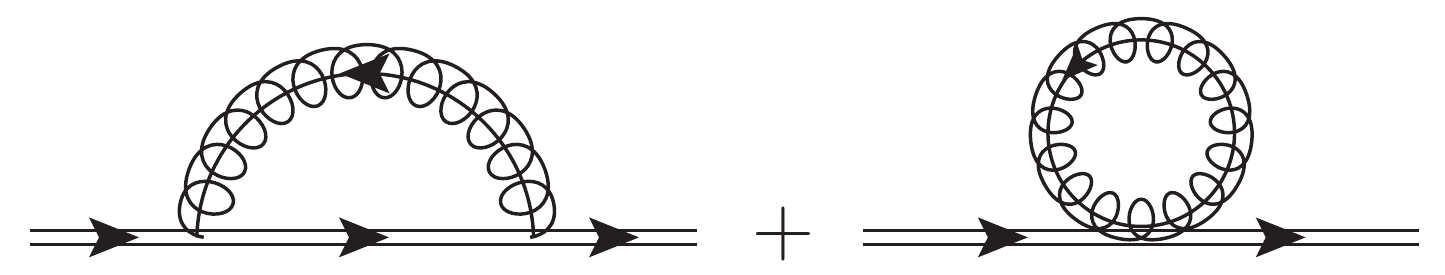}
\caption{One-loop diagrams contributing to the wave-function renormalization at $\mathcal{O}(\alpha_s)$.}\label{fig:self-energy}
\end{figure}
It is important to remember that $\chi_n$ in Eq.~\eqref{eq:Jet-cut} is composed of bare SCET (quark and gluon) fields, and that it
is convenient to carry out our computations using perturbation theory ``around'' those (that is, we will not use the so-called renormalized perturbation theory). For the
jettiness computation through the discontinuity of the forward matrix element, this entails that the wave-function renormalization factor
\begin{equation}\label{eq:Zfactor}
Z_{\xi}=1 + \frac{\alpha_sC_F}{4\pi}\Biggl[-\frac{3}{\varepsilon}+6\log\!{\biggl(\frac{m}{\mu}\biggr)}-4\Biggr] + \mathcal{O}(\alpha_s^2)\,,
\end{equation}
computed with the diagrams shown in Fig.~\ref{fig:self-energy} (the soft-gluon contribution vanishes), never appears directly. The mass in Eq.~\eqref{eq:Zfactor} should be understood as the
pole scheme. When using Eq.~\eqref{eq:Jet-cut} one needs to account for $Z_{\xi}^{1/2}$ since this factor is precisely the overlap between the quantum (bare) collinear field $\xi_n$ and the physical collinear
state $\ket{q_n}$: $\bra{0}\xi_n \ket{q_n(\vec{p},s)} = Z_{\xi}^{1/2} u_s(\vec{p}\,)$, with $u$ a particle spinor in the collinear limit, and $s$, $\vec{p}$ the spin and $3$-momentum of the
on-shell collinear quark. On the other hand, when using Eq.~\eqref{eq:Jet-cut} self-energy diagrams on external legs are not included, since their effect is already accounted for in the
$Z_{\xi}^{1/2}$ factor, and it is in this way that one has a one-to-one correspondence with the computation through the imaginary part of the forward matrix element.

\begin{figure}[t!]\centering
\includegraphics[width=0.3\textwidth]{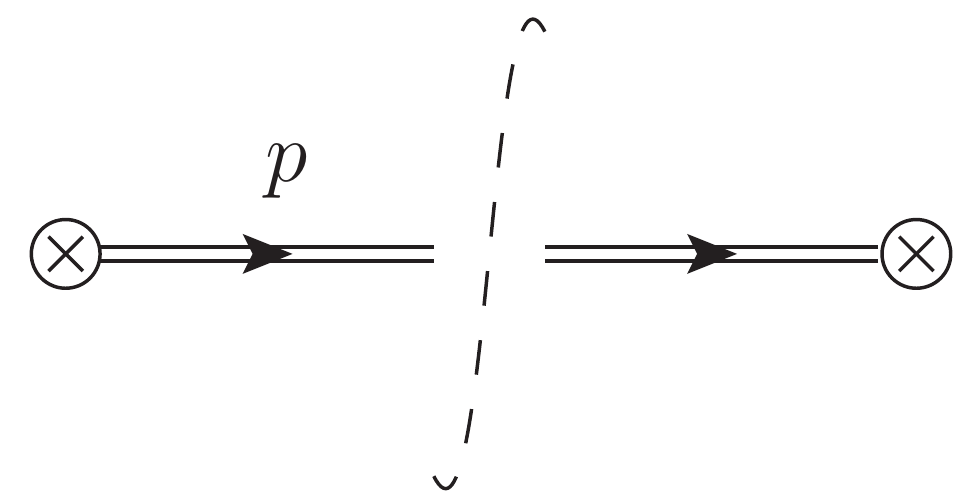}
\caption{Lowest order diagram for the jet function.}\label{fig:Tree}
\end{figure}

The computation at leading order is simple enough that can be carried out for the two massive schemes simultaneously. The corresponding tree-level diagram is shown in Fig.~\ref{fig:Tree}, where
the double line represents a heavy quark and the dashed line marks which particles are on-shell. To compute the phase-space integration it is convenient to use the following
parametrization
\begin{align}\label{eq:phase-space}
\frac{{\rm d}^{d - 1} \vec{p}}{2 E_p} =\,& \frac{{\rm d} p^-}{2 p^-}\,\theta(p^-) \,{\rm d}^{d - 2} \vec{p}_{\perp}\,,\\
\int\!{\rm d}^{d - 2} \vec{p}_{\perp} =\,& \frac{2\pi^{1-\varepsilon}}{\Gamma(1-\varepsilon)} |\vec{p}_{\perp}|^{1-2\varepsilon}\!\!\int\!{\rm d} |\vec{p}_{\perp}|\,,\nonumber
\end{align}
which implies that $p^+$ has to be expressed in terms of the minus and perpendicular components through the on-shell condition $p^+ = (m^2 + |\vec{p}_{\perp}|^2)/p^-$, and since the mass appears through
on-shell kinematic relations it corresponds always to the pole scheme. In the second line we have carried out the angular integrals for the perpendicular momentum, assuming that matrix elements depend only
on its magnitude. We then obtain
\begin{equation}
J_n^{\rm tree}(s)\! = \!\!\int\!\frac{{\rm d} p^-}{2 p^-} {\rm d}^{d-2} \vec{p}_{\perp}
\delta^{(d-2)} (\vec{p}_{\perp}) \delta (p^- - Q) \delta [s - Q^2e_{n\!} (X)] \sum_s
{\rm Tr}\! \biggl[ \frac{\overline{n} \! \! \! /}{2} u_s (p)
\overline{u}_s (p) \biggr] \!= \delta(s - s_{\min}),
\end{equation}
where we have used that the trace of the polarization sum equals $2p^-$ and have integrated all delta functions except the one with the measurement. The color trace
cancels the $1/N_c$ prefactor, and the on-shell condition implies $p^+=m^2/Q$, such that for the $2$-particle collinear measurement we get
\begin{equation}
e_J (X) = \frac{p^+}{Q} = \frac{m^2}{Q^2}\,, \qquad e_{\tau} (X) = \frac{p^+}{Q} - \frac{m^2}{p^-} = 0\,,
\end{equation}
which correspond to $e_{\rm min}$. To include the wave-function renormalization at $\mathcal{O}(\alpha_s)$ one only needs to multiply this result by $Z_{\xi}$.

\subsection{Virtual Radiation}
\begin{figure}[t!]\centering
\includegraphics[width=0.7\textwidth]{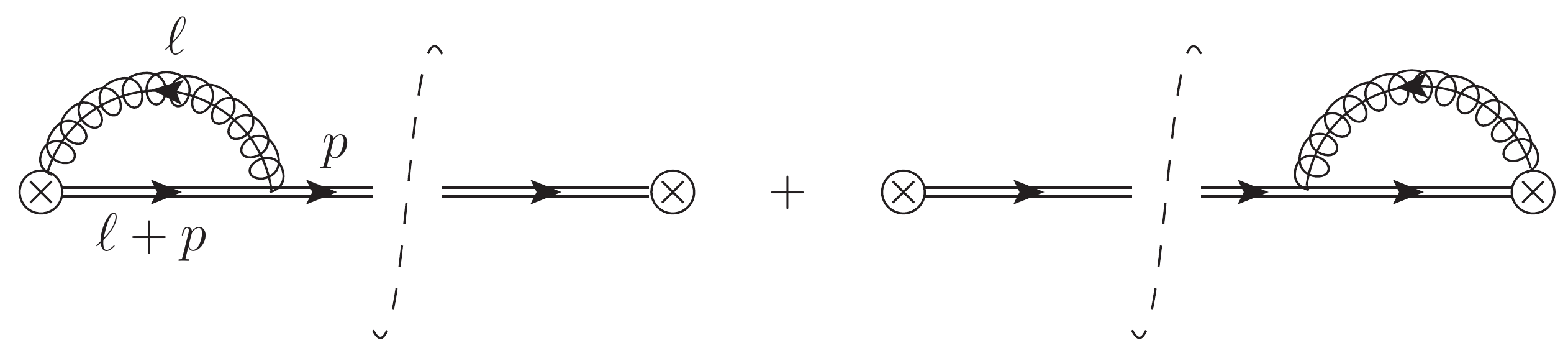}
\caption{Virtual diagrams contributing to the jet function at $\mathcal{O}(\alpha_s)$.}\label{fig:1-loop}
\end{figure}

The contribution from virtual gluons can be carried out for the two massive schemes simultaneously since the phase-space integration is identical to the tree-level computation. There are two
diagrams contributing, as shown in Fig.~\ref{fig:1-loop}, which yield the same result, so we will compute only one of them which will be multiplied by a factor of $2$. Pulling out a collinear
gluon field from the Wilson line and using the Feynman rules for massive collinear quarks we obtain the following integral for the leftmost diagram:
\begin{equation}
J_1^{\rm virt}= -2 i C_F g_s^2\tilde\mu^{2\varepsilon}\!\!\!\int\!\! \frac{{\rm d}^d\ell}{(2\pi)^d}\frac{\bar n(p+\ell)}{(\bar n \ell)\ell^2[(p+\ell)^2-m^2]}\equiv -2 i C_F g_s^2( I_1+p^- I_2) \,,
\end{equation}
where $\tilde\mu^2=\mu^2e^{\gamma_E}/(4\pi)$, the factor of $2$ comes from the product $\bar n n$ and the Casimir $C_F$ from the color trace with two Gell-Mann matrices. We are left
with two master integrals $I_1$ and $I_2$ that can be solved using Feynman parameters for the former
\begin{equation}
\!\!\! I_1 =\!\!\int_0^1\!\!{\rm d}x \!\!\int\!\! \frac{{\rm d}^d\ell}{(2\pi)^d}\frac{\tilde\mu^{2\varepsilon}}{(\ell^2-x^2 m^2)^2}=\!
\frac{i\, \Gamma(\varepsilon)e^{\varepsilon\gamma_E}}{4\pi}\!\!\int_0^1\!\!{\rm d}x \biggl(\frac{x m}{\mu}\biggr)^{\!\!-2\varepsilon}\!\!
\!=\! \frac{i}{(4\pi)^2} \frac{\Gamma(\varepsilon)e^{\varepsilon\gamma_E}}{(1-2\varepsilon)} \biggl(\frac{m}{\mu}\biggr)^{\!\!-2\varepsilon}\!\!.
\end{equation}
and with a combination of Feynman and Georgi parameters for the latter
\begin{align}
I_2=\,&2\!\!\int_0^\infty\!\!{\rm d}\lambda\!\!\int_0^1\!\!{\rm d}x \!\!\int\!\! \frac{{\rm d}^d\ell}{(2\pi)^d}\frac{(1-x)\tilde\mu^{2\varepsilon}}{[\ell^2-x^2 m^2 -x (1-x)\lambda]^3}\nonumber\\
=\, & \!-\!\frac{i\, \mu^{2\varepsilon}e^{\gamma_E}}{(4\pi)^2}\,\Gamma(1+\varepsilon)\!\!\int_0^1\!\!{\rm d}x \,x^{-1-\varepsilon}\!\!\int_0^\infty\!\!{\rm d}\lambda
\biggl\{x(1-x)p^- \!\biggl[\lambda+\frac{m^2 x}{(1-x)p^-}\biggr]\biggr\}^{\!-1-\varepsilon}\\
=\, & \!-\!\frac{i}{(4\pi)^2}\frac{\Gamma(\varepsilon)e^{\gamma_E}}{p^-}\biggl(\frac{m}{\mu}\biggr)^{\!\!-2\varepsilon}\!\!\!\int_0^1\!\!{\rm d}x \,x^{-1-2\varepsilon}
= \frac{i}{(4\pi)^2}\frac{\Gamma(\varepsilon)}{2\varepsilon p^-}\biggl(\frac{m}{\mu}\biggr)^{\!\!-2\varepsilon} \,.\nonumber
\end{align}
Adding those two results we find a closed expression for $J_1^{\rm virt}$:
\begin{align}\label{eq:J1virt}
J_1^{\rm virt}=\,&\frac{\alpha_s C_F}{4\pi}\frac{e^{\varepsilon\gamma_E}\Gamma(1+\varepsilon)}{\varepsilon^2(1-2\varepsilon)}\biggl(\frac{m}{\mu}\biggr)^{\!\!-2\varepsilon}\\
=\,& \frac{\alpha_s C_F}{4\pi}\biggl\{\frac{1}{\varepsilon^2} + \frac{2}{\varepsilon}\biggl[1-\log\biggl(\frac{m}{\mu}\biggr)\biggr] + 4 + \frac{\pi^2}{12}
- 4 \log\biggl(\frac{m}{\mu}\biggr) + 2 \log^2\biggl(\frac{m}{\mu}\biggr)\biggr\}
\,.\nonumber
\end{align}
Interestingly, this result is zero in the massless limit (which has to be taken before expanding in $\varepsilon$). Therefore, using dimensional regularization, only real-radiation diagrams
contribute. The $m$ appearing in Eq.~\eqref{eq:J1virt} is strictly speaking bare, but since we limit our computation to $\mathcal{O}(\alpha_s)$ we can safely take it as the pole
mass, as the difference between these two is a higher order correction. Implementing this result to the jet function computation and integrating the real momentum results in adding a
factor of $\delta (e_n - e_{\min})$. Multiplying by $2$, expanding in $\varepsilon$ and adding the wave-function renormalization, which is obviously a virtual contribution, we obtain
\begin{equation}\label{eq:virtual}
J_n^{\rm virt}(s,\mu) = \frac{\alpha_s C_F}{4\pi}\,\delta (s - s_{\min})\!\biggl[\frac{2}{\varepsilon^2}+\frac{1}{\varepsilon}-\frac{4}{\varepsilon} \log\Bigl(\frac{m}{\mu}\Bigr)
+ 4 +\frac{\pi^2}{6}-2\log\Bigl(\frac{m}{\mu}\Bigr)+4\log^2\Bigl(\frac{m}{\mu}\Bigr)\!\biggr] ,
\end{equation}
which is the final result of this section. Equation~\eqref{eq:virtual} should be valid also for SCET-II type observables since the 1-particle phase space is not yet afflicted by rapidity divergences.

\subsection{Real Radiation}
\begin{figure}[t!]\centering
\includegraphics[width=0.9\textwidth]{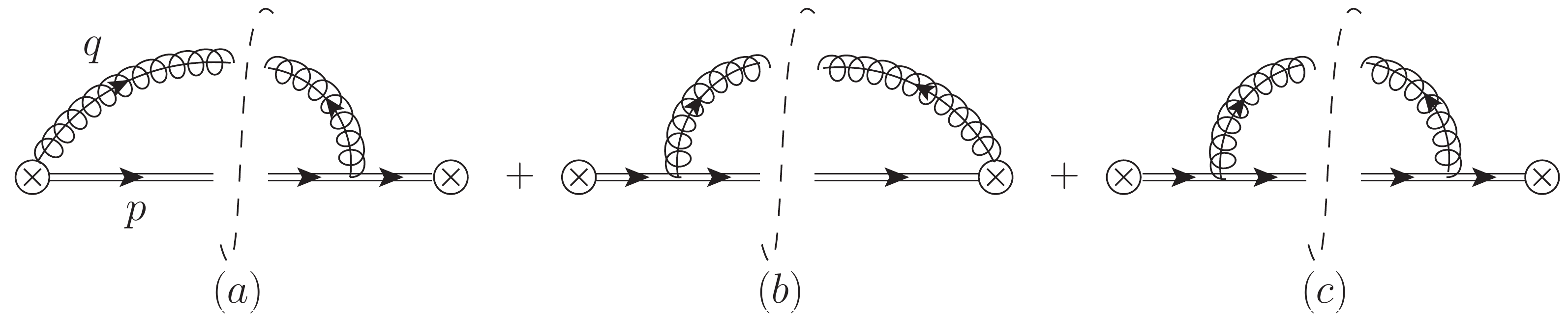}
\caption{Real-radiation diagrams contributing to the jet function at $\mathcal{O}(\alpha_s)$.}\label{fig:Real}
\end{figure}
Since the phase-space integrals with two particles do not fully collapse with the Dirac delta functions, the real radiation contributions differ depending on the collinear measurement.
The diagrams that contribute at $\mathcal{O}(\alpha_s)$ are shown in Fig.~\ref{fig:Real}, where we have omitted the term in which both gluons are radiated from the Wilson line since
it vanishes. Diagrams (a) and (b) give identical contributions and therefore we will compute one of them which will be multiplied by a factor of $2$. For all the real contributions
label-momentum conservation implies $\vec{q}_\perp=-\vec{p}_\perp$ and $q^-=Q-p^-$, which together with the Heaviside function in Eq.~\eqref{eq:phase-space} sets the integration limits for
$p^-$ between $0$ and $Q$. For the first diagram, after integrating the gluon momenta with the delta functions and carrying out the angular perpendicular integration one gets
\begin{equation}\label{eq:Ja}
J_a^{\rm real}(s,\mu)=\frac{4 \alpha_s C_F \,Q\tilde \mu^{2\varepsilon}}{(4\pi)^{1-\varepsilon}\Gamma(1-\varepsilon)}\!\!\int_0^Q\!{\rm d} p^-\frac{p^-}{Q-p^-}
\frac{|\vec{p}_\perp|^{1-2\varepsilon}{\rm d}|\vec{p}_\perp|}{m^2(Q-p^-)^2+Q^2|\vec{p}_\perp|^2}\,
\delta(s-Q^2 e_n)\,.
\end{equation}
Since this diagram involves gluons radiated from Wilson lines, one expects $1/\varepsilon^2$ poles, implying also harder integrals. On the other hand, since the corresponding Feynman
rule is simpler, the result is also shorter. For the third (symmetric) diagram, which does not need a factor of two, the result reads
\begin{align}\label{eq:Jb}
J_c^{\rm real}(s,\mu)=\,& \frac{\alpha_s C_F \,Q^2 e^{\varepsilon\gamma_E}\mu^{2\varepsilon}}{(2\pi)\Gamma(1-\varepsilon)}\!\!\int_0^Q\!{\rm d} p^-
\delta(s-Q^2 e_n)\,\frac{(Q-p^-)(p^-)^2|\vec{p}_\perp|^{1-2\varepsilon}{\rm d}|\vec{p}_\perp|}{m^2(Q-p^-)^2+Q^2|\vec{p}_\perp|^2} \\
&\times\biggl[\frac{2(1-\varepsilon)(|\vec{p}_\perp|^2+m^2)}{(p^-)^2} - 2(1-\varepsilon)\frac{m^2}{Q^2}+\frac{4(2-\varepsilon)m^2}{Qp^-}\biggr].\nonumber
\end{align}
Since Feynman rules are more cumbersome for gluons radiated from a massive quark, the result is lengthier. On the other hand, since these correspond to boosted QCD
processes single $1/\varepsilon$ poles are expected to appear, meaning that no special treatment for the integral is necessary.

\subsubsection{P-scheme Thrust}
To solve the integrals in the previous section, the thrust measurement must be expressed in terms of $p^-$ and $|\vec{p}_\perp|$. Since the gluon is massless and the quark has
mass $m$ we have
\begin{equation}
Q^2\tau = Q\biggl(p^++q^+ - \frac{m^2}{p^-}\biggr)=Q \,|\vec{p}_\perp|^2\biggl(\frac{1}{p^-}+\frac{1}{Q-p^-}\biggr) = \frac{Q^2 |\vec{p}_\perp|^2 }{(Q-p^-)p^-} \,,
\end{equation}
which is mass independent. With this result we can use the measurement delta function to integrate $|\vec{p}_\perp|$
\begin{equation}
\delta(s-Q^2\tau) = \frac{(Q-p^-)p^-}{2|\vec{p}_\perp|Q^2}\, \delta\biggl(\!|\vec{p}_\perp| -\frac{ \sqrt{(Q-p^-)s\,p^-}}{Q}\,\biggr) \,.
\end{equation}
Switching variables to $p^- = Q x$ we find the following 1-dimensional integral for $J_a$:
\begin{align}\label{eq:Ja-thrust}
J_{a,P}^{\rm real}(s,\mu)=\,&\frac{C_F\alpha_s\,e^{\varepsilon\gamma_E}}{2\pi m^2\Gamma(1-\varepsilon)}\biggl(\frac{s}{\mu^2}\biggr)^{\!\!-\varepsilon}\!\!\!\!
\int_0^1\!{\rm d}x\,\frac{x^{2-\varepsilon}(1-x)^{-1-\varepsilon}}{1-x\bigl(1-\frac{s}{m^2}\bigr)}\,.
\end{align}
The complication arises because the integral diverges as $1/\varepsilon^2$ and contains distributions. The divergence comes from the $(1-x)^{-1-\varepsilon}$ factor, but the
subtraction around $x=1$ behaves as $1/s$, which combined with the $s^{-\varepsilon}$ prefactor implies a new divergence and invalidates the subtraction. This
pathological behavior is usual in two-loop computations involving double integrals, and the standard way of solving it is using sector decomposition~\cite{Binoth:2000ps}. To do so, one
needs to get rid of distributions by considering the cumulative jet function, which converts Eq.~\eqref{eq:Ja-thrust} into a double integral. In Appendix~\ref{sec:secDec} we show how to use this
general method to solve the integral, and follow in this section an easier, albeit less general, procedure. Before that, we solve the integral for the massless case, which is not affected by the
problem just described, and is valid for 2-jettiness and thrust:
\begin{align}
J_{a,m=0}^{\rm real}(s,\mu)=\,&\!-\!\frac{C_F\alpha_s e^{\gamma_E}}{(2\pi)\mu^2}\biggl(\frac{s}{\mu^2}\biggr)^{\!\!-1-\varepsilon}
\frac{\Gamma(2-\varepsilon)}{\varepsilon\Gamma(2-2\varepsilon)}\\
=\,&
\frac{C_F\alpha_s}{(2\pi)}\!\Biggl[\!\biggl(\frac{1}{\varepsilon^2}+\frac{1}{\varepsilon}+2-\frac{\pi^2}{4}\biggr)\delta(s)
- \frac{1}{\mu^2}\!\biggl(\frac{\mu^2}{s}\biggr)_{\!\!+}\!\biggl(\frac{1}{\varepsilon} + 1\biggr) \!+\!
\frac{1}{\mu^2}\!\biggl(\frac{\mu^2\log(s/\mu^2)}{s}\biggr)_{\!\!+} \,\Biggr].\nonumber
\end{align}
In the second line we have expanded the result around $\varepsilon=0$ to obtain distributions using the identity
\begin{equation}\label{eq:dist-expansion}
x^{- 1 + \varepsilon} = \frac{1}{\varepsilon} \delta (x) + \sum_{n = 0}
\frac{\varepsilon^n}{n!} \biggl[\frac{\log^n (x)}{x} \biggr]_+ .
\end{equation}

For the $m>0$ case we can transform the integral using hypergeometric function identities. Since these special functions will appear also in Sec.~\eqref{eq:evolution}, we
remind here its integral definition (the hypergeometric function is symmetric with respect to its first two arguments):
\begin{align}\label{eq:2F1}
_2 F_1 (a, b , c , z) =\,& \frac{\Gamma (c)}{\Gamma (b) \Gamma (c - b)} \!\int_0^1\! \text{d} x\, x^{b - 1} (1 - x)^{c - b - 1} (1 - z x)^{- a} \,\\
=\,&(1 - z)^{c - b - a} {}_2 F_1 (c - a, c - b, c , z)\,,\nonumber
\end{align}
where the second line is the so called Euler transformation.\footnote{This property can be easily shown as follows: switching variables $x\to1-x$
and rearranging terms one finds
\begin{equation}\label{eq:Euler-1}
_2 F_1 (a, b , c , z) = (1 - z)^{-a}{}_2 F_1 \!\Bigl( a, c - b , c ,\frac{z}{z - 1} \Bigr) = (1 - z)^{- b}{}_2 F_1 \!\Bigl( c - a, b , c ,\frac{z}{z - 1} \Bigr) ,
\end{equation}
where the second term is obtained using the symmetry $a\leftrightarrow b$ on the first equality. Using the first relation followed from the second one arrives to
the second line of Eq.~\eqref{eq:2F1}.} The integral in Eq.~\eqref{eq:Ja-thrust} is already in this canonical form and therefore we can write
\begin{align}\label{eq:Ja-P}
J_{a,P}^{\rm real}(s,\mu)=&-\!\frac{\Gamma(3-\varepsilon)\,C_F\alpha_s}{(2\pi)m^2\varepsilon\Gamma(3-2\varepsilon)}\biggl(\frac{s}{\mu^2}\biggr)^{\!\!-\varepsilon}
{}_2F_1\!\biggl(1,3-\varepsilon,3-2\varepsilon,1-\frac{s}{m^2}\biggr)\\
= &-\!\frac{\Gamma(3-\varepsilon)\,C_F\alpha_s}{(2\pi)m^2\varepsilon\Gamma(3-2\varepsilon)}\biggl(\frac{s}{m^2}\biggr)^{\!\!-1-2\varepsilon}\!\!
\biggr(\frac{\mu^2}{m^2}\biggl)^{\!\!\varepsilon} {}_2F_1\!\biggl(2-2\varepsilon,-\varepsilon,3-2\varepsilon,1-\frac{s}{m^2}\biggr)\nonumber\\
=\,&\frac{C_F\alpha_s}{(2\pi)m^2\Gamma(1-\varepsilon)}\biggl(\frac{s}{m^2}\biggr)^{\!\!-1-2\varepsilon}\!\!
\biggr(\frac{\mu^2}{m^2}\biggl)^{\!\!\varepsilon}\!\int_0^1\!\!{\rm d}x (1-x)^{2-\varepsilon}x^{-1-\varepsilon}\!
\biggl[1 - x \!\biggl(1-\frac{s}{m^2}\biggr)\!\biggr]^{-2+2\varepsilon} \!,\nonumber
\end{align}
where in the second step we have used Euler's identity and in the third we write the hypergeometric function back as an integral. The integration in the last term can be easily
expanded in $\varepsilon$ using Eq.~\eqref{eq:dist-expansion}, and defining $\tilde s\equiv s/m^2$ we have\,\footnote{Alternatively one can use the Mathematica
package HypExp~\cite{Huber:2007dx} to expand directly the hypergeometric function.}
\begin{align}\label{eq:I3}
&I_3(\tilde s) \equiv \int_0^1\!{\rm d}x \,(1-x)^{2-\varepsilon}x^{-1-\varepsilon}[1 - x (1-\tilde s)]^{-2+2\varepsilon} = -\frac{1}{\varepsilon} - \tilde s \!\!\int_0^1\!{\rm d}x\, \frac{[2-(2-\tilde s) x]}{[1-(1-\tilde s) x]^2} \\
&+\varepsilon \!\!\int_0^1\!{\rm d}x\frac{2 (1-x)^2 \log[1 - x (1-\tilde s)]+\tilde s x [2-(2-\tilde s) x+2] \log (x)-(1-x)^2\log(1-x)}{x [1-(1-\tilde s) x]^2} \nonumber\\
& = -\frac{1}{\varepsilon} + \frac{\tilde s \, [1 - \tilde s + \!(2 - \tilde s) \log(\tilde s)]}{(1-\tilde s)^2} + \varepsilon f_1(\tilde s)
\equiv -\frac{1}{\varepsilon} + \tilde s\, f_0(\tilde s) + \varepsilon f_1(\tilde s)+\mathcal{O}(\varepsilon^2)\,,\nonumber
\end{align}
with $f_1(s)$ a function involving a dilogarithm. This result can be reexpanded in $\varepsilon$ together with the prefactor ${\tilde s}^{\,-1-2\varepsilon}$,
responsible for the appearance of distributions, finding then
\begin{equation}
{\tilde s}^{\,-1-2\varepsilon}I_3(\tilde s) = \frac{2}{\varepsilon^2}\,\delta(\hat s) -\frac{1}{\varepsilon}\biggl(\frac{1}{\hat s}\biggr)_{\!\!+}
+ 2 \biggl(\frac{\log(\hat s)}{\hat s}\biggr)_{\!\!+} - \frac{1}{2}f_1(0)\,\delta(\hat s) + f_0(\hat s)+\mathcal{O}(\varepsilon)\,.
\end{equation}
Therefore one only needs $f_1(0)$, which takes a simple form
\begin{equation}
f(0) = \!\int_0^1 \!{\rm d}x\,\frac{\log (1-x)}{x} = -\frac{\pi^2}{6}\,.
\end{equation}
Putting all partial results together and expanding in $\varepsilon$ we get the following expression
\begin{align}\label{eq:Ja-P Result}
J_{a,P}^{\rm real}(s,\mu)=\,& \frac{C_F\alpha_s}{2\pi}\biggl\{\delta(s)\biggl[\frac{1}{2\varepsilon^2}+\frac{1}{\varepsilon}\log\Bigl(\frac{m}{\mu}\Bigr)
+ \log^2\Bigl(\frac{m}{\mu}\Bigr) + \frac{\pi^2}{24}\biggr] + \frac{2}{\mu^2}\biggl[\frac{\mu^2 \log(s/\mu^2)}{s}\biggr]_{\!+}\nonumber\\
&\!-\!\biggl[\frac{1}{\varepsilon}+2 \log\Bigl(\frac{m}{\mu}\Bigr) \biggr]\frac{1}{\mu^2}\biggl(\frac{\mu^2}{s}\biggr)_{\!\!+}
-\frac{1}{s-m^2}-\frac{s-2m^2}{(s-m^2)^2}\log\biggl(\frac{s}{m^2}\biggr)\biggr\}.
\end{align}
The result is divergent for $s\to m^2$, although at that kinematic point there is no physical phenomenon that implies a singularity. We therefore expect that the
singularity will cancel when adding together all real-radiation diagrams.

For the cut self-energy diagram in Fig.~\ref{fig:Real}~(c), performing the same change of variable as in Eq.~\eqref{eq:Ja-thrust} we arrive at
\begin{align}\label{eq:Jb-P}
J_{c,P}^{\rm real}(s,\mu)=&\,\frac{C_F\alpha_s\,e^{\varepsilon\gamma_E}}{(2\pi)\Gamma(1-\varepsilon)}\biggl(\frac{s}{\mu^2}\biggr)^{\!\!-\varepsilon}\\
&\times\!\int_0^1\!{\rm d}x\,\frac{x^{-\varepsilon}(1-x)^{1-\varepsilon}}{[s(1-x)+x m^2]^2}\{(1-\varepsilon)(1-x)x s-m^2[2(1-x)-(1-\varepsilon)x^2]\}\,.\nonumber
\end{align}
To see how the $1/\varepsilon$ divergence occurs, we compute first the massless limit of $J_c^{\rm real}$, for which we get
\begin{align}
J_{c,m=0}^{\rm real}(s,\mu)=& \,\frac{C_F\alpha_s\,(1-\varepsilon)\,e^{\varepsilon\gamma_E}}{(2\pi)\mu^2\Gamma(1-\varepsilon)}\biggl(\frac{s}{\mu^2}\biggr)^{\!\!-1-\varepsilon}\!\!\!
\!\int_0^1\!{\rm d}x\, x^{1-\varepsilon}(1-x)^{-\varepsilon}\\
=& \,\frac{C_F\alpha_s\,e^{\varepsilon\gamma_E}\Gamma(2-\varepsilon)}{(4\pi)\mu^2\Gamma(2-2\varepsilon)}\biggl(\frac{s}{\mu^2}\biggr)^{\!\!-1-\varepsilon}
= \frac{C_F\alpha_s}{4\pi}\biggl[ \frac{1}{\mu^2}\biggl(\frac{\mu^2}{s}\biggr)_{\!\!+} -\biggl(\frac{1}{\varepsilon}+1\!\biggr)\delta(s)\biggr].\nonumber
\end{align}
At the light of this result one can realize that switching variables to $x=y s/m^2$ exposes the divergence, factoring it out front the integral:
\begin{align}
J_{c,P}^{\rm real}(s,\mu)=&\,\frac{C_F\alpha_s\,}{(2\pi)\Gamma(1-\varepsilon)\mu^2}\biggl(\frac{m\,e^{\gamma_E}}{\mu}\biggr)^{\!\!2\varepsilon}
\biggl(\frac{s}{\mu^2}\biggr)^{\!\!-1-2\varepsilon}\!\!\!\int_0^{\frac{m^2}{s}}{\rm d}y\,
\frac{y^{-\varepsilon}\bigl(1-\frac{s}{m^2}\bigr)^{\!1-\varepsilon}}{\bigl[1+y\big(1-\frac{s}{m^2}\bigr)\bigr]^2} \\
&\times\biggl[(1-\varepsilon)y(1+y)\frac{s^2}{m^4} - (1-\varepsilon)y^2\frac{s^3}{m^6} +\frac{2sy}{m^2}-2\biggr]\nonumber\\
=&\,\frac{C_F\alpha_s\,}{2\pi}\biggl\{
\frac{\delta(s)}{\varepsilon\Gamma(1-\varepsilon)}\biggl(\frac{m\,e^{\gamma_E}}{\mu}\biggr)^{\!\!2\varepsilon}\!\!\int_0^\infty{\rm d}y\,\frac{y^{-\varepsilon}}{(1+y)^2}
+ \frac{1}{\mu^2}\biggl(\frac{\mu^2}{s}\biggr)_{\!\!+} \nonumber\\
& \times \int_0^{\frac{m^2}{s}}{\rm d}y\,\frac{1-\frac{s}{m^2}}{\bigl[1+y\big(1-\frac{s}{m^2}\bigr)\bigr]^2}\biggl[y(1+y)\frac{s^2}{m^4} - y^2\frac{s^3}{m^6} +\frac{2sy}{m^2}-2\biggr]\nonumber
\biggr\}\\
=\,&\frac{C_F\alpha_s}{2\pi}\biggl\{\!\biggl[\frac{1}{\varepsilon}-\!2\log\Bigl(\frac{m}{\mu}\Bigr)\!\biggr]\delta(s)
- \!\frac{2}{\mu^2}\!\biggl(\frac{\mu^2}{s}\biggr)_{\!\!+} \nonumber\\
&+\frac{1}{2(s-m^2)^3}\biggl[5s^2-16m^2s+11m^4-2m^2(s-4m^2)\log\Bigl(\frac{s}{m^2}\Bigr)\biggr]\biggr\} .\nonumber
\end{align}
In the one-to-last step we have used Eq.~\eqref{eq:dist-expansion} to partially expand in $\varepsilon$ and in the last step the following relation is used:
\begin{equation}
f(x) \biggl[\frac{1}{x} \biggr]_+ = f(0) \biggl[\frac{1}{x} \biggr]_+ +\frac{f (x) - f (0)}{x}\, .
\end{equation}
The Dirac delta function $\delta(s)$ sets the upper integration limit to infinity and $s=0$ in the integrand. This makes the integral so simple that
no further expansion in $\varepsilon$ is necessary for this term. For the contribution proportional to
the plus distribution we can set $\varepsilon=0$ right away, and solve the integral with standard methods. In the last step we have consistently expanded in $\varepsilon$
the full result. The expression is again divergent as $s\to m^2$, but as anticipated, the full real-radiation contribution is regular in this limit:
\begin{align}
J_P^{\rm real}(s,\mu)=\,& \frac{C_F\alpha_s}{2\pi}\biggl\{\!\biggl[\frac{1}{\varepsilon^2}+\frac{1}{\varepsilon}+\frac{2}{\varepsilon}\log\Bigl(\frac{m}{\mu}\Bigr)+\frac{\pi^2}{12}-
2\log\Bigl(\frac{m}{\mu}\Bigr)+2\log^2\Bigl(\frac{m}{\mu}\Bigr)\!\biggr]\delta(s)\nonumber \\
&\!- \!\biggl[\frac{1}{\varepsilon}+1+2 \log\Bigl(\frac{m}{\mu}\Bigr) \biggr]\frac{1}{\mu^2}\biggl(\frac{\mu^2}{s}\biggr)_{\!\!+}
+ \frac{4}{\mu^2}\biggl[\frac{\mu^2 \log(s/\mu^2)}{s}\biggr]_{\!+} \\
&+ \frac{s - 7 m^2}{2(s - m^2)^2} - \frac{s (2 s - 5 m^2)}{(s - m^2)^3} \log\Bigl( \frac{s}{m^2} \Bigr)\biggr\}.\nonumber
\end{align}

For completeness, we also provide the real-radiation contribution for the massless case, which coincides with the full jet function. Adding the tree-level result we recover
the known result
\begin{align}\label{eq:Jm0}
J_{m=0} =\,& \delta(s)-\frac{C_F\alpha_s e^{\gamma_E}}{(4\pi)\mu^2}\biggl(\frac{s}{\mu^2}\biggr)^{\!\!-1-\varepsilon}
\frac{(4-\varepsilon)\Gamma(2-\varepsilon)}{\varepsilon\Gamma(2-2\varepsilon)}\\
=\,& \delta(s) + \frac{C_F\alpha_s}{4\pi}\biggl\{ \delta (s) \biggl(\frac{4}{\varepsilon^2} + \frac{3}{\varepsilon} + 7 - \pi^2\biggr) -
\frac{3+\frac{4}{\varepsilon}}{\mu^2} \biggl( \frac{\mu^2}{s}
\biggr)_+ + \frac{4}{\mu^2} \biggl[ \frac{\mu^2 \log( s/\mu^2)}{s} \biggr]_+ \biggr\}. \nonumber
\end{align}

\subsubsection{Jettiness}
Let us express the jettiness measurement in terms of minus and perpendicular components:
\begin{equation}
Q^2\tau = Q(p^++q^+)=Q\biggl(\frac{ |\vec{p}_\perp|^2+m^2}{p^-}+\frac{|\vec{p}_\perp|^2}{Q-p^-}\biggr) = \frac{Q^2 |\vec{p}_\perp|^2 + m^2 Q (Q-p^-) }{(Q-p^-)p^-} \,,
\end{equation}
which can be used to solve the measurement delta function for the magnitude of the perpendicular momentum
\begin{equation}
\delta(s-Q^2\tau_J) = \frac{(Q-p^-)p^-}{2|\vec{p}_\perp|Q^2}\, \delta\biggl(\!|\vec{p}_\perp| -\frac{ \sqrt{(Q-p^-)(s p^--Q m^2)}}{Q}\,\biggr) .
\end{equation}
The argument of this delta function can be zero only if $sp^- > Q\,m^2$, what sets the lower limit of integration. Therefore, changing variables to
$p^- = Q (1-x)$ we obtain for the diagram in which the gluons are radiated from the Wilson line and the quark particle the following result
\begin{align}\label{eq:Jb-2J}
J_{a,J}^{\rm real}(s,\mu)=\,&\frac{C_F\alpha_s}{(2\pi)\Gamma(1-\varepsilon)}\frac{\mu^{2\varepsilon}e^{\varepsilon\gamma_E}}{s-m^2}\!
\int_0^{1-\frac{m^2}{s}}\!{\rm d}x\,(1-x)x^{-1-\varepsilon}[(1-x)s-m^2]^{-\varepsilon}\nonumber \\
=\,&\frac{C_F\alpha_s}{(2\pi)\Gamma(1-\varepsilon)}\frac{s^\varepsilon\mu^{2\varepsilon}e^{\varepsilon\gamma_E}}{(s-m^2)^{1+2\varepsilon}}\!
\int_0^1{\rm d}y\,(1-y)^{-\varepsilon}\biggl(y^{-1-\varepsilon}-y^{-2\varepsilon}\frac{s-m^2}{s}\biggr)\nonumber\\
=\,&\!-\!\frac{C_F\alpha_se^{\varepsilon\gamma_E}}{(2\pi)\mu^2}\frac{\Gamma(1-\varepsilon)}{\varepsilon\,\Gamma(2-2\varepsilon)}
\biggl(\frac{s}{\mu^2}\biggr)^{\!\!-1+\varepsilon}\!\biggl(\frac{s-m^2}{\mu^2}\biggr)^{\!\!-1-2\varepsilon}
[\,s(1-\varepsilon)-\varepsilon\, m^2\,]\\
=\,& \frac{C_F\alpha_s}{2\pi}\biggl\{\delta(s-m^2)\biggl[\frac{1}{2\varepsilon^2}+\frac{1}{\varepsilon}\log\Bigl(\frac{m}{\mu}\Bigr)
+ \log^2\Bigl(\frac{m}{\mu}\Bigr) - \frac{\pi^2}{8}\biggr] \!-\frac{1}{s}-\frac{\log\bigl(\frac{s}{m^2}\bigr)}{(s-m^2)}\nonumber\\
&\!-\!\biggl[\frac{1}{\varepsilon}+2 \log\Bigl(\frac{m}{\mu}\Bigr) \biggr]\frac{1}{\mu^2}\biggl(\frac{\mu^2}{s-m^2}\biggr)_{\!\!+}
+ \frac{2}{\mu^2}\biggl[\frac{\mu^2 \log[(s-m^2)/\mu^2]}{s-m^2}\biggr]_{\!+}\biggr\},\nonumber
\end{align}
where in the second line we have switched variables to $x = y (1-m^2/s)$. Performing the same change of variables in the diagram in which the gluons are radiated
from both quark lines one gets
\begin{align}
J_{c,J}^{\rm real}(s,\mu) =\,& \frac{C_F\alpha_s}{(2\pi)\Gamma(1-\varepsilon)}\frac{\mu^{2\varepsilon}e^{\varepsilon\gamma_E}}{(s-m^2)^2}
\int_0^{1-\frac{m^2}{s}}\!{\rm d}x\,x^{-\varepsilon}[(1-x)s-m^2]^{-\varepsilon} \\
&\bigl\{(1-\varepsilon)\bigl[x s+(2-x)m^2\bigr]-(2-\varepsilon)m^2\bigr\}\nonumber\\
=\,&\frac{C_F\alpha_s}{2\pi}\frac{(s-m^2)^{-1-2\varepsilon}}{s^2\Gamma(1-\varepsilon)}\biggl(\frac{s\,e^{\gamma_E}}{\mu^2}\biggr)^{\!\!\varepsilon}
\int_0^1{\rm d}y\,y^{-\varepsilon}(1-y)^{-\varepsilon}\bigl[(s-m^2)^2y(1-\varepsilon)-2m^2s\bigr]\nonumber\\
=\,&\frac{C_F\alpha_s}{4\pi}\frac{(s-m^2)^{-1-2\varepsilon}}{s^2}\biggl(\frac{s\,e^{\gamma_E}}{\mu^2}\biggr)^{\!\!\varepsilon}
\frac{\Gamma (1 - \varepsilon)}{\Gamma (2 - 2 \varepsilon)} [(1 -\varepsilon) (s - m^2)^2 - 4 m^2 s]\nonumber\\
=\,&\frac{C_F\alpha_s}{2\pi}\biggl\{\!\biggl[\frac{1}{\varepsilon}+2+\!2\log\Bigl(\frac{m}{\mu}\Bigr)\!\biggr]\delta(s-m^2)
- \!\frac{2}{\mu^2}\biggl(\frac{\mu^2}{s-m^2}\biggr)_{\!\!+}\!+\frac{5s-m^2}{2s^2}\biggr\} .\nonumber
\end{align}
where we have carried out the same manipulations as in Eq.~\eqref{eq:Jb-2J}. Adding the two results we obtain the total contribution for 2-jettiness:
\begin{align}
J_J^{\rm real}(s,\mu)=\,& \frac{C_F\alpha_s}{2\pi}\biggl\{\!\biggl[\frac{1}{\varepsilon^2}+\frac{1}{\varepsilon}+\frac{2}{\varepsilon}\log\Bigl(\frac{m}{\mu}\Bigr)+2-\frac{\pi^2}{4}+
2\log\Bigl(\frac{m}{\mu}\Bigr)+2\log^2\Bigl(\frac{m}{\mu}\Bigr)\!\biggr]\delta(s)\nonumber \\
&\!- \!\biggl[\frac{1}{\varepsilon}+1+2 \log\Bigl(\frac{m}{\mu}\Bigr) \biggr]\frac{1}{\mu^2}\biggl(\frac{\mu^2}{s}\biggr)_{\!\!+}
+ \frac{4}{\mu^2}\biggl[\frac{\mu^2 \log(s/\mu^2)}{s}\biggr]_{\!+} \\
&+ \frac{s-m^2}{2s^2} - \frac{2}{s-m^2} \log \Bigl(\frac{s}{m^2} \Bigr)\biggr\}.\nonumber
\end{align}
\subsection{Final Result for the Jet Function}\label{sec:finalSCET}
\begin{figure*}[t!]
\subfigure[]
{
\includegraphics[width=0.46\textwidth]{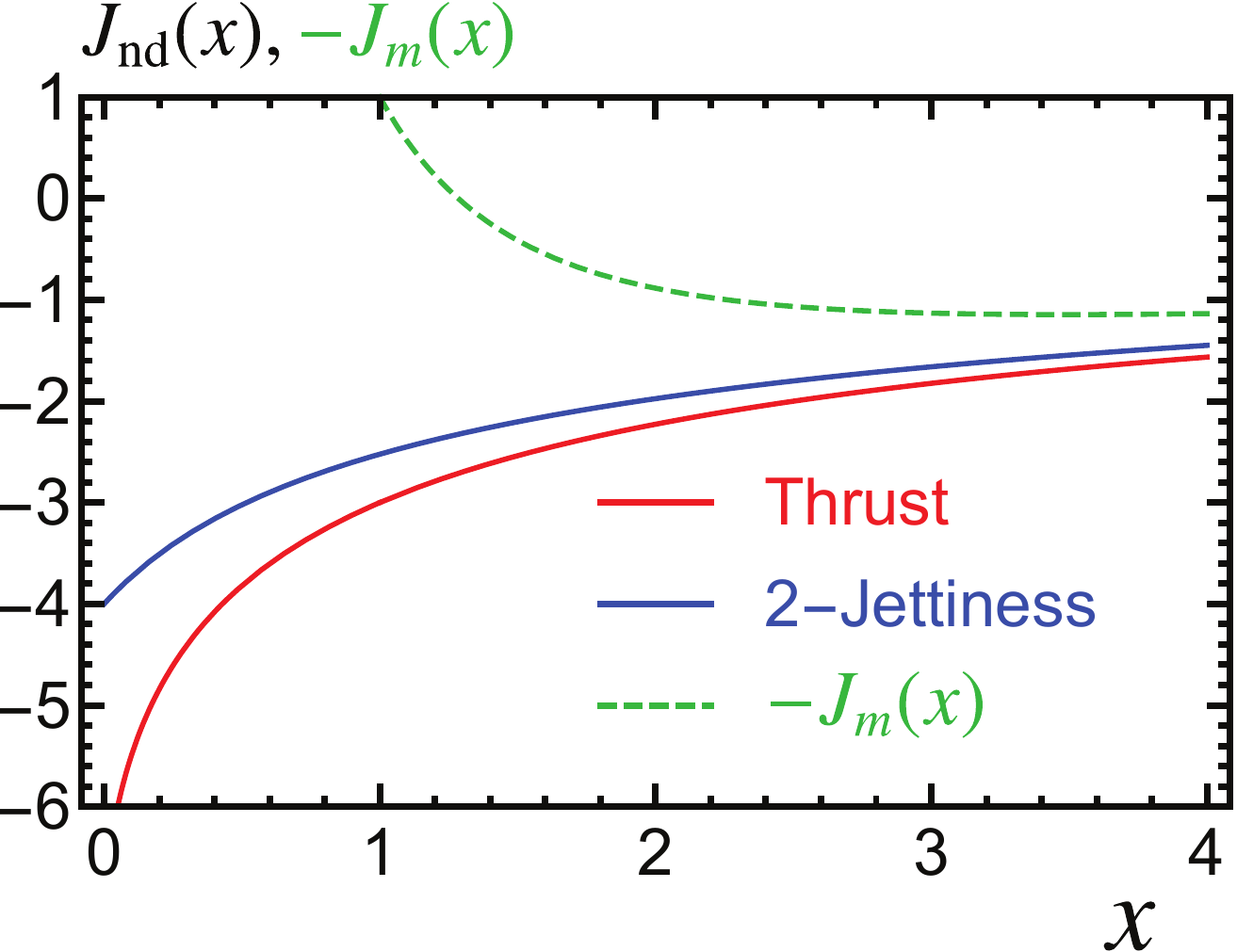}
\label{fig:Jnd}
}
\subfigure[]{
\includegraphics[width=0.48\textwidth]{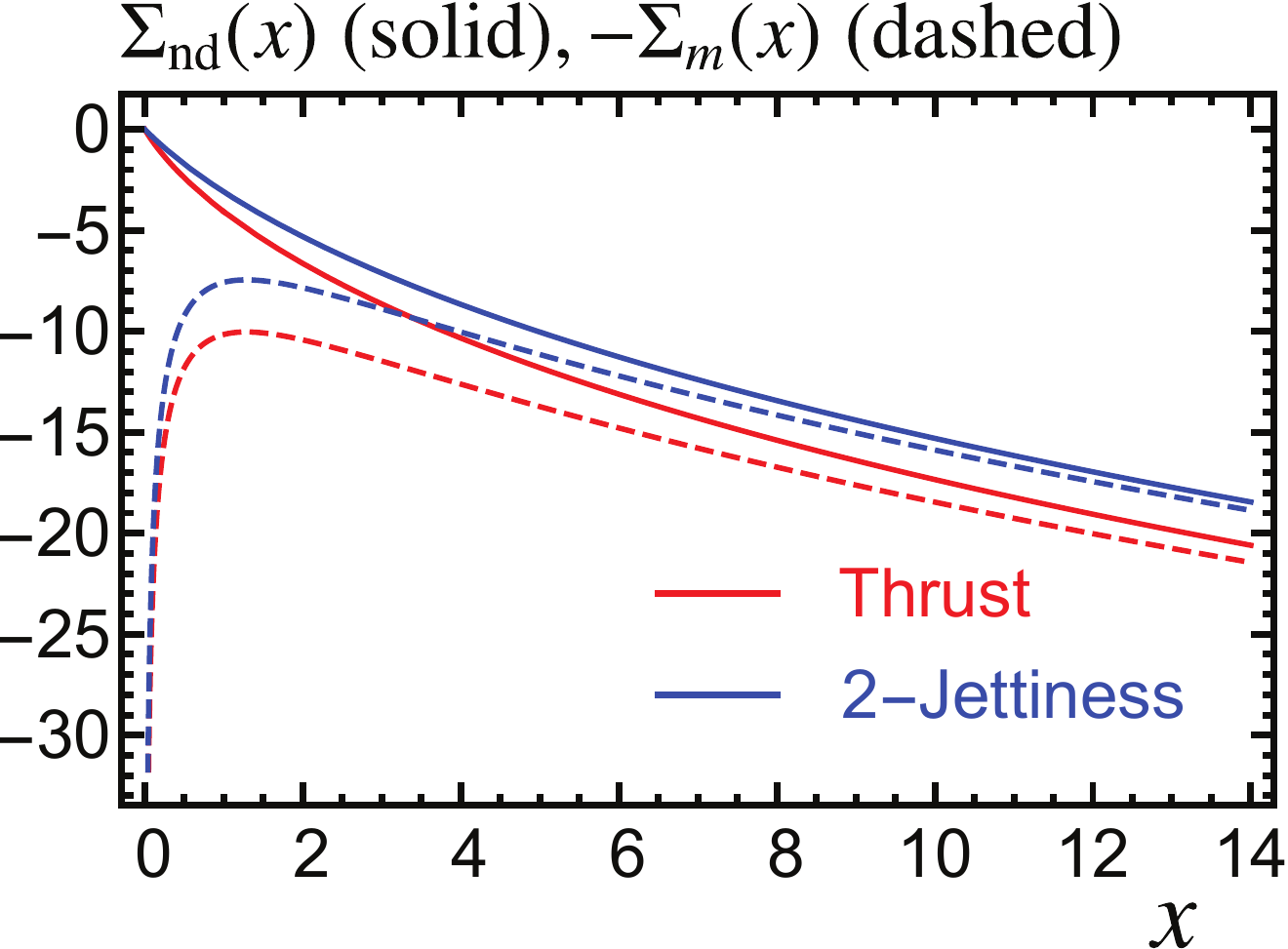}
\label{fig:SigmaNd}
}
\caption{Massive corrections to the jet function. Panels (a) and (b) show the differential and cumulative jet functions, respectively. We show with solid lines the non-distributional functions $J_{\rm nd}$
and $\Sigma_{\rm nd}$ for P- (red) and M- (blue) schemes. The differential $J_{m}$ function is shown multiplied $-1$ as a green dashed
line in panel (a) (for $x>0$ it is common to both schemes), while $-\Sigma_m$ is shown in panel (b) with red and blue dashed lines for P- and M-schemes, respectively.}
\label{fig:JetND}
\end{figure*}
Adding together the contributions from real and virtual corrections one obtains the complete jet function. The divergences are now entirely of UV origin and can be renormalized
multiplicatively (with a convolution). Since in either massive scheme these are the same as for massless quarks, the renormalization factor is identical, along with the anomalous dimension.
Therefore we quote directly the result for the renormalized jet functions \mbox{[\,$\alpha_s\equiv \alpha_s(\mu)$\,]}:
\begin{align}
J_n^P (s, \mu) =\, & \delta (s) + \frac{\alpha_s C_F}{4 \pi} \biggl\{\!
\biggl[2 \log \Bigl( \frac{m}{\mu} \Bigr) + 8 \log^2 \Bigl(\frac{m}{\mu} \Bigr) \!+ 4 + \frac{\pi^2}{3} \biggr] \delta (s) \! +
\frac{8}{\mu^2} \biggl[ \frac{\log(s/\mu^2)}{s / \mu^2}
\biggr]_+ \nonumber\\
& -\! \frac{4}{\mu^2} \biggl[ 1 + 2\log \Bigl( \frac{m}{\mu} \Bigr) \biggr]\! \biggl( \frac{\mu^2}{s} \biggr)_{\!\!+}\! +
\frac{s - 7m^2}{(s - m^2)^2} - \frac{2 s (2 s - 5 m^2)}{(s - m^2)^3} \log \Bigl(\frac{s}{m^2} \Bigr)\! \biggr\},\nonumber\\
J_n^J(s+m^2, \mu) =\, & \delta (s) + \frac{\alpha_s C_F}{4 \pi} \biggl\{\!
\biggl[2 \log \Bigl( \frac{m}{\mu} \Bigr) + 8 \log^2 \Bigl(\frac{m}{\mu} \Bigr) \!+ 8 - \frac{\pi^2}{3} \biggr] \delta (s) \! +
\frac{8}{\mu^2} \biggl[ \frac{\log(s/\mu^2)}{s / \mu^2}
\biggr]_+ \nonumber\\
&-\! \frac{4}{\mu^2} \biggl[ 1 + 2\log \Bigl( \frac{m}{\mu} \Bigr) \biggr]\! \biggl( \frac{\mu^2}{s} \biggr)_{\!\!+}\!
+ \frac{s}{(m^2 + s)^2} - \frac{4}{s} \log \Bigl( 1 + \frac{s}{m^2} \Bigr)\! \biggr\}.
\end{align}
From these equations one can easily read out the functional form of $J_{\rm nd}(x)$ defined in Eq.~\eqref{eq:jetDecomposition}. With some manipulations one can also figure out
expressions for $J_m(x)$ defined in the same equation:
\begin{align}
J_m(x)=\,& A_S\, \delta(x) - \biggl( \frac{1}{x} \biggr)_{\!\!+} + 4\biggl[\frac{\log( x )}{x} \biggr]_+ \,,\nonumber\\
J_{\rm nd}^J(x) = \,&\frac{x}{(x + 1)^2} - \frac{4}{x} \log(1 + x)\,,\\
J_{\rm nd}^P(x) =\,& \frac{x - 7}{(x - 1)^2} - \frac{2 x (2 x - 5)}{(x - 1)^3} \log(x)\,,\nonumber
\end{align}
with $A_J=2 \pi^2/3 + 1$ and $A_P=4 \pi^2/3 - 3$. We shall see that $J_{\rm nd}(x\to\infty) = -J_m(x)$ for both schemes, and show this behavior graphically
in Fig.~\ref{fig:Jnd}. This is expected since it corresponds to taking the massless limit, and therefore mass corrections should vanish such that the jet function becomes equal to the
(renormalized) massless result of Eq.~\eqref{eq:Jm0}. Since $J_{\rm nd}$ contains distributions in this limit, it is advantageous to work with the cumulative jet function
\begin{equation}
\Sigma_n(s_c) \equiv \!\int_0^{s_c}\!{\rm d}s\, J_n(s)\,,
\end{equation}
which is a regular function. Likewise, one can define the cumulative functions for $J_{\rm nd}$ and $J_m$, which are also shown in Fig.~\ref{fig:SigmaNd}.
The result can be obtained easily and involves polylogarithms:
\begin{align}
\Sigma_n^J(s + m^2,\mu) =\, & 1 + \frac{\alpha_s C_F}{4 \pi} \biggl\{
2\log \Bigl( \frac{m}{\mu} \Bigr) + 8 \log^2 \Bigl( \frac{m}{\mu} \Bigr) + 8 - \frac{\pi^2}{3} + \log \Bigl( 1 + \frac{s}{m^2}
\Bigr) - \frac{s}{s + m^2}\nonumber \\
& + \,4 \log^2 \Bigl( \frac{s}{\mu^2} \Bigr) - 4 \biggl[ 1 +2 \log \Bigl( \frac{m}{\mu} \Bigr) \biggr]\!
\log \Bigl( \frac{s}{\mu^2} \Bigr) + 4\, \text{Li}_2 \Bigl( - \frac{s}{m^2}\Bigr) \biggr\} , \\
\Sigma_n^P(s,\mu) =\, & 1 + \frac{\alpha_s C_F}{4 \pi} \biggl\{
2\log \Bigl( \frac{m}{\mu} \Bigr) + 8 \log^2 \Bigl( \frac{m}{\mu} \Bigr)+ 4 + \frac{\pi^2}{3} -
4 \biggl[ 1 +2 \log \Bigl( \frac{m}{\mu} \Bigr) \biggr] \!\log \Bigl( \frac{s}{\mu^2} \Bigr) \nonumber\\
& + 4 \log^2 \Bigl( \frac{s}{\mu^2} \Bigr) + 4 \,\text{Li}_2 \Bigl( 1 - \frac{s}{m^2} \Bigr) + \frac{3 s}{s- m^2} + \frac{(s - 4 m^2) s }{(s - m^2)^2}\log \Bigl( \frac{s}{m^2} \Bigr)
\biggr\} . \nonumber
\end{align}
As expected, in the $m\to 0$ limit both results reduce to the (same) massless cumulative jet function
\begin{equation}
\Sigma_n^{m=0}(s,\mu) = 1 + \frac{\alpha_s (\mu) C_F}{4 \pi} \biggl[ 7 -
\pi^2 - 3 \log \Bigl( \frac{s}{\mu^2} \Bigr) + 2 \log^2 \Bigl(
\frac{s}{\mu^2} \Bigr) \biggr] .
\end{equation}
To take the derivative one needs to recall that the jet function has support only for positive~$s$, such that it is effectively proportional to an (implicit) Heaviside function $\theta(s)$.
Using the following relations:
\begin{equation}\label{eq:ThetaDer}
\frac{\text{d} \theta (x)}{\text{d} x} = \delta (x)\;, \qquad
\frac{\text{d}}{\text{d} x} [\theta (x) \log^n (x)] = n \biggl[ \frac{\log^{n - 1} (x)}{x} \biggr]_+ ,
\end{equation}
one readily arrives at Eq.~\eqref{eq:Jm0}. For $s>0$ one can expand around $m=0$ to find the following compact series
\begin{align}
J_n^P (s>0, \mu) =\, &\frac{\alpha_s C_F}{4 \pi} \frac{1}{s} \biggl\{ 4 \log \Bigl(\frac{s}{\mu^2} \Bigr)\! - 3 + \sum_{i = 1} \biggl[ 1 - 6 i - (4 + i - 3 i^2)
\log \Bigl( \frac{s}{m^2} \Bigr) \!\biggr]\! \biggl( \frac{m^2}{s} \biggr)^{\!\!i}\,
\biggr\}\, ,\nonumber\\
J_n^J (s>0, \mu) =\, & \frac{\alpha_s C_F}{4 \pi} \frac{1}{s} \biggl\{ 4 \log \Bigl(\frac{s}{\mu^2} \Bigr)\! - 3 + \sum_{i = 1}(-1)^i
\biggr( i + 1 + \frac{4}{i} \biggl)\! \biggl( \frac{m^2}{s} \biggr)^{\!\!i}\,
\biggr\}\,,
\end{align}
with similar results for the cumulative jet functions. Since individual pieces of the P-scheme thrust jet function have divergences at $s= m^2$ it is convenient to compute the
expansion of $J^P_{\rm nd}(x)$ around $x=1$, which can be cast as
\begin{equation}\label{eq:nd1}
J_{\rm nd}^P(x) = - 2 \sum_{i = 0} (1- x)^i \frac{9 + 5 i}{(i + 1) (i + 2) (i + 3)}\,.
\end{equation}

\section{Fixed-order Prediction in SCET}\label{sec:FO-SCET}
\begin{figure}[t]\centering
\includegraphics[width=0.5\textwidth]{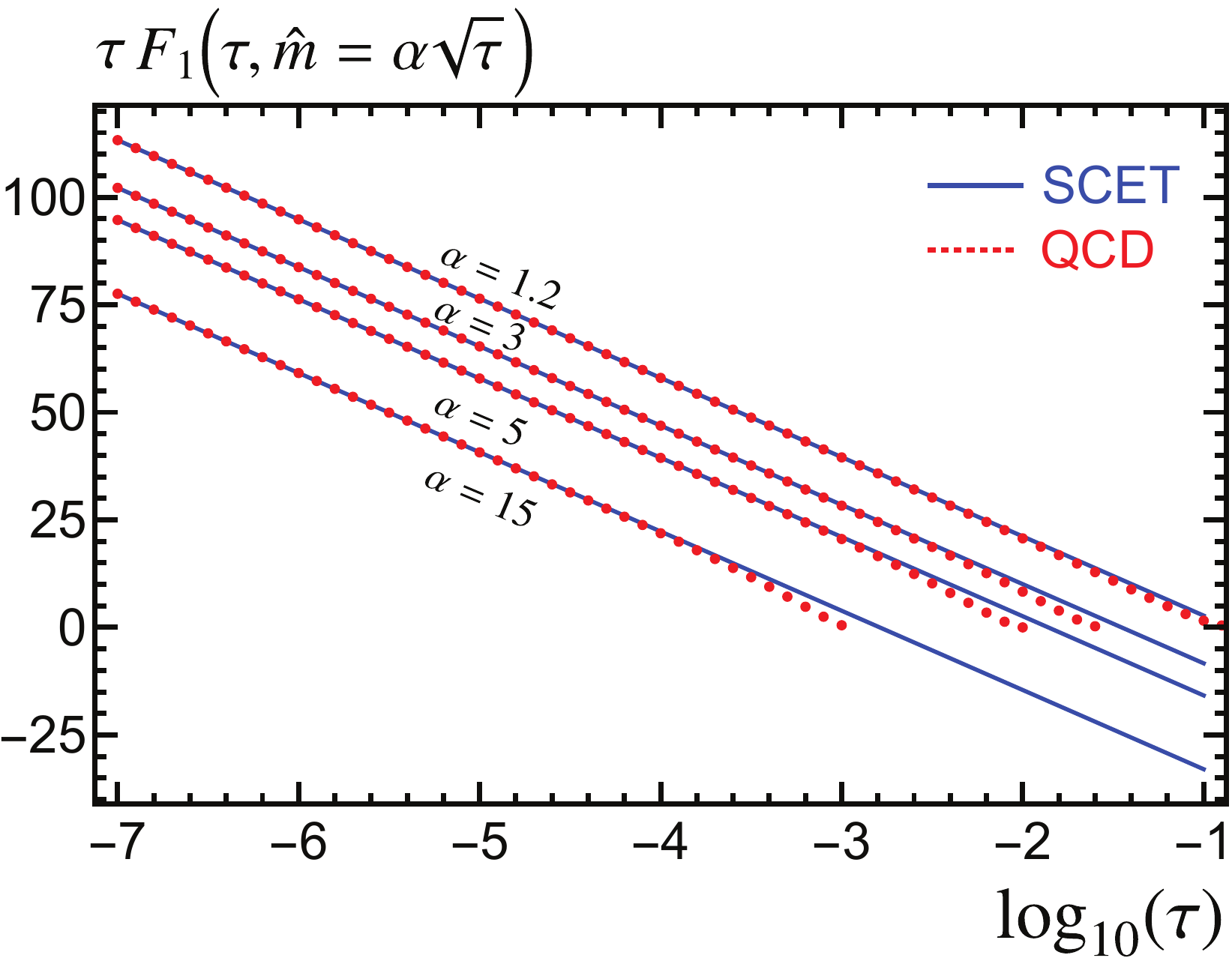}
\caption{Comparison of the $\mathcal{O}(\alpha_s)$ correction to the differential cross sections in QCD [\,$F^{\rm QCD}_V(\tau,\hat m)$\,] and SCET
[\,$F^{\rm SCET}_1(\tau,\hat m)$\,]. We enforce the
SCET power counting by scaling the reduced mass as $\hat m=\alpha\sqrt{\tau}$, with $\alpha=\mathcal{O}(1)$. Solid blue lines show SCET analytic results, while red
dots correspond to QCD numerical predictions obtained from the computations in Ref.~\cite{Lepenik:2019jjk}. The numerical values of $\alpha$ are shown in the
figure.\label{fig:QCD-SCET}}
\end{figure}

Inserting our result for the jet function into the SCET factorization theorem of Eq.~\eqref{eq:FacMom}, setting all renormalization scales equal and using the known results for the hard and
soft function at one loop
\begin{align}
H (Q, \mu) =\, & 1 + \frac{\alpha_s (\mu) C_F}{4 \pi} \biggl[ \frac{7\pi^2}{3} - 16 + 12 \log \biggl( \frac{Q}{\mu} \biggr) - 8 \log^2 \biggl(\frac{Q}{\mu} \biggr) \biggr], \\
S (\ell, \mu) = \,& \delta (\ell) + \frac{\alpha_s (\mu) C_F}{4 \pi} \biggl\{\frac{\pi^2}{3} \delta (\ell) - \frac{16}{\mu} \biggl[ \frac{\mu \log (\ell /\mu)}{\ell} \biggr]_+ \biggr\} , \nonumber
\end{align}
one arrives at the fixed-order prediction for the partonic singular terms of the P-scheme thrust differential cross
section:\,\footnote{The partonic fixed-order bHQET cross section is identical to the SCET one dropping $F_{\rm NS}^{\rm SCET}$.}
\begin{align}
\frac{1}{\sigma_0} \frac{\text{d} \hat\sigma^{\rm SCET}_{\rm FO}}{\text{d} \tau} =\,
& \delta (\tau) + \frac{\alpha_s (\mu) C_F}{4 \pi} F^{\rm SCET}_1(\tau, \hat m)+ \Ord(\alpha_s^2)\\
F^{\rm SCET}_1(\tau, \hat m) =\,& \delta (\tau)\!
\biggl[ \frac{10 \pi^2}{3} - 8 + 4 \log (\hat{m}) + 16 \log^2 (\hat{m})\biggr]-8 [1 + 2 \log (\hat{m})] \biggl( \frac{1}{\tau} \biggr)_{\!\!+} \nonumber\\
& \!\!\! + \frac{2(\tau - 7 \hat{m}^2)}{(\tau - \hat{m}^2)^2} - \frac{4
\tau (2 \tau - 5 \hat{m}^2)}{(\tau - \hat{m}^2)^3} \log \Bigl(\frac{\tau}{\hat{m}^2} \Bigr) \nonumber\\
\equiv \,& A^{\rm SCET}({\hat m})\delta(\tau) + B^{\rm SCET}_{\rm plus}({\hat m})\biggl(\frac{1}{\tau} \biggr)_{\!\!+}
+ F_{\rm NS}^{\rm SCET} (\tau, \hat{m})\,.\nonumber
\end{align}
In the same way, one can get a similar expression for the cumulative distribution $\Sigma_P^{\rm SCET}$, which is among other things useful to take the $m\to 0$ limit.
The differential cross section has a similar structure in full QCD, although it is different for vector and axial-vector currents as discussed in Ref.~\cite{Lepenik:2019jjk}, and
for P-scheme thrust takes the following form\,\footnote{In the threshold limit one gets the same result as in full QCD dropping $F_{\rm NS}^C (\tau, \hat{m})$, which is a power
correction.}
\begin{align}\label{eq:general-diff}
\frac{1}{\sigma^C_0} \frac{\dd \hat \sigma^C_{\rm FO}}{\dd \tau} =\,&R_0^C(\hat m)\,\delta (\tau) + C_F \frac{\alpha_s}{\pi} F^{\rm QCD}_C(\tau, \hat m) + \Ord(\alpha_s^2)\,,\\
F^{\rm QCD}_C(\tau, \hat m) =\, & A^C({\hat m})\delta(\tau) + B^C_{\rm plus}({\hat m})\biggl(\frac{1}{\tau} \biggr)_{\!\!+}
+ F_{\rm NS}^C (\tau, \hat{m})\,,\nonumber
\end{align}
where $C=V,A$ labels the type of current and with $R^C_0$ the tree-level massive R-ratio. Analytic results for $A^C$ and $B_C^{\rm plus}$ can be found in
Ref.~\cite{Lepenik:2019jjk} and we quote here the universal value for the latter:
\begin{equation}
B^C_{\rm plus} (\hat{m}) = \binom{3 - \beta^2}{2\beta^2}\! \biggl[(1 +\beta^2) \log\biggl( \frac{1 + \beta}{2 \hat{m}} \biggr) - \beta \biggl] ,
\end{equation}
with $\beta = \sqrt{1 - 4 \hat{m}^2}$, and where the first and second line of the expression in big parentheses correspond to the vector and axial-vector currents, respectively.
One recovers the SCET result for small masses, $A^C(\hat m \to 0)=A^{\rm SCET}(\hat m)$ and $B_{\rm plus}^C(\hat m \to 0)=B_{\rm plus}^{\rm SCET}(\hat m)$, and also
\begin{equation}\label{eq:alphaLim}
\lim_{\tau\to 0} F_{\rm NS}^C \Bigl(\tau, \hat m=\alpha \sqrt{\tau}\,\Bigr) \to F_{\rm NS}^{\rm SCET}\Bigl(\tau, \hat m=\alpha \sqrt{\tau}\,\Bigr) ,
\end{equation}
with $\alpha\sim\mathcal{O}(1)$. Since for thrust $F_{\rm NS}^C$ is only known numerically, in Fig.~\ref{fig:QCD-SCET} we show a comparison of QCD and SCET results for
the NLO corrections scaling the mass as indicated
in Eq.~\eqref{eq:alphaLim}. Excellent numerical agreement is found as $\tau\to 0$ for various values of $\alpha$ between $1.2$ and $15$. We show only the vector
current as for small values of $\tau$ it is indistinguishable from the axial-vector one.

\section{bHQET Jet Function Computation}\label{sec:jet-bHQET}
The computation of the bHQET jet function is significantly simpler than for the SCET counterpart since in this EFT the mass is no longer a dynamical scale
and we are left with tadpole-like integrals. As an immediate consequence of that, much as it happened for the massless SCET jet function, all virtual graphs are automatically
zero in dimensional regularization since they are scaleless (this includes the wave-function renormalization factor). We are then left with the tree-level, which is common for
both massive schemes, and real-radiation diagrams. The collinear event-shape measurements are the same in SCET and bHQET, 
although the contribution of massive particles needs to be power expanded, such that using $p=mv + k$ we obtain for thrust and 2-jettiness the following results
\begin{align}\label{eq:bHQET-measurements}
Q (\tau^J_n-\hat m^2) =\,& p^+ \!- \frac{m^2}{Q}= m v_+ \!- \frac{m^2}{Q} + k^+=k^+\,,\\
Q \tau^P_n =\,& p^+ \!- \frac{m^2}{p^-} = k^+ \!+ \frac{m^2}{Q} - \frac{m^2}{Q+k^-} = k^+ \!+ \hat m^2 k^- = 0\,,\nonumber
\end{align}
where to get to the third equality of the second line we have used $Q\gg k^-$ [\,as can be seen from Eq.~\eqref{eq:bhqet-count}\,], and to obtain
$\tau_n^P=0$ we use the on-shell condition for heavy quarks $v\cdot k=0$, to be discussed later in this section.
The field-theoretical definition of the bHQET jet function can be obtained from the expression given in Eq.~\eqref{eq:Jet-cut}, and taking into account
that applying the bHQET power counting to the minus component of momenta one obtains $\delta(\mathcal{\bar P} - Q)\to\delta(k^- + \sum q^-)$, with
$q$ representing momenta of particles other than the heavy quark, the jet functions can be written as
\begin{equation}\label{eq:bJet-cut}
B_n(\hat s) = \frac{(2 \pi)^{d - 1}Q}{2m^2N_C} \mathrm{Tr} \langle 0 | W_{v_+}^\dagger\!(0) h_{v_+}\!(0) \delta\! \biggl[\hat s - \frac{Q^2}{m}(\hat{e}_n-e_{\rm min})\biggr]
\delta^{(d - 2)} (\vec{\mathcal{K}}^{\perp}) \delta (\mathcal{K}^-)
\bar{h}_{v_+}\!(0)W_{v_+}\!(0) | 0 \rangle .
\end{equation}
Here $\vec{\mathcal{K}}$ is an operator that pulls out the residual momenta of the heavy quarks and the (full) momenta of ultra-collinear particles. We have also
used $\vec{v}_\perp = \vec{0}$ and $W_{v_+}$ has the same functional form as $W_n$ replacing the collinear gluons by ultra-collinear fields: $A_n \rightarrow A_+$:
\begin{align}
W_{v_+}(x)&\equiv\bar{P}\exp\biggl[-ig\!\!\int_0^{\infty}\!\dd s\,\,\,\bar{n}\cdot \!A_+(\bar{n}\,s+x)\biggr],\\
W_{v_+}^\dagger(x)&\equiv P\exp\biggl[ig\!\!\int_0^{\infty}\!\dd s\,\,\,\bar{n}\cdot\! A_+(\bar{n}\,s+x)\biggr].\nonumber
\end{align}
The bHQET phase-space integration involving a heavy quark gets also simplified, and using again $p=mv+k$ one has that
\begin{align}
\frac{{\rm d}^dp}{(2\pi)^{d-1}}\,\delta(p^2-m^2)\theta(p^0)
=\,&\frac{{\rm d}p^+{\rm d}p^-{\rm d}^{d-2}\vec p_{\perp}}{2(2\pi)^{d-1}}\,\delta\bigl[p^-p^+-\vert \vec p_{\perp}\vert^2-m^2\bigr]\theta(p^-+p^+)\\
=\,&\frac{{\rm d}k^+{\rm d}k^-{\rm d}^{d-2}\vec k_{\perp}}{2(2\pi)^{d-1}}\,\delta\Bigl(Qk^++\frac{m^2}{Q}k^-+k^2\Bigr)\theta\Bigl(Q+k^-\!+\frac{m^2}{Q}+k^+\Bigr)\nonumber\\
=\,&\frac{{\rm d}k^+{\rm d}k^-{\rm d}^{d-2}\vec k_{\perp}}{2(2\pi)^{d-1}}\,\delta\Bigl(Qk^++\frac{m^2}{Q}k^-\!\Bigr)\theta(Q)
=\frac{{\rm d}k^-{\rm d}^{d-2}\vec k_{\perp}}{2Q(2\pi)^{d-1}}\,,\nonumber
\end{align}
where in the second line we have used Eq.~\eqref{eq:bhqet-count} and in the third we power-count away the $k^2$ in the delta function argument along with all terms but
$Q$ inside the Heaviside function. The on-shell condition for heavy quarks written in light-cone coordinates implies
\mbox{$v^-k^+ + k^- v^+=0$}, in agreement with the argument of the delta function. When comparing to Eq.~\eqref{eq:phase-space} we observe that the $p^-$ in the denominator
got replaced by $Q$ and that the $p^-$ integration is not limited to positive values only. The phase-space integration for ultracollinear particles stays the same as in SCET.

Feynman diagrams look exactly the same in SCET and bHQET, with the replacement $p\to k$ for the heavy quark momenta. Let us compute the tree-level contribution
as given in Fig.~\ref{fig:Tree}, which is analogous to the corresponding SCET calculation:
\begin{equation}
m B_n^{\rm tree}(s) = \!\!\int\! \frac{{\rm d} k^-\!}{4m}{\rm d}^{d-2} \vec{k}_{\perp}
\delta^{(d-2)} (\vec{k}_{\perp}) \delta (k^-) \delta\! \biggl[\hat s - \frac{Q^2}{m}(\hat{e}_n-e_{\rm min})\biggr] \!\sum_s
{\rm Tr}\bigl[u_s (p)\overline{u}_s (p) \bigr] = \delta(\hat s)\,,
\end{equation}
where we have used that the trace of the polarization sum equals $4m$ and have integrated all delta functions except the one with the measurement. The
condition $k^-=0$ imposed by the Dirac delta function makes both (shifted) measurements coincide at tree-level, see Eq.~\eqref{eq:bHQET-measurements}. There are some generic features to be learned from this
diagram: since there is no Dirac structure in the diagram, the trace of the polarization sum will be always $4m$ at any loop order, and since there is always one heavy quark
which brings an inverse power of $Q$ through its phase space one has the following combination:
\begin{equation}
\frac{Q}{2m}\frac{1}{2Q} {\rm Tr}\bigl[u_s (p)\overline{u}_s (p) \bigr] = 1\,,
\end{equation}
which eliminates the spurious dependence on $m$ and $Q$, ultraviolet scales that should not appear in EFT computations. To make this non-dependence explicit at higher orders
one can rescale the minus component of ultracollinear real particles as $q_i^-=(Q/m)\ell_i$, as we shall do in the rest of the section.

We turn our attention now to real-radiation contributions, for which we can simplify the heavy-quark propagator using $v\cdot k=0$.
We start with diagram (a) of Fig.~\ref{fig:Real}, that after applying the bHQET Feynman rules becomes
\begin{equation}
m B_a^{\rm real}(\hat s, \mu) = \frac{2\alpha_s C_F (\mu^2 e^{\gamma_E})^\varepsilon}{\pi\Gamma(1-\varepsilon)}\!
\int\!\! \frac{{\rm d}\ell^-}{(\ell^-)^2}\frac{|\vec q_\perp|^{1-2\varepsilon}{\rm d}|\vec q_\perp|}{v\!\cdot\! q}\theta(\ell^-) \delta\! \biggl[\hat s - \frac{Q^2}{m}(\hat{e}_n-e_{\rm min})\biggr].
\end{equation}
The on-shell condition on the ultra-collinear gluon momenta implies in light-cone coordinates: $2\,v\!\cdot\! q = |\vec q_\perp|^2Q/(m q^-) + mq^-/Q= [\,|\vec q_\perp|^2 + (\ell^-)^2\,]/\ell^-$.
For diagram (c) we get instead
\begin{equation}
m B_c^{\rm real}(\hat s, \mu) = -\frac{\alpha_s C_F (\mu^2 e^{\gamma_E})^\varepsilon}{2\pi\Gamma(1-\varepsilon)}\!
\int\! \frac{{\rm d}\ell^-}{\ell^-}\frac{|\vec q_\perp|^{1-2\varepsilon}{\rm d}|\vec k_\perp|}{(v\!\cdot \!q)^2}\theta(\ell^-) \delta\! \biggl[\hat s - \frac{Q^2}{m}(\hat{e}_n-e_{\rm min})\biggr].
\end{equation}
Let us work out the measurements for thrust and 2-jettiness
\begin{align}
\frac{Q^2}{m} (\tau^J_n-\hat m^2)=\,&\frac{Q}{m}(q^++k^+)=\frac{Q}{m}\frac{|\vec{q}_\perp|^2}{q^-} +\frac{m}{Q}q^-
=\frac{|\vec{q}_\perp|^2}{\ell^-} +\ell^-\,,\\
\frac{Q^2}{m} \tau^P_n=\, &\frac{Q}{m}q^+=\frac{Q}{m} \frac{|\vec{q}_\perp|^2}{q^-}=\frac{|\vec{q}_\perp|^2}{\ell^-} \,,\nonumber
\end{align}
where we have used Eq.~\eqref{eq:bHQET-measurements}, the on-shell condition for heavy quarks and ultra-collinear massless gluons, and the fact that label momentum
conservation implies $k^-=-q^-$. With this result it is very simple to solve the measurement delta function in terms of the perpendicular gluon momenta
\begin{align}\label{eq:delta-bHQET}
\delta\!\biggl[\hat s - \frac{Q^2}{m}(\tau^J_n-\hat m^2)\biggr]=\, & \frac{\ell^-}{2|\vec{q}_\perp|}\,
\delta\!\Bigl[|\vec{q}_\perp|-\sqrt{\hat s\, \ell^- - (\ell^-)^2}\,\Big] ,\\
\delta\biggl(\hat s - \frac{Q^2}{m}\tau^P_n\biggr)=\, & \frac{\ell^-}{2|\vec{q}_\perp|}\,\delta\Bigl(|\vec{q}_\perp|-\sqrt{\hat s\, \ell^-}\,\Bigr),\nonumber
\end{align}
and we will use these results to compute the jet functions in the next two sub-sections.

\subsection{Thrust}
We start with the diagram in which the gluon is radiated from the Wilson line. Switching variables to $\ell^-=\hat s x$ we arrive at
\begin{align}\label{eq:BHQET-a}
m B_{a,P}^{\rm real}(\hat s,\mu) =\,&\frac{\alpha_s(\mu^2e^{\gamma_E})^\varepsilon C_F}{2\pi}\frac{\hat s^{\,-1-2\varepsilon}}{\Gamma(1-\varepsilon)}\int_0^\infty
{\rm d}x\,\frac{x^{-1-\varepsilon}}{1+x}
=-\frac{\alpha_s\Gamma(1+\varepsilon) C_F\,e^{\varepsilon \gamma_E}}{2\pi\mu^2\varepsilon}\biggl(\frac{\hat s}{\mu^2}\biggr)^{\!\!-1-2\varepsilon}\nonumber\\
=\, & \frac{\alpha_s C_F}{4\pi}\biggl[\biggl(\frac{1}{\varepsilon^2}+\frac{\pi^2}{12}\biggr)\delta(\hat s)
-\frac{2}{\varepsilon\mu}\biggl(\frac{\mu}{\hat s}\biggr)_{\!\!+} + \frac{4}{\mu}\biggl(\frac{\mu\log(\hat s/\mu)}{\hat s}\biggr)_{\!\!+} \,\biggr] .
\end{align}
With an identical change of variables we arrive at the following result for diagram (c):
\begin{align}
m B_{c,P}^{\rm real}(\hat s,\mu) =&-\!\frac{\alpha_s(\mu^2e^{\gamma_E})^\varepsilon C_F}{\pi}\frac{\hat s^{\,-1-2\varepsilon}}{\Gamma(1-\varepsilon)}\int_0^\infty
{\rm d}x\,\frac{x^{-\varepsilon}}{(1+x)^2}\\
= & -\!\frac{\alpha_s\Gamma(1+\varepsilon) C_F\,e^{\varepsilon \gamma_E}}{\pi\mu^2}\biggl(\frac{\hat s}{\mu^2}\biggr)^{\!\!-1-2\varepsilon}
=\frac{\alpha_s C_F}{4\pi}\biggl[\frac{2}{\varepsilon}\,\delta(\hat s)-\frac{4}{\mu}\biggl(\frac{\mu}{\hat s}\biggr)_{\!\!+} \,\biggr] .\nonumber
\end{align}
Adding both diagrams with the appropriate factors we obtain the final expression for the P-scheme hemisphere jet function:
\begin{align}
m B_n^P(\hat s,\mu) =\,& \!-\!\frac{\alpha_s\Gamma(2+\varepsilon) C_F\,e^{\varepsilon \gamma_E}}{\pi\mu^2\varepsilon}\biggl(\frac{\hat s}{\mu^2}\biggr)^{\!\!-1-2\varepsilon}\\
=\, & \frac{\alpha_s C_F}{4\pi}\biggl[\biggl(\frac{2}{\varepsilon^2}+\frac{2}{\varepsilon}+\frac{\pi^2}{6}\biggr)\delta(\hat s)
-\frac{4}{\mu}\biggl(\frac{1}{\varepsilon}+1\biggr)\!\biggl(\frac{\mu}{\hat s}\biggr)_{\!\!+} + \frac{8}{\mu}\biggl(\frac{\mu\log(\hat s/\mu)}{\hat s}\biggr)_{\!\!+} \,\biggr] \nonumber.
\end{align}

\subsection{Jettiness}
The Dirac delta function in Eq.~\eqref{eq:delta-bHQET} implies that there is a solution for $|\vec q_\perp|$ only if \mbox{$\ell^- < \hat s$}, which can be implemented through a
Heaviside function and bounds the upper integration limit for $\ell^-$. With the change of variables implemented in the previous section the integration limits are mapped to the
interval $(0,1)$ and we get the following result for diagram (a):
\begin{align}
m B_{a,J}^{\rm real}(\hat s,\mu) =\,& \frac{\alpha_s(\mu^2e^{\gamma_E})^\varepsilon C_F}{2\pi}\frac{\hat s^{\,-1-2\varepsilon}}{\Gamma(1-\varepsilon)}\!\!\int_0^1\!\!
{\rm d}x\,x^{-1-\varepsilon}(1-x)^{-\varepsilon}\nonumber\\
=& -\!\frac{\alpha_s\Gamma(1-\varepsilon) C_Fe^{\varepsilon \gamma_E}}{2\pi\mu^2\varepsilon\Gamma(1-2\varepsilon)}
\biggl(\frac{\hat s}{\mu^2}\biggr)^{\!\!-1-2\varepsilon}\\
=\, & \frac{\alpha_s C_F}{4\pi}\biggl[\biggl(\frac{1}{\varepsilon^2}-\frac{\pi^2}{4}\biggr)\delta(\hat s)
-\frac{2}{\varepsilon\mu}\biggl(\frac{\mu}{\hat s}\biggr)_{\!\!+} + \frac{4}{\mu}\biggl(\frac{\mu\log(\hat s/\mu)}{\hat s}\biggr)_{\!\!+} \,\biggr] ,\nonumber
\end{align}
that, as expected, differs from the expression in Eq.~\eqref{eq:BHQET-a} only in the non-divergent term of the delta-function coefficient. Similarly, we obtain for diagram (c)
\begin{align}
m B_{c,J}^{\rm real}(\hat s,\mu) =\,&\!-\!\frac{\alpha_s(\mu^2e^{\gamma_E})^\varepsilon C_F}{\pi}\frac{\hat s^{\,-1-2\varepsilon}}{\Gamma(1-\varepsilon)}\!\int_0^1\!
{\rm d}x\,[x(1-x)]^{-\varepsilon}\!
\\
=\, & \!-\frac{\alpha_s \Gamma(1-\varepsilon)C_F\,e^{\varepsilon \gamma_E}}{\pi\mu^2\Gamma(2-2\varepsilon)}\biggl(\frac{\hat s}{\mu^2}\biggr)^{\!\!-1-2\varepsilon}=\frac{\alpha_s C_F}{4\pi}\biggl[2\biggl(\frac{1}{\varepsilon}+2\biggr)\,\delta(\hat s)-\frac{4}{\mu}\biggl(\frac{\mu}{\hat s}\biggr)_{\!\!+} \,\biggr] ,\nonumber
\end{align}
again almost identical to the corresponding P-scheme computation. Adding twice the first diagram plus the second we recover the known result for the 2-jettiness bHQET jet
function:
\begin{align}
m B_n^J(\hat s,\mu) =\,& \!-\!\frac{\alpha_s\Gamma(2-\varepsilon) C_F\,e^{\varepsilon \gamma_E}}{\pi\mu^2\varepsilon\Gamma(2-2\varepsilon)}\biggl(\frac{\hat s}{\mu^2}\biggr)^{\!\!-1-2\varepsilon}\\
=\, & \frac{\alpha_s C_F}{4\pi}\biggl[\biggl(\frac{2}{\varepsilon^2}+\frac{2}{\varepsilon}+4-\frac{\pi^2}{2}\biggr)\delta(\hat s)
-\frac{4}{\varepsilon\mu}\biggl(\frac{1}{\varepsilon}+1\biggr)\!\biggl(\frac{\mu}{\hat s}\biggr)_{\!\!+} + \frac{8}{\mu}\biggl(\frac{\mu\log(\hat s/\mu)}{\hat s}\biggr)_{\!\!+}\, \biggr] \nonumber.
\end{align}
Both schemes have the same divergent structure and hence their anomalous dimension, as expected, are identical. Furthermore, the difference between the respective delta coefficients
is the same as that in the SCET jet functions. This result was also expected since both theories should smoothly match in the bHQET limit.

\section{RG Evolution of the SCET Jet Function}\label{eq:evolution}
In this section we solve the renormalization group equation for the non-distributional part of the jet function for thrust and 2-jettiness. This amounts to finding an analytic
expression for the function $I_{\rm nd}$ defined in the last line of Eq.~\eqref{eq:ndRun}. Even though the result for $I^J_{\rm nd}$ has been already worked out in
Ref.~\cite{Fleming:2007xt}, we present here the main steps to find the solution as they are illustrative. Using the rightmost integral expression of the bottom line in
Eq.~\eqref{eq:ndRun} we find
\begin{equation}\label{eq:int-J}
I^J_{\rm nd}(\tilde \omega, y) = \frac{1}{\Gamma(-\tilde\omega)}\int_0^1 {\rm d} z\, (1 - z)^{- 1 - \tilde{\omega}}
\biggl[ \frac{z y}{(1 +z y)^2}-\frac{4 \log (1+z y)}{z y} \biggr].
\end{equation}
While the first term on the right-hand side of Eq.~\eqref{eq:int-J} is already in a canonical form such that Eq.~\eqref{eq:2F1} can be directly applied, the second contains a
logarithm. Expressing it as an integral
\begin{equation}\label{eq:log-int}
\frac{\log (1 + z y)}{z y} = \int_0^1 \dd x \,\frac{1}{1 + x z y}\,,
\end{equation}
brings the second term also into a canonical form that we can easily integrate, finding
\begin{align}
&\int_0^1 \dd z (1 - z)^{- 1 - \tilde{\omega}} \, \frac{\log (1 + z y)}{z y} =\! \int_0^1 \!\dd x \!\int_0^1 \!\dd z (1 - z)^{- 1 - \tilde{\omega}} \frac{1}{1 + x z y} \\
&\qquad\qquad = - \frac{1}{\tilde{\omega}} \!\int_0^1 \!\dd x\,{}_2 F_1 (1, 1, 1 -\tilde{\omega}, - x y) = - \frac{1}{\tilde{\omega}} \,_3 F_2 (1, 1, 1, 2, 1 -\tilde{\omega}, - y), \nonumber
\end{align}
where in the last step we have used the integral representation of the $_3 F_2$ function:
\begin{equation}\label{eq:3F2-inte}
_3 F_2 (a_1, a_2, a_3, b_1, b_2, z) = \frac{\Gamma (b_2)}{\Gamma (a_3)
\Gamma (b_2 - a_3)}\! \int_0^1\! \dd t\, t^{a_3 - 1} (1 - t)^{b_2 - 1 - a_3}
\,{}_2 F_1 (a_1, a_2, b_1, tz) \,,
\end{equation}
with $a_1 = a_2 = a_3 = 1$, $b_1 = 1 - \tilde{\omega}$ and $b_2 = 2$. After adding the result for the first term we find an expression slightly simpler than that quoted in
Ref.~\cite{Fleming:2007xt}, although fully equivalent:
\begin{equation}
I^J_{\rm nd}(\tilde \omega, y) = \frac{1}{\Gamma(2-\tilde{\omega})}\bigl[y\,_2 F_1 (2, 2, 2 -\tilde{\omega}, - y) -
(1-\tilde\omega) \,_3 F_2 (1, 1, 1, 2, 1 -\tilde{\omega}, - y)\bigr]\,,
\end{equation}
which has a smooth $\tilde\omega\to 0$ limit. For a numerical implementation, one can use standard routines to evaluate ${}_2F_1$ hypergeometric functions in programming
languages such as Mathematica, Fortran, Python, or \texttt{C++}. For $_3F_2$ there are built-in routines in Mathematica and Python, while for other languages one can
use a numerical integration over $_2F_1$ as shown in Eq.~\eqref{eq:3F2-inte}.

For P-scheme thrust we can write the logarithm as a derivative to bring all terms into a canonical form:
\begin{align}\label{eq:int-P}
I^P_{\rm nd}(\tilde \omega, y) =\, &\frac{1}{\Gamma(-\tilde\omega)}\! \int_0^1 \dd z (1 - z)^{- 1 - \tilde{\omega}} \left[ \frac{z y - 7}{(1 - y
z)^2} + \frac{2 z y (2 z y - 5)}{(1 - y z)^3} \log (z y) \right]\\
& \frac{1}{\Gamma(-\tilde\omega)}\!\int_0^1\! {\rm d} z\, (1 - z)^{- 1 - \tilde{\omega}} \biggl\{\! \frac{z y - 7}{(1 -	y z)^2}
+ \frac{2 y z (2 z y - 5)}{(1 - y z)^3} \biggl[\log (y) + \frac{\rm d}{{\rm d} \varepsilon}z^{\varepsilon}\biggr] \! \biggr\}_{\varepsilon\to 0}.\nonumber
\end{align}
For $y>1$ each one of the terms in the integral diverges when $z=1/y$. We can regularize the divergence adding a small imaginary part $y\to y+i\epsilon$.
This makes each integral complex, although the sum is real when $\epsilon\to 0$. To express our result in terms of a minimal set of hypergeometric functions,
we use the following identity:
\begin{align}\label{eq:id1}
(c-b){}_2F_1(a,b-1,c,z)&+(c-1)(z-1) {}_2F_1(a,b,c-1,z)\\
&+[z(a-c+1)+b-1] {}_2F_1(a,b,c,z)=0.\nonumber
\end{align}
Furthermore, one can use an additional identity to make the final result manifestly real also for the case $y > 1$
\begin{align}\label{eq:id2}
_2 F_1 (a, b ; c ; z) = \,& \frac{\Gamma (c) (1 - z)^{c - a - b} \Gamma(a+ b - c) \,{}_2 F_1 (c - a, c - b , 1 + c - a - b , 1 - z)}{\Gamma (a) \Gamma(b)} \\
& + \frac{\Gamma (c) \Gamma (c - a - b)\,{}_2 F_1 (a, b , a + b - c + 1, 1 - z)}{\Gamma (c - a) \Gamma (c - b)} . \nonumber
\end{align}
After solving all integrals, recursively applying Eq.~\eqref{eq:id1} and transforming the hypergeometric functions using Eq.~\eqref{eq:id2}, one arrives at
the second important result of this article:\,\footnote{The result as given in this equation is very convenient for a numerical implementation, since one
only needs to evaluate two hypergeometric functions (which might be numerically expensive) using the following approximations:
\begin{align}
\frac{{\rm d}}{{\rm d} \varepsilon}\, {}_2 F_1 (1, 1 + \varepsilon, 2+\tilde{\omega} , 1 - y)\Bigl|_{\varepsilon=0} \simeq \,&\frac{1}{2\varepsilon}
[{}_2 F_1 (1, 1 + \varepsilon, 2+\tilde{\omega} , 1 - y) - {}_2 F_1 (1, 1 - \varepsilon, 2+\tilde{\omega} , 1 - y) ]\,,\\
{}_2 F_1 (1, 1, 2+\tilde{\omega} , 1 - y) \simeq \,&\frac{1}{2}
[{}_2 F_1 (1, 1 + \varepsilon, 2+\tilde{\omega} , 1 - y) + {}_2 F_1 (1, 1 - \varepsilon, 2+\tilde{\omega} , 1 - y) ]\,,\nonumber
\end{align}
with a value of $\varepsilon$ which can be safely taken as small as $10^{-6}$.}
\begin{align}\label{eq:res-der}
I^P_{\rm nd}(\tilde \omega, y)
=\, & \frac{4 y^2 - 2 (\tilde{\omega} + 5) y - \tilde{\omega} (3\,	\tilde{\omega} + 7)}{(1 - y)^2 (1 + \tilde{\omega})\Gamma(1-\tilde \omega)} \biggl\{ \tilde{\omega}
\frac{{\rm d}}{{\rm d} \varepsilon}\, {}_2 F_1 (1, 1 + \varepsilon, 2+\tilde{\omega} , 1 - y)\nonumber \\
&\!\!\!\!\! +\! \bigg( \tilde{\omega} \log (y) - \tilde{\omega} H_{1 -
\tilde{\omega}} - \frac{1 - 2 \tilde{\omega}}{1 - \tilde{\omega}} \biggr) {}_2F_1 (1, 1 ,2 + \tilde{\omega} , 1 - y) \biggr\}_{\!\varepsilon \to 0} \\
& \!\!\!\!-\! \frac{(1 - \tilde{\omega}) \tilde{\omega} (3\, \tilde{\omega} - y +
7) [H_{1 - \tilde{\omega}} - \log (y)] - \tilde{\omega} [3\, \tilde{\omega}
(y + 1) - 5 y + 14] - y + 7}{\Gamma(2 - \tilde{\omega}) (1 - y)^2}\,, \nonumber
\end{align}
with $H_a$ the harmonic number, which for non-integer values of $a$ can be expressed in terms of the digamma function: $H_a = \psi^{(0)} (1 + a) + \gamma_E$.
Equation~\eqref{eq:res-der} has been cast in a way in which the no-resummation limit $\tilde \omega \to 0$ is smooth. The singularities that appear in individual
terms of $J^P_{\rm nd}$ for $x=1$ manifest themselves now as a double pole in $I^P_{\rm nd}$ at $y=1$, which is however fictitious, as the result is indeed smooth at this value. To solve this
problem in numerical implementations we provide in Sec.~\ref{sec:y1Expansion} an expansion of this result around $y=1$ at an arbitrarily high order. The result in Eq.~\eqref{eq:res-der}
is adequate for a numerical implementation since the derivative with respect to $\varepsilon$ can be taken numerically through finite differences. It can be also performed analytically,
using Eq.~\eqref{eq:id2} in ${}_2 F_1 (1, 1 + \varepsilon, 2+\tilde{\omega} , 1 - y)$ and the following identity
\begin{equation}\label{eq:derivative}
\frac{\text{d}}{\text{d} \varepsilon} {}_2 F_1 (1, 1 + \varepsilon , 1 - \tilde{\omega} + \varepsilon, y) \biggl|_{\varepsilon \rightarrow 0} \!\!=
- \frac{\tilde{\omega}\, y (1 - y)^{- 1 - \tilde{\omega}}}{(1 - \tilde{\omega})^2}
{}_3 F_2 (1 - \tilde{\omega}, 1 - \tilde{\omega}, 1 - \tilde{\omega} , 2 - \tilde{\omega}, 2 - \tilde{\omega}, y) ,
\end{equation}
to arrive at the equivalent expression:
\begin{align}\label{eq:realy}
& I^P_{\rm nd}(\tilde \omega, y) = \\
& \frac{\tilde{\omega} \,y (1 - y)^{-3 - \tilde{\omega}} [\,2\,y\,(\tilde{\omega} + 5) + \tilde{\omega} (3 \tilde{\omega} + 7) - 4 y^2\,]
\,{}_3F_2 (1 - \tilde{\omega}, 1 - \tilde{\omega}, 1 - \tilde{\omega} , 2 -	\tilde{\omega}, 2 - \tilde{\omega} , y)}{\Gamma(2-\tilde\omega)(1 - \tilde{\omega})} \nonumber\\
& - \frac{(1 - \tilde{\omega})\, \tilde{\omega} \,(3\, \tilde{\omega} - y +7) [\,H_{1 - \tilde{\omega}} - \log (y)\,] - \tilde{\omega}\, [\,3 \,\tilde{\omega}\,
(y + 1) - 5 \,y + 14\,] - y + 7}{\Gamma(2-\tilde\omega) (1 - y)^2} \nonumber\\
& +\frac{[\,2\,y\, (\tilde{\omega} + 5) + \tilde{\omega}\, (3\, \tilde{\omega} +
7) - 4 y^2\,] \{ (1 - \tilde{\omega}) [\,H_{1 - \tilde{\omega}} - \log (y)\,] - 1
\}_2 F_1 (1, 1, 1 - \tilde{\omega} , y)}{\Gamma(2-\tilde\omega) (1 - y)^2} .\nonumber
\end{align}
The $(1 - y)^{-3 - \tilde{\omega}}$ factor and both hypergeometric functions are complex for $y>1$ but the combination is real. To have all terms explicitly real for $y>1$
one can use the following relation
\begin{align}\label{eq:alejandro}
& \frac{\text{d}}{\text{d} \varepsilon} {}_2 F_1 (1, 1 + \varepsilon , 2
+ \tilde{\omega} , 1 - y) \Bigr|_{\varepsilon \rightarrow 0} =
\frac{1 - y}{y^2 (2 + \tilde{\omega})^2} \,{}_3 F_2 \biggl(\! 2, 2 +
\tilde{\omega}, 2 + \tilde{\omega} , 3 + \tilde{\omega}, 3 + \tilde{\omega}
, 1 - \frac{1}{y} \biggr)\nonumber\\
& \qquad + y^{\tilde{\omega}} (1 - y) \frac{1 + \tilde{\omega}}{(2 + \tilde{\omega})^2}\,
{}_3 F_2 (2 + \tilde{\omega}, 2 + \tilde{\omega}, 2 + \tilde{\omega} , 3 +
\tilde{\omega}, 3 + \tilde{\omega} , 1 - y)\,,
\end{align}
which does not rely on numerical derivatives, is manifestly real for all positive values of $y$ but is numerically unstable if $y\to 0$. This poses no problem in practice,
since for $y < 1$ one can simply switch to Eq.~\eqref{eq:realy}. To derive the result in Eq.~\eqref{eq:alejandro} we proceed as follows:
\begin{align}\label{eq:der-int}
&\frac{\text{d}}{\text{d} \varepsilon} {}_2 F_1 (1, 1 + \varepsilon , 2
+ \tilde{\omega} , 1 - y) \Bigr|_{\varepsilon \rightarrow 0} = -\frac{1}{y} \frac{\text{d}}{\text{d} b}
{} _2 F_1\! \biggl( 1, b , 2 + \tilde{\omega} , 1 - \frac{1}{y} \biggr) \biggl|_{b \rightarrow 1 + \tilde\omega}\nonumber\\
&\qquad=\frac{1}{y}\frac{\text{d}}{\text{d} b}\biggl[ {} _2 F_1 \biggl( 1, 1 + \tilde{\omega} , b + 1 , 1 - \frac{1}{y} \biggr) -
{} _2 F_1 \biggl( 1, b , b + 1 , 1 -\frac{1}{y} \biggr)\biggr]_{b \rightarrow 1 + \tilde{\omega}}\\
&\qquad=\frac{1}{y}\frac{\text{d}}{\text{d} b}\biggl[ y^{1 + \tilde{\omega}} {}_2 F_1 (b, 1 + \tilde{\omega} , b + 1 , 1 - y) -
{} _2 F_1 \biggl( 1, b , b + 1 , 1 -\frac{1}{y} \biggr)\biggr]_{b \rightarrow 1 + \tilde{\omega}}\nonumber
\end{align}
where in the first step we have used Eq.~\eqref{eq:Euler-1}, in the second we apply the chain rule on derivatives, and in the third line we use again Eq.~\eqref{eq:Euler-1} on the
first term. Using the identity
\begin{equation}
\frac{\text{d}}{\text{d} a}\,{}_2 F_1 (a, b , a + 1 , z) = \frac{b z}{(a + 1)^2} \,{} _3 F_2 (a + 1, a + 1, b + 1 , a + 2, a + 2 , z)\,,
\end{equation}
in the two terms of the last line in Eq.~\eqref{eq:der-int} we arrive at the result quoted in Eq.~\eqref{eq:alejandro}. In Appendix~\ref{sec:yReal} we present an alternative (although more
complicated) expression for $I^P_{\rm nd}$ which does not involve numerical derivatives and with every term manifestly real for $y>1$. We use this result as an additional
cross check of our analytic derivations. In any case, we shall see that for numerical implementations one never needs to use expressions involving hypergeometric functions.

\subsection[Expansion around $s=0$]
{\boldmath Expansion around $s=0$}\label{sec:exp-s0}
For numerical implementation purposes, it might be convenient to obtain an analytic expansion of $I_{\rm nd}(\tilde \omega, y)$ around $y=0$. One can do so by using the known
expansions for the hypergeometric functions, e.g.
\begin{equation}\label{eq:hyper-series}
_2 F_1 (a, b , c , z) = \frac{\Gamma (c)}{\Gamma (a) \Gamma (b)} \sum_{i= 0}^{\infty} \frac{\Gamma (a + i) \Gamma (b + i)}{\Gamma (c + i) \Gamma (i+ 1)}\, z^i\,,
\end{equation}
but in order to have a relation valid at arbitrarily high orders it is simpler to use the expansion of $J_{\rm nd}(x)$ around $x=0$
\begin{align}
J_{\rm nd}^P(x) =\,& - \!\sum_{i = 0} [\,6\, i + 7+ i (7 + 3 i) \log(x)\,]\, x^i\,,\\
J_{\rm nd}^J(x) =\,&- \!\sum_{i = 0} \biggl( i + \frac{4}{i + 1} \biggr) (- x)^i\,,\nonumber
\end{align}
on the leftmost expression of the bottom line in Eq.~\eqref{eq:ndRun} and integrate analytically term by term. 
It turns out that one can sum up the corresponding series using Eq.~\eqref{eq:hyper-series} to recover an expression analytically equivalent to Eq.~\eqref{eq:realy}. The master integrals that
we will need are
\begin{align}
\frac{y^{\tilde{\omega}}}{ \Gamma (-\tilde{\omega})} \int_0^y \text{d} x\, (y - x)^{- 1 - \tilde{\omega}} x^i =\,& y^i \frac{\Gamma (1 + i)}{\Gamma (1 + i - \tilde{\omega})}\,,\\
\frac{y^{\tilde{\omega}}}{ \Gamma (-\tilde{\omega})} \int_0^y \text{d} x\, (y - x)^{- 1 - \tilde{\omega}} x^i \log(x) =\,& y^i \frac{\Gamma (1 + i)}{\Gamma (1 + i - \tilde{\omega})}
[H_i-H_{i-\tilde\omega}+\log(y)]\,,\nonumber
\end{align}
where the bottom line can be obtained from the top one acting with a derivative with respect to $i$. We then arrive at
\begin{align}
I^P_{\rm nd}(\tilde \omega, y) =\,& \!-\!\frac{1}{\Gamma(1-\tilde\omega)}\! \sum_{i = 0} \frac{i!}{(1 - \tilde{\omega})_i}\{(6 i + 7)+ i (7 + 3 i) [H_i-H_{i-\tilde\omega}+\log(y)]\,\}\, y^i\,,\\
I^J_{\rm nd}(\tilde \omega, y) =\,& \!-\! \frac{1}{\Gamma(1-\tilde\omega)}\!\sum_{i = 0} \biggl( i + \frac{4}{i + 1} \biggr)\frac{i!}{(1 - \tilde{\omega})_i} (-y)^i\,.\nonumber
\end{align}
where we have used the Pochhammer symbol $(a)_n\equiv\Gamma(a+n)/\Gamma(a)$ since it is convenient for an optimized numerical implementation. Both series converge well
for $|y| < 1$, and therefore apply mainly in the peak of the distribution. For 2-jettiness the series can be easily summed up using Eq.~\eqref{eq:hyper-series} and the series
expansion for the $_3F_2$ hypergeometric function:
\begin{equation}
_3 F_2 (a, b, c , d, e , z) = \frac{\Gamma (c)}{\Gamma (a) \Gamma (b)}
\sum_{i = 0}^{\infty} \frac{\Gamma (a + i) \Gamma (b + i) \Gamma (c + i)}{\Gamma (d + i) \Gamma (e + i) \Gamma (i + 1)} \,z^i \,.
\end{equation}
For P-scheme thrust one can convert the term involving harmonic numbers into the derivative of ratios of gamma functions to use Eq.~\eqref{eq:hyper-series} and
recover the result we already obtained with a direct integration. The numerical implementation of Ref.~\cite{Butenschoen:2016lpz} (which dealt with 2-jettiness) did not use
this expansion and the evaluation of the non-distributional jet function running was the most severe performance bottleneck for the code.

\subsection[Expansion around $s=m^2$]{\boldmath Expansion around $s=m^2$}\label{sec:y1Expansion}
The results obtained in Eqs.~\eqref{eq:res-der} and \eqref{eq:realy} are not useful for a numerical implementation in the vicinity of $y=1$. When $y$ is sufficiently close to unity one
can switch to a series expansion to arbitrary high power using the change of variables $z\to 1- z$ in the rightmost expression at the bottom of Eq.~\eqref{eq:ndRun} and the following
expansion:
\begin{align}
&J_{\rm nd}^P[(1 - z) (1 + y)] = - \biggl[ \frac{2 (1 - z) (3 + 2 z)}{z^3} \log (1 - z) + \frac{z + 6}{z^2} \biggr] - y (1 - z) \biggl[ 2\frac{9 - 2 z - 2
z^2}{z^4} \nonumber\\
& \times \log (1 - z) + \frac{5 z + 18}{z^3} \biggr] \!- y^2 (1 - z) \biggl\{ \frac{2(1 - z) (18 - 3 z - 2 z^2)}{z^5} \log (1 - z) \nonumber\\
& + \frac{36 - 24 z - 7 z^2}{z^4} \biggr\} \!- \log (1 - z) \sum_{i = 3} y^i \frac{(1 - z)^i}{z^{i + 3}} [3 (i + 1) (i + 2) - 2 (i + 1) z - 4 z^2] \\
& - \frac{1}{2} \sum_{i = 3} y^i \frac{(1 - z)^{i - 1}}{z^{2 + i}} \bigl\{
6 (i + 1) (i + 2) - (i + 1) (3 i + 10) z - 2 [1 - (i - 5) i] z^2 \bigr\}
\nonumber\\
& + \sum_{i = 3} y^i \sum_{k = 0}^{i - 3} (- 1)^{k + i} \frac{(1 -z)^{k + 1}}{z^{k + 3}} \frac{(k + 1) (k + 2) (6 - 5 i + 5 k + 4 z)}{(i - k- 2) (i - k - 1) (i - k)}. \nonumber
\end{align}
Terms have been combined such that the coefficient of each power in $y$ has a well-defined $z\to 0$ limit and therefore we can integrate coefficient by coefficient.
In practice one can integrate each piece assuming a non-integer value of $i$ and subsequently convert the gamma functions that would become divergent if $\tilde \omega=0$
using the identity
\begin{equation}
\frac{\Gamma (\varepsilon - n)}{\Gamma (1 + \varepsilon)} = \frac{(- 1)^{n -
1} \Gamma (- \varepsilon)}{\Gamma (n + 1 - \varepsilon)} .
\end{equation}
As expected, there are large cancellations among the various terms for a given power of $y$, but when adding all contributions one gets the following nicely convergent series:
\begin{align}
I^P_{\rm nd}(\tilde \omega, y) =\,& \!-\!\frac{2 (5\, \tilde \omega+9) \tilde \omega H_{-\tilde \omega}+7\, \tilde \omega^3+19\, \tilde \omega^2+
22\, \tilde \omega+18}{(\tilde \omega+1) (\tilde \omega+2) (\tilde \omega+3) \Gamma (1-\tilde \omega)}\\
&\! -\!\frac{\tilde \omega\,\Gamma(1+\tilde \omega)}{\Gamma(1-\tilde \omega)}\sum_{i=3}(1 - y)^i
\sum_{k=0}^{i-3} \frac{(k+1) (k+2)! [(k+2) (5 \,\tilde \omega+9)-5 i (\tilde \omega+1)]}{(i-k-2) (i-k-1) (i-k) \Gamma (k+\tilde \omega+4)}\nonumber\\
& \!+\! \frac{\Gamma(1+\tilde \omega)}{2\,\Gamma(1-\tilde \omega)}\sum_{i=1}\frac{i! (1-y)^i}{\Gamma (i+\tilde \omega+4)}\big\{
\tilde \omega (i+\tilde \omega+1) [2 i^3+i^2 (\tilde \omega-3)+i (\tilde \omega (5 \,\tilde \omega-8)-39) \nonumber\\
&\!-\!(\tilde \omega+6) (5\, \tilde \omega+9)] +[20 i-2 \,i \, \tilde \omega (i (3 \,\tilde \omega+7)+7 \, \tilde \omega+9)+4 (5 \, \tilde \omega+9)]\nonumber\\
&\!\times\! [ \tilde \omega\, \psi ^{(0)}(i+1)- \tilde \omega\, \psi ^{(0)}(1- \tilde \omega)-1] \big\},\nonumber
\end{align}
where again special care has been taken to write the expression in a manner in which one can set $\tilde \omega = 0$ without any worries. The series converges well for
$|1-y| < 1$, and therefore, combined with the expansion worked out in the previous section, for P-scheme thrust one can use expansions if $y< 2$.

\subsection[Expansion around $s = \infty$]{\boldmath Expansion around $s = \infty$}
Since the numerical evaluation of hypergeometric functions is slow, it is convenient to figure out another series expression (in this case, of asymptotic type) around $s= \infty$, which is of
course tantamount to $m= 0$. This limit is very relevant, since it can be applied in the tail of the distribution and almost everywhere if the heavy quark mass is much smaller than the
center-of-mass energy, as is the case for bottom quarks at LEP. Such asymptotic expansion was not known by the time in which Refs.~\cite{Abbate:2010xh,Abbate:2012jh,Butenschoen:2016lpz}
were published, and was significantly affecting the performance of the respective numerical codes. Even though one could, in principle, use known results for the asymptotic expansions of $_2F_1$
and $_3F_2$ hypergeometric functions, it is in practice simpler and more efficient to compute the series directly on its integral form. This is complicated since, as we shall see, the expansions involve
powers of $\log(y)$, and so one cannot simply expand the integrand and integrate term by term, as we did in Secs.~\ref{sec:exp-s0} and \ref{sec:y1Expansion}. We found out that the Mellin-Barnes representation
\begin{equation}\label{eq:MB}
\frac{1}{(1 + X)^{\nu}} = \frac{1}{2 \pi i} \int_{c - i \infty}^{c + i
\infty} \dd t \,(X)^{- t} \frac{\Gamma (t) \Gamma (\nu - t)}{\Gamma (\nu)}\,,
\end{equation}
with $0<c<\nu$, is optimal to achieve our goal~\cite{Friot:2005cu}.\footnote{This representation can also be used to solve the RG equation exactly. Applying a Mellin transformation to the first line of Eq.~\eqref{eq:int-P} in the $y$-variable, solving the $z$-integral and transforming back one gets a closed (and rather short) analytic expression for $I_{\rm nd}^P$ in terms of MeijerG functions,
\begin{equation}
I_{\rm nd}^P(\tilde{\omega}, y) =3\, G_{3,3}^{2,3}\Biggl(y\Biggl|
\begin{array}{c}
0,0,0 \\
1,1,\tilde{\omega} \\
\end{array}
\Biggr)-7\, G_{3,3}^{2,3}\Biggl(y\Biggl|
\begin{array}{c}
0,0,0 \\
0,1,\tilde{\omega} \\
\end{array}
\Biggr),
\end{equation}
which are not very convenient for a direct numerical evaluation, but can be related to hypergeometric functions.}
After applying Eq.~\eqref{eq:MB}, the asymptotic expansion around $X\gg 1$ is obtained integrating by residues the poles
that appear on the real axis for $t>\nu$ (the poles for $t\leq 0$ correspond to the expansion \mbox{$X\ll1$}). We work out this expansion for thrust and 2-jettiness in the rest of this section,
but before that we note that the asymptotic expansion is well convergent if $1/y < 1$, which for \mbox{P-scheme} thrust means that in numerical evaluations one can always
use one of the three expansions discussed in this section and never needs to evaluate hypergeometric functions with dedicated routines. For jettiness the same statement is almost
true, except in a small vicinity of $y=1$ in which, to the best of our knowledge, no expansion can be used.

\subsubsection*{2-Jettiness}
We start from the integral form given in Eq.~\eqref{eq:int-J}. The only complication in this case is that we have to deal with a logarithm, which does not have
the form in Eq.~\eqref{eq:MB}. However, it can be brought to the standard form using Eq.~\eqref{eq:log-int}
\begin{align}
\frac{1}{y z} \log (1 + z y) =\, & \frac{1}{2 \pi i} \int_0^1 \dd x \!\int_{d - i
\infty}^{d + i \infty} \!\dd t \,(x y z)^{- t} \Gamma (t) \Gamma (1 - t) \\
= \, &\frac{1}{2 \pi i} \!\int_{d - i \infty}^{d + i \infty} \!\dd t\, (zy)^{- t} \frac{\Gamma (t) \Gamma (1 - t)}{1 - t}, \nonumber
\end{align}
with $0<d<1$. Since the denominator of the first term in Eq.~\eqref{eq:int-J} is squared, when applying the Mellin-Barnes representation \eqref{eq:MB} the first pole appears at $t=2$.
This is accompanied by an extra power of $y$, such that we can nicely map the poles of the first term into those of the second by shifting the integration variable $t\to t + 1$ in the former.
After integrating over $z$ we obtain
\begin{equation}
I^J_{\rm nd}(\tilde \omega, y) = \frac{1}{2 \pi i} \!\int_{c - i \infty}^{c +
i \infty}\! \dd t \,y^{ - t} \frac{\Gamma (1 - t)^2 \Gamma (t)}{\Gamma(1 - t - \tilde{\omega})} \frac{t^2 - t + 4}{t - 1} \,, \nonumber
\end{equation}
with $0<c<1$. The integrand has a triple pole at $t=1$ and double poles at natural values of $t$ larger than $1$. We compute the triple pole by itself and treat the rest generically
using
\begin{equation}\label{eq:G2-exp}
\Gamma (\varepsilon-n)^2 = \frac{1}{(n!)^2} \biggl[\frac{1}{\varepsilon^2} + \frac{2 (H_n - \gamma_E)}{\varepsilon} \biggr] +\mathcal{O} (\varepsilon^0)\,.
\end{equation}
With this result we obtain the following asymptotic expansion:
\begin{align}
I^J_{\rm nd}(\tilde \omega, y) =\,&\frac{1}{\Gamma(1-\tilde \omega)}\sum_{n = 1} (- y)^{- n} c_n[\tilde{\omega}, \log (y)]\,,\\
c_{1} (\tilde{\omega}, L) = \,& -1 - 2 \tilde{\omega} [H_{- \tilde{\omega}} - L]^2 - (4 + \tilde{\omega} )
[H_{- \tilde{\omega}} - L] - [1 + \pi^2 - 2 \psi^{(1)} (1 -\tilde{\omega})] \tilde{\omega}\,,\nonumber\\
c_{n>1} (\tilde{\omega}, L) =\, & \frac{(1 + \tilde{\omega})_{n - 1}}{(n - 1)^2 (n - 1) !} \{ (n - 1) [(n -
2) (n + 1) + 6] [\tilde{\omega} (H_{n - 1} - H_{n + \tilde{\omega} - 1} + L)\nonumber \\
&- \cos (\pi \tilde{\omega}) \Gamma (1 - \tilde{\omega})
\Gamma (1 + \tilde{\omega})] - (n - 3) (n + 1) \,\tilde{\omega} \}\,,\nonumber
\end{align}
using again the Pochhammer symbol. We have written each coefficient in a form such that the $\tilde\omega\to 0$
limit, relevant in the far tail of the distribution, is smooth.

\subsubsection*{P-scheme thrust}
Applying the Mellin-Barnes representation in Eq.~\eqref{eq:MB} to the first line of Eq.~\eqref{eq:int-P} and integrating over $z$ we arrive at an expression that involves
different powers of $y$ with poles shifted accordingly. Therefore, using the same strategy as in the previous section, we can shift the integration variable by one or two
units such that poles and powers of $y$ in each term exactly match. This is very important, since the expansion in $1/y$ must be carried out consistently given the large
cancellations that take place among the various terms due to the divergence at $x=1$ of individual terms in $J_{\rm nd}^P$ (exactly as it happened for the expansion around
$s=m^2$). After some work we arrive at the following expression:
\begin{align}\label{eq:MB-J}
I_{\rm nd}^P(\tilde{\omega}, y) =\, & \frac{1}{2 \pi i} \int_{c - i \infty}^{c +i \infty} \dd t\, y^{- t}\frac{ \cos (\pi t) \Gamma (1 - t)^2 \Gamma (t) }{\Gamma (1 - t -
\tilde{\omega})}\\
& \times \{ (7 - 3 t) t [H_{- t} - H_{- t - \tilde{\omega}} + \log (y)] + 6 t -
7 \}\, . \nonumber
\end{align}
We have already implemented a few simplifications because we assume the result is real, and therefore discarded the imaginary parts that would arise from $(- y)^{-t}$. We have
checked that indeed this is the case as long as one expands strictly in $y$ without mixing any powers. Harmonic numbers are caused by the term in $J^P_{\rm nd}$ proportional to
$\log(z)$. The integrand has now double and triple poles, located at natural values of $t$, the latter arising precisely because of the harmonic numbers. There are no poles arising
from $H_{- t - \tilde{\omega}}$ because the corresponding gamma function in the denominator has poles at the same values, making the ratio regular. To compute the
residues of the poles we need, on top of Eq.~\eqref{eq:G2-exp}, the following expansion
\begin{align}
H_{\varepsilon- n} \Gamma (\varepsilon- n)^2 =\, & \frac{1}{(n!)^2}
\biggl\{ - \frac{1}{\varepsilon^3} + \frac{1}{\varepsilon^2} [\psi^{(0)} (n)
- 2 H_n + 3 \gamma_E] \\
&\, + \frac{2 \gamma_E n + 2 (n \gamma_E - 1) n \psi^{(0)} (n) -
3}{n^2 \varepsilon} \biggr\} + \mathcal{O} (\varepsilon^0) \,, \nonumber
\end{align}
which can be obtained from the relation between harmonic numbers and the digamma function and a bit of algebra. Using these results we arrive at the following expression, in which again
special care has been taken to make the $\tilde\omega \to 0$ limit smooth:
\begin{align}
I^P_{\rm nd}(\tilde \omega, y) =\,&\frac{1}{\Gamma(1-\tilde \omega)} \sum_{n = 1} \frac{c_n [\tilde{\omega}, \log
(y)]}{y^n}\,, \\
c_n (\tilde{\omega}, L) = \,&\frac{(1+\tilde{\omega})_{n - 1}}{2 (n -
1) !} \biggl\{ \frac{2}{n + \tilde{\omega}} [L (3 n - 7) n (n +
\tilde{\omega}) - 3 n^2 - 6 n \tilde{\omega} + 7 \tilde{\omega}]
\nonumber\\
& \times [\cos (\pi \tilde{\omega}) \Gamma (1 - \tilde{\omega}) \Gamma
(1 + \tilde{\omega}) - \tilde{\omega} (\psi^{(0)} (n) - \psi^{(0)} (n +
\tilde{\omega} + 1))] \nonumber\\
& + \frac{\tilde{\omega}}{n (n + \tilde{\omega})} [2 Ln (3 n (n + 2
\tilde{\omega}) - 7 \tilde{\omega}) - 3 n (\tilde{\omega} - 3 n +
7) - 7 \tilde{\omega}] \nonumber\\
& - (7 - 3 n) n \biggl[ \tilde{\omega}\, \psi^{(1)} (n + 1) -
\tilde{\omega} \psi^{(1)} (n + \tilde{\omega}) - \tilde{\omega} L^2 -
\frac{\tilde{\omega}}{(n + \tilde{\omega})^2} \nonumber\\
& + [\psi^{(0)} (n) - \psi^{(0)} (n + \tilde{\omega} + 1)]
[\tilde{\omega} (\psi^{(0)} (1 + n + \tilde{\omega}) - \psi^{(0)} (n)) \nonumber\\
&+ 2 \cos (\pi \tilde{\omega}) \Gamma (1 - \tilde{\omega}) \Gamma (\tilde{\omega}
+ 1)] \vphantom{\frac{\tilde{\omega}}{(n + \tilde{\omega})^2}} \biggr]
\biggr\}. \nonumber
\end{align}

\section{Kinematic, Mass and Hadronization Power Corrections}\label{sec:power}
The resummed SCET cross section can be matched to full QCD such that its validity is extended beyond the peak and tail into the far tail. The usual procedure
is to add in fixed-order those terms not included in the factorization theorem. For massless quarks these are usually denoted
as non-singular contributions, since singular terms (that is, delta or plus functions) are fully accounted for in SCET. For massive quarks, terms not contained in the
factorization theorem can be singular, and therefore these will be referred to as non-SCET. In the far tail one sets all renormalization scales equal due to the large cancellations
that take place between SCET and non-SCET terms around $\tau\sim 1/3$ that would be spoiled by resummation. To ensure this cancellation when including
hadronization effects one usually convolves the added terms with the same shape function.

As already explained in the introduction and discussed in further detail in Sec.~\ref{sec:finalSCET}, the fixed-order QCD prediction contains terms which are singular as $\tau$
approaches $0$. The reason is that when including quark masses there are two kind of power corrections to the leading-order EFT prediction: kinematic and massive.
Both are power-counted equally in SCET and therefore, when considering the leading-power cross section one necessarily neglects higher powers of $m$. Mass power corrections
can be singular, while kinematic power corrections are genuinely non-singular. It is in general desirable to absorb mass power corrections into the EFT description (although this
is not strictly speaking necessary), since resummation turns $\log^n(\tau)/\tau$ into an integrable singularity. This prescription has been adopted
already for 2-jettiness in Refs.~\cite{Abbate:2010xh,Abbate:2012jh,Butenschoen:2016lpz}, and here we succinctly explain how this is implemented.

One can modify the SCET matrix elements to absorb all singular pieces. Since at tree-level all matrix elements are either $1$ or Dirac delta functions, one only needs
to multiply the tree-level SCET results by $R^C_0$ to fully account for massive power corrections at this order. The $\mathcal{O}(\alpha_s)$ massive corrections can be implemented
modifying the hard function, that we write as
\begin{equation}
H^C\!(Q,\hat m,\mu) = R_0^C(\hat m) \biggl\{ 1 + \frac{\alpha_s(\mu) C_F}{4 \pi} \biggl[ \frac{7
\pi^2}{3} - 16 + h^C_m(\hat m) + 12 \log \biggl( \frac{Q}{\mu} \biggr) - 8 \log^2 \!\biggl(\frac{Q}{\mu} \biggr)\! \biggr]\! \biggr\},
\end{equation}
such that it includes the tree-level mass modification,\footnote{Therefore one has to rescale the non-distributional jet function $J_{\rm nd}\to J_{\rm nd}/R^C_0(\hat m)$ as well.} and the jet
function (at this order one cannot have mass corrections from soft dynamics). For the latter one only needs to modify $J_m$, the mass correction with distributions, which we write as
\begin{equation}
J^C_m(x,\hat m) = [\,j^C_m(\hat m) + A_P\,] \delta(x) + 4 \biggl[ \frac{ \log(x)}{x} \biggr]_+ - [\,1-y^C_m(\hat m)\,] \biggl( \frac{1}{x}\biggr)_+.
\end{equation}
Corrections coming from $B^C_{\rm plus}$ are easy to implement, since they only come from the jet function (the hard function does not contain any distributions). It is convenient
to define $A^C(\hat m)= A^{\rm SCET}(\hat m) + A_C^{\rm NS}(\hat m)$ and $B^C_{\rm plus}(\hat m) = B^{\rm SCET}(\hat m) + B^C_{\rm NS}(\hat m)$, with
\begin{equation}
A^{\rm SCET}(\hat{m}) = \frac{2 \pi^2}{3} - 2 + 2 A_P + 4 \log (\hat{m}) + 16\log^2 (\hat{m})\,.
\end{equation}
Implementing these modifications into the SCET factorization theorem with fixed scales one arrives at
\begin{align}
y^C_m(\hat m) =\,&\frac{1}{2 R^C_0(\hat m)} \{[1 - R^C_0(\hat m)] B_S(\hat m) + B^C_{\rm NS}(\hat m)\}\,,\nonumber\\
R^C_0(\hat m) \,[\,h^C_m(\hat m) + 2 j^C_m(\hat m)\,] =\,& 2 \,\{B_S(\hat m) [1 - R^C_0(\hat m)] + B^C_{\rm NS}(\hat m)\} \log (\hat{m}) \\
&\!\!+ [1 - R^C_0(\hat m)] A_S(\hat m) + A^C_{\rm NS}(\hat m) \equiv H^C_{\rm corr}(\hat m)\,.\nonumber
\end{align}
Since one cannot resolve separately $h^C_m$ and $j^C_m$, we make an ansatz and split it according to a parameter $\xi$
\begin{equation}\label{eq:xi}
h^C_m(\hat m) = \frac{1 - \xi}{R^C_0(\hat m)} H^C_{\rm corr}(\hat m)\,, \qquad j^C_m(\hat m) = \frac{\xi}{2R^C_0(\hat m)} H^C_{\rm corr}(\hat m) \,.
\end{equation}
This parameter reflects our lack of knowledge on the structure of mass power corrections. To estimate the associated uncertainty we vary it between $0$ and $1$,
such that for the extreme values the correction is fully contained in the hard or jet functions. Once all singular corrections have been absorbed into the SCET matrix elements
we can incorporate the truly non-singular corrections as an additive term
\begin{align}\label{eq:partonic-final}
\frac{\dd\hat \sigma^C}{\dd \tau} =\,& \frac{\dd \hat\sigma^C_{\rm SCET}}{\dd \tau} + \frac{\dd \hat\sigma^C_{\rm NS}}{\dd \tau} \,,\\
\frac{1}{\sigma^C_0} \frac{\dd \hat\sigma^C_{\rm NS}}{\dd \tau} =\,& \frac{\alpha_s(\mu)C_F}{4\pi}[F_{\rm NS}^C (\tau, \hat{m})-F_{\rm NS}^{\rm SCET} (\tau, \hat{m})]\,,\nonumber
\end{align}
where $\dd\hat\sigma^C_{\rm SCET}/\dd \tau$ refers to the mass-corrected (resummed) partonic SCET cross section, that is, to Eq.~\eqref{eq:FacMom} with the substitutions
$H\to H^C$ and $J_m\to J_m^C$ and with resummation kernels implemented. A similar strategy can be carried out to add power corrections to the bHQET cross section, but
these will be discussed elsewhere.

So far we have dealt with partonic cross sections. Although for infrared- and collinear-safe observables partonic predictions are already a good description of the full result, for
a precision analysis hadronization cannot be ignored. Here we will be concerned with the dominant effect of hadronization, that comes from soft dynamics and is already contained in the
leading-power SCET factorization theorem. It is well known that for $Q\tau \gg \Lambda_{\rm QCD}$ the main effect of soft hadronization is shifting the distribution to the right
$\tau \to \tau - \Omega_1/Q$ (due to the operator product expansion or OPE), with $\Omega_1$ a non-perturbative parameter that can be defined in terms of field-theory matrix
elements. In the peak of the distribution, hadronization is more complex and must be taken into account by convolving the partonic result with a hadronic shape function $F(p)$:
\begin{equation}\label{eq:hadronic-final}
\frac{\dd \sigma^C(\tau)}{\dd \tau} = \!\int_0^{Q\tau}\!\!\dd p\, \frac{\dd\hat \sigma^C}{\dd \tau}\!\biggl(\tau- \frac{p}{Q}\biggr)F(p)\,.
\end{equation}
The shape function has support for $p>0$ and is normalized $\int_0^\infty \dd p F(p)=1$. It is strongly peaked at $p\sim \Lambda_{\rm QCD}$ and has an exponential tail extending
towards infinity that ensures any of its moments is well defined. This behavior enforces the OPE and one has \mbox{$\int_0^\infty \dd p\,p F(p)=\Omega_1$}. As discussed in
Ref.~\cite{Hoang:2007vb}, $\Omega_1$ is afflicted by an $u=1/2$ renormalon that can be removed with appropriate subtractions defined on the partonic soft
function~\cite{Hoang:2008fs}. Since at the order we are working these effects are not yet relevant they will not be discussed any longer.

So far we have presented all our results in the pole scheme for the heavy quark mass $m\equiv m_p$. Expressing our cross section in the $\MSb$ scheme is trivial at this order, since
for P- and E-scheme thrust there is no mass dependence at lowest order except in $R_0^C$. Therefore one simply has to substitute $m\to\overline m(\mu)$ everywhere the mass
appears: jet function, mass power corrections and fixed-order kinematic power corrections, and add an $\alpha_s$ correction from the conversion of $R_0^C$ to the $\MSb$ scheme.
We associate the $\MSb$ mass renormalization scale to $\mu_J$ since the jet function is the main responsible for mass effects at the order we are working
(at higher orders one can have mass effects coming from gluon splitting in the soft function).

\section{Numerical Analysis}\label{eq:numerics}
We have implemented our result for the cross section as given in Eqs.~\eqref{eq:partonic-final} and \eqref{eq:hadronic-final} in two independent numerical codes, which use
Mathematica~\cite{mathematica} and Python~\cite{Rossum:1995:PRM:869369}, respectively, that agree with each other within more than $8$ significant digits. For the evaluation of
dilogarithms ${\rm Li}_2$, hypergeometric ${}_2F_1$, ${}_3F_2$ and polygamma functions $\psi^{(n)}$, as well as for interpolations and numerical integration in Python we use
the \texttt{scipy} module~\cite{Virtanen:2019joe}, that builds on the \texttt{numpy} package~\cite{oliphant2006guide}, which is also used for numerical constants such
as $\pi$ or $\gamma_E$. In Mathematica we simply use built-in native functions.

While for the partonic SCET cross section all ingredients are analytic, the partonic non-SCET cross section is only known numerically through the results of
Ref.~\cite{Lepenik:2019jjk}. The algorithm used in that article allows to determine the fixed-order cross section with high-precision, and in practice numerical errors are negligibly
small. Our strategy to parametrize the fixed-order cross section is based on a combination of fit functions and interpolations. In a first step we make the curve less
divergent at threshold by subtracting the known singular structures. This leaves integrable log-type singularities that cannot be described with an interpolation. To make the
curve smoother as $m\to 0$ we also subtract the SCET non-distributional contribution. We split this doubly-subtracted cross section into two regions that meet at
$\tau=\tau_{\,\rm lim} = 0.0016661$, and use a fit function below $\tau_{\,\rm lim}$ and an interpolation above, constructed in a way such that the curve is smooth at the
junction. The fit function is the sum of a term linear in $\log(\tau)$ multiplied by a degree-$7$ polynomial in $\tau$ plus a second polynomial of the same degree. The logarithm
contains the expected behavior of non-singular terms. The coefficients of these two polynomials are functions of the reduced mass, and each one of them is parametrized with a
fit function of $\hat m$, which again consists on the sum of a $10^{\rm th}$-degree polynomial in $\hat m$, and $\log(\hat m)$ times another polynomial of identical degree. For the
interpolation we take an evenly spaced grid with $0.033$ bin-size for $\tau > 0.026$ and a finer grid below. While the values of $\tau$ in the grid do not depend on the mass, the
height of each node does, and we use fit functions of $\hat m$ to parametrize this dependence with the same functional form used for $\tau <\tau_{\,\rm lim}$: a logarithm of the
mass times a polynomial of degree $10$, plus another polynomial of the same order. To code this parametrization in Python in a way which is flexible and efficient we use object-oriented
programming.

The convolution of the (now fully analytic) partonic cross sections with the shape function is performed numerically. Although our code is completely general,
for the plots shown in this section we use the simplest shape function built from the basis proposed in Ref.~\cite{Ligeti:2008ac}:
\begin{equation}
F(\ell)= \frac{128\, \ell^3\, e^{-\frac{4 \ell}{\lambda }}}{3 \lambda ^4}\,,
\end{equation}
with $\lambda=0.5\,$GeV. To ensure that resummation is properly implemented in the peak
and tail of the distribution, being smoothly switched away in the far-tail, 
we employ the profile functions introduced in Ref.~\cite{Hoang:2014wka}. 
It is reasonable to think that the presence of a non-zero quark mass should modify the profile functions, but since we consider here physical situations in which
the mass is still small, we stick to mass-independent parametrizations. More sophisticated profile parametrizations depending on the value of $m$ were employed e.g.\ in
Ref.~\cite{Butenschoen:2016lpz}. All plots and analyses carried out in the rest of this section use the $\MSb$ scheme for the heavy quark mass. Furthermore, we do not implement
gap subtractions since they are not very relevant when matrix elements are used at the one-loop level. Unless stated otherwise, we take $n_\ell=4$ massless quarks and a massive
bottom with $\mbar_b(\mbar_b)=4.2\,$GeV. For the strong coupling we use 4-loop running with the boundary condition \mbox{$\alpha_s^{(n_f=5)}(m_Z)=0.1181$}.

\begin{figure*}[t!]
\subfigure[]
{
\includegraphics[width=0.43\textwidth]{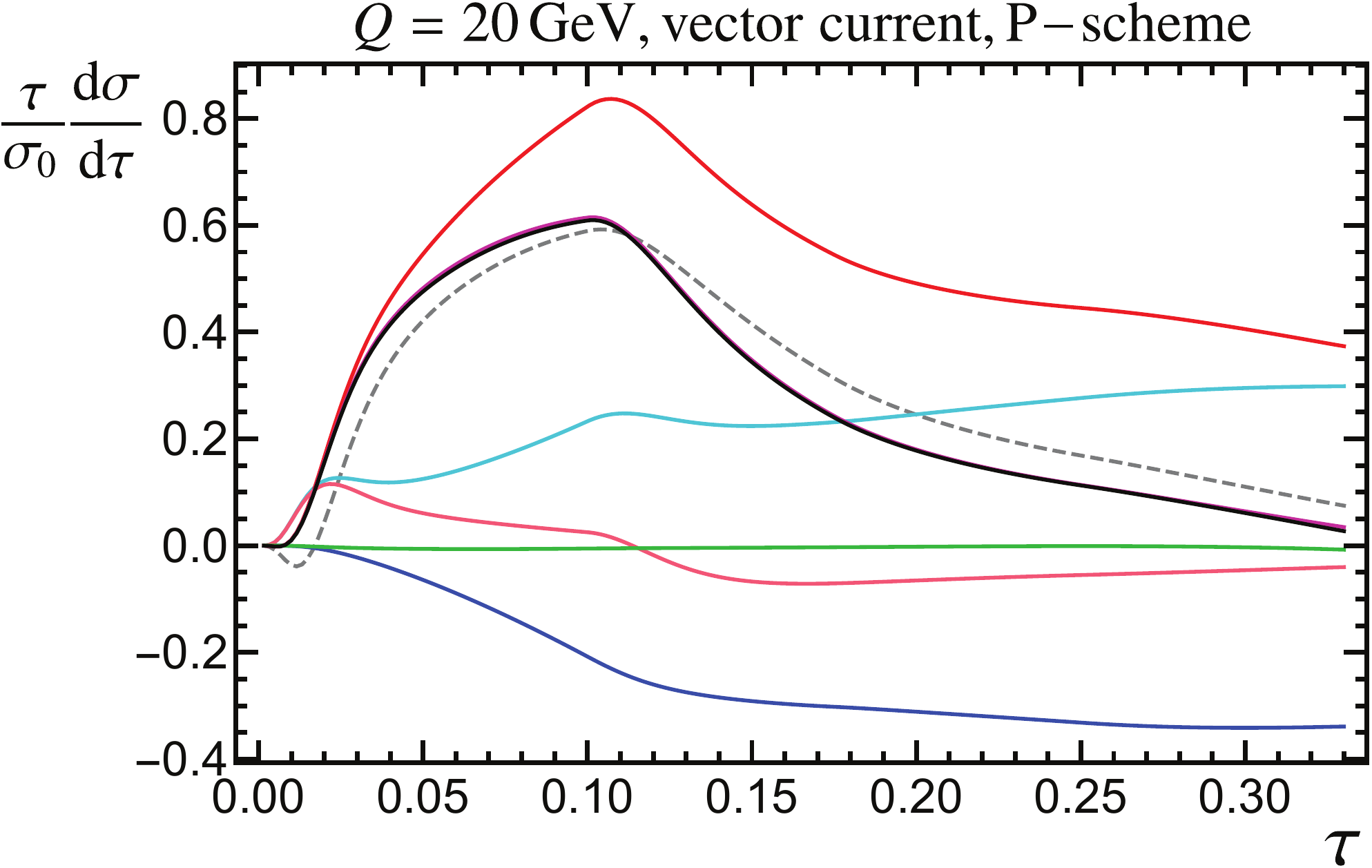}
\label{fig:C-20-vec}
}~~~
\subfigure[]{
\includegraphics[width=0.58\textwidth]{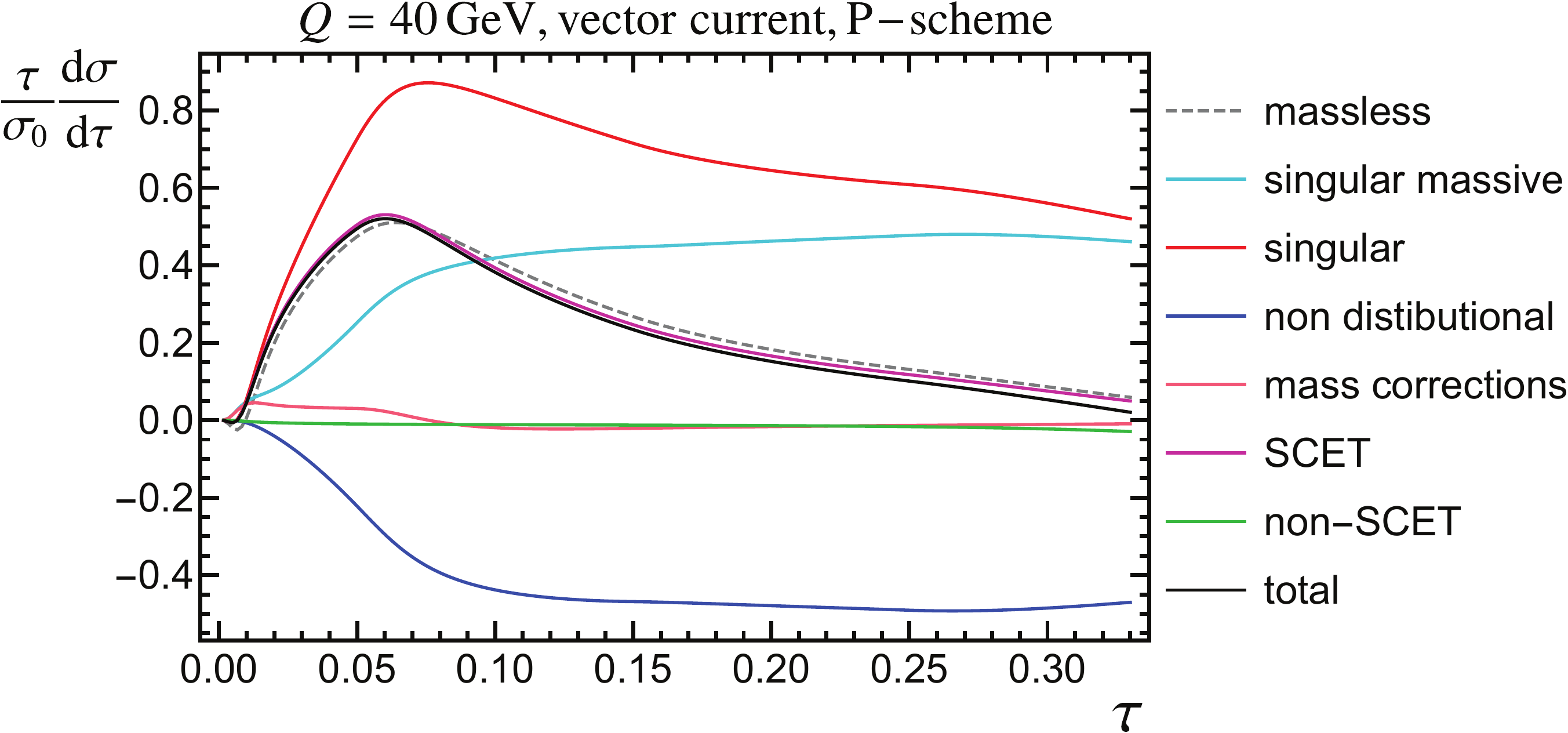}
\label{fig:C-40-vec}
}	\subfigure[]
{
\includegraphics[width=0.43\textwidth]{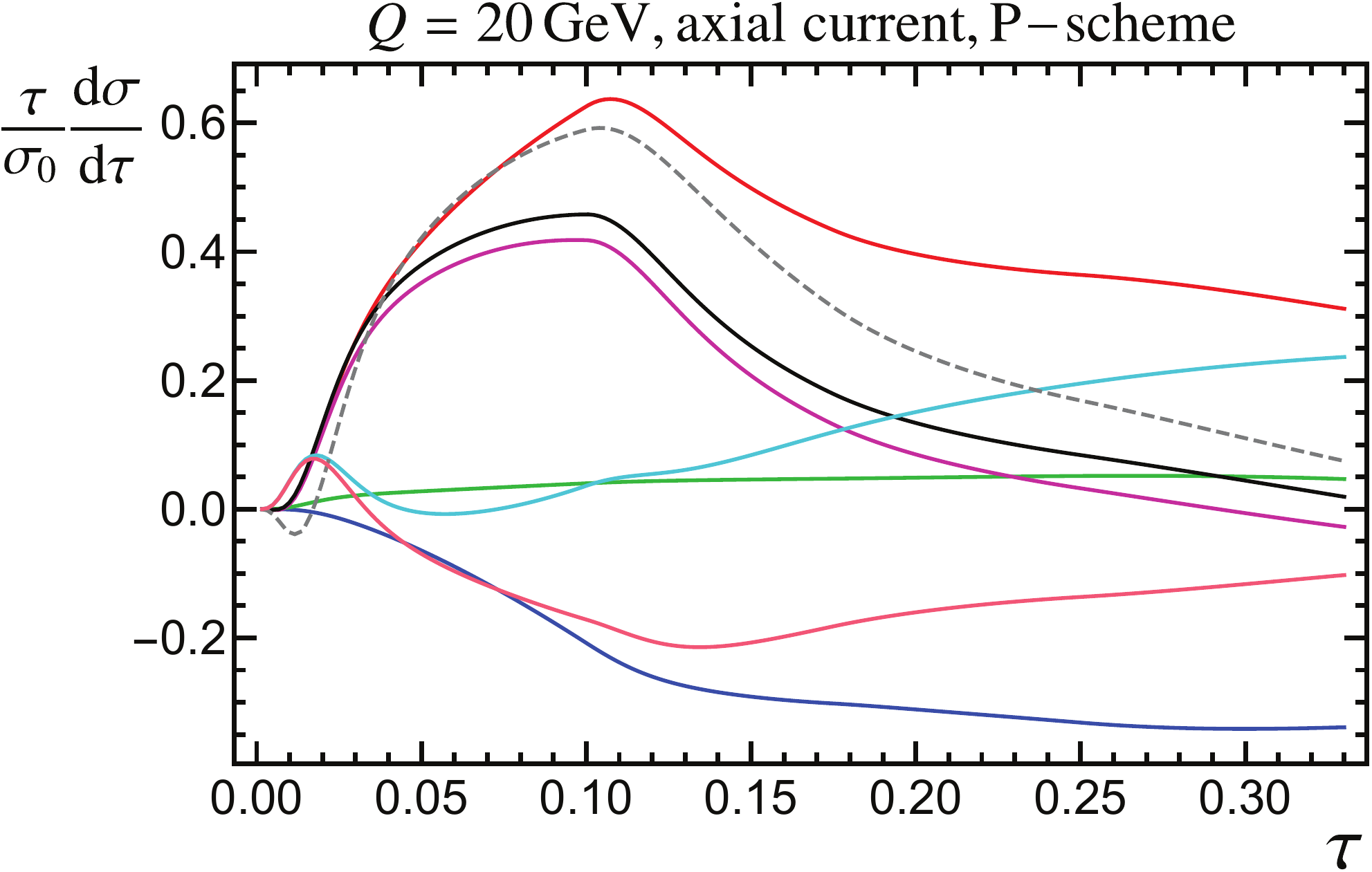}
\label{fig:C-20-ax}
}~~~
\subfigure[]{
\includegraphics[width=0.43\textwidth]{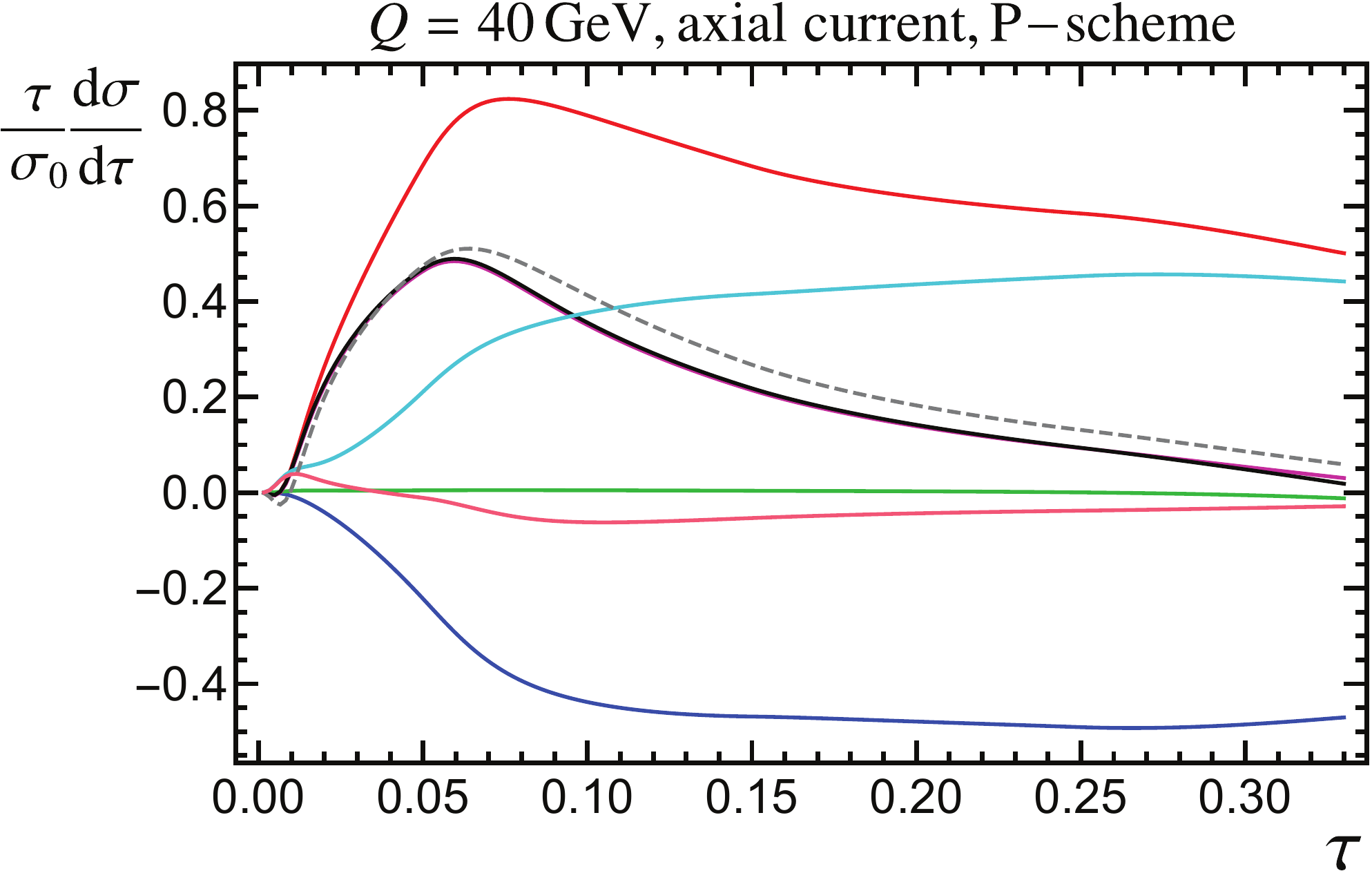}
\label{fig:C-40-ax}
}
\caption{Decomposition of the differential cross section at $Q=20\,$GeV (left panels) and $40\,$GeV (right panels) in various components for the vector (upper plots) and
axial-vector (lower plots) currents. Red and blue correspond to the singular and non-distributional terms, respectively, while their sum defines the SCET cross section,
shown in magenta. The massless approximation is shown as a dashed gray line, while massive singular corrections are depicted in cyan. The massive corrections to the
SCET cross section (massive singular plus non-distributional) are shown in pink. Finally, the black solid line is the sum of all contributions. }
\label{fig:components}
\end{figure*}
We start our numerical discussion by analyzing the size of each term in Fig.~\ref{fig:components}, which shows differential cross sections for $Q=20\,$GeV and $40\,$GeV for
vector and axial-vector currents. We use only the default parameters for the profiles and set the parameter $\xi$ defined in Eq.~\eqref{eq:xi} to its canonical value $0.5$. We split
the distribution as follows (to alleviate notation, in the remainder of this section we drop the superscript $C$ that indicates the current):
\begin{align}
\frac{\dd \sigma}{\dd \tau} =\,& \frac{\dd \sigma_{\rm SCET}}{\dd \tau} + \frac{\dd \sigma_{\rm NS}}{\dd \tau} \equiv
\frac{\dd \sigma_{\rm sing}}{\dd \tau} + \frac{\dd \sigma_{\rm nd}}{\dd \tau} + \frac{\dd \sigma_{\rm NS}}{\dd \tau} \\
\equiv \,& \frac{\dd \sigma^{\rm sing}_{m=0}}{\dd \tau} + \frac{\dd \sigma^{\rm sing}_m}{\dd \tau} +
\frac{\dd \sigma_{\rm nd}}{\dd \tau} + \frac{\dd \sigma_{\rm NS}}{\dd \tau} \nonumber
\equiv \frac{\dd \sigma^{\rm sing}_{m=0}}{\dd \tau} + \frac{\dd \sigma^{\rm SCET}_m}{\dd \tau} + \frac{\dd \sigma_{\rm NS}}{\dd \tau} \,,\nonumber
\end{align}
\begin{figure*}[t!]
\subfigure[]
{
\includegraphics[width=0.47\textwidth]{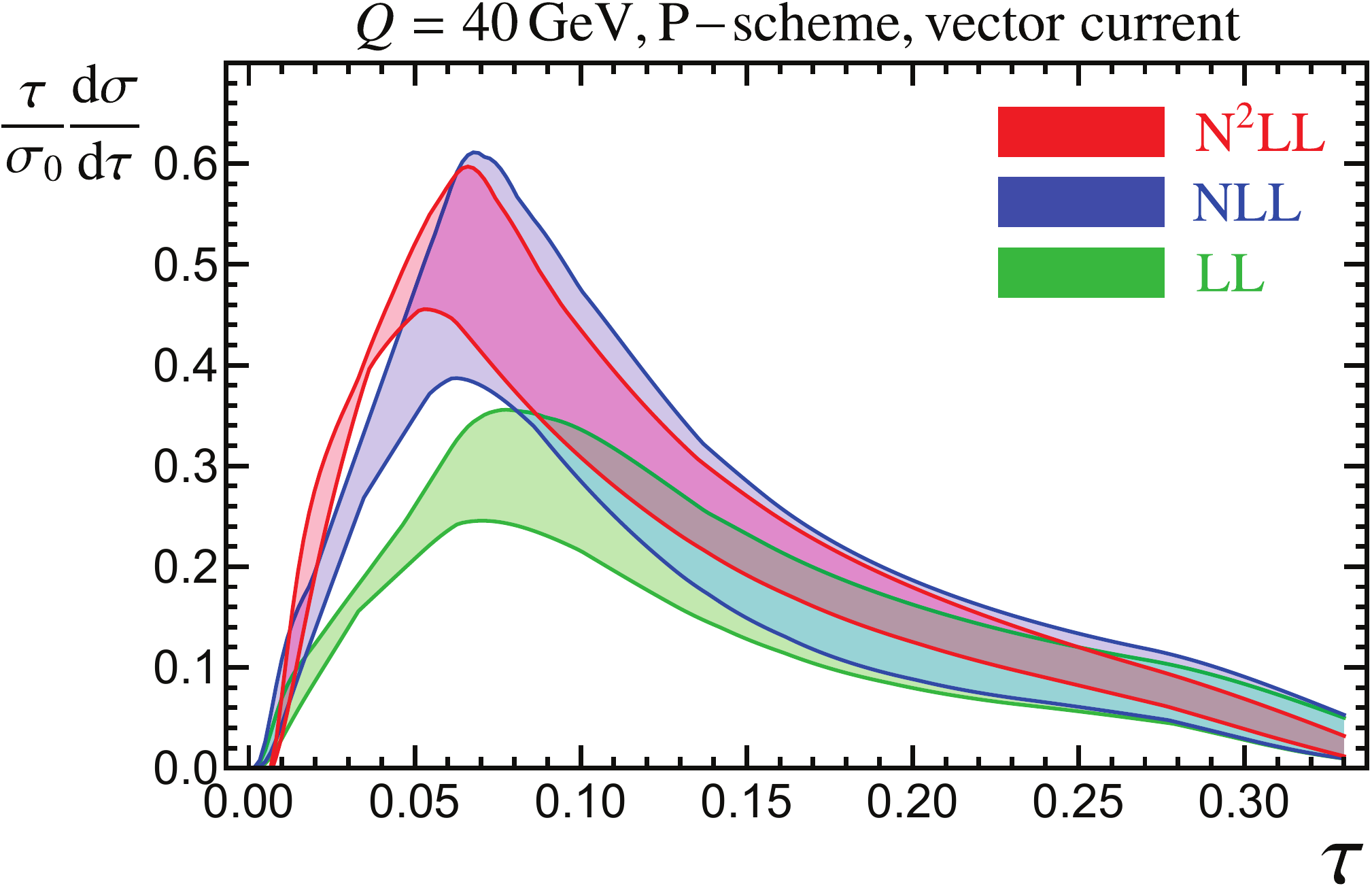}
\label{fig:40-vec}
}~~~
\subfigure[]{
\includegraphics[width=0.48\textwidth]{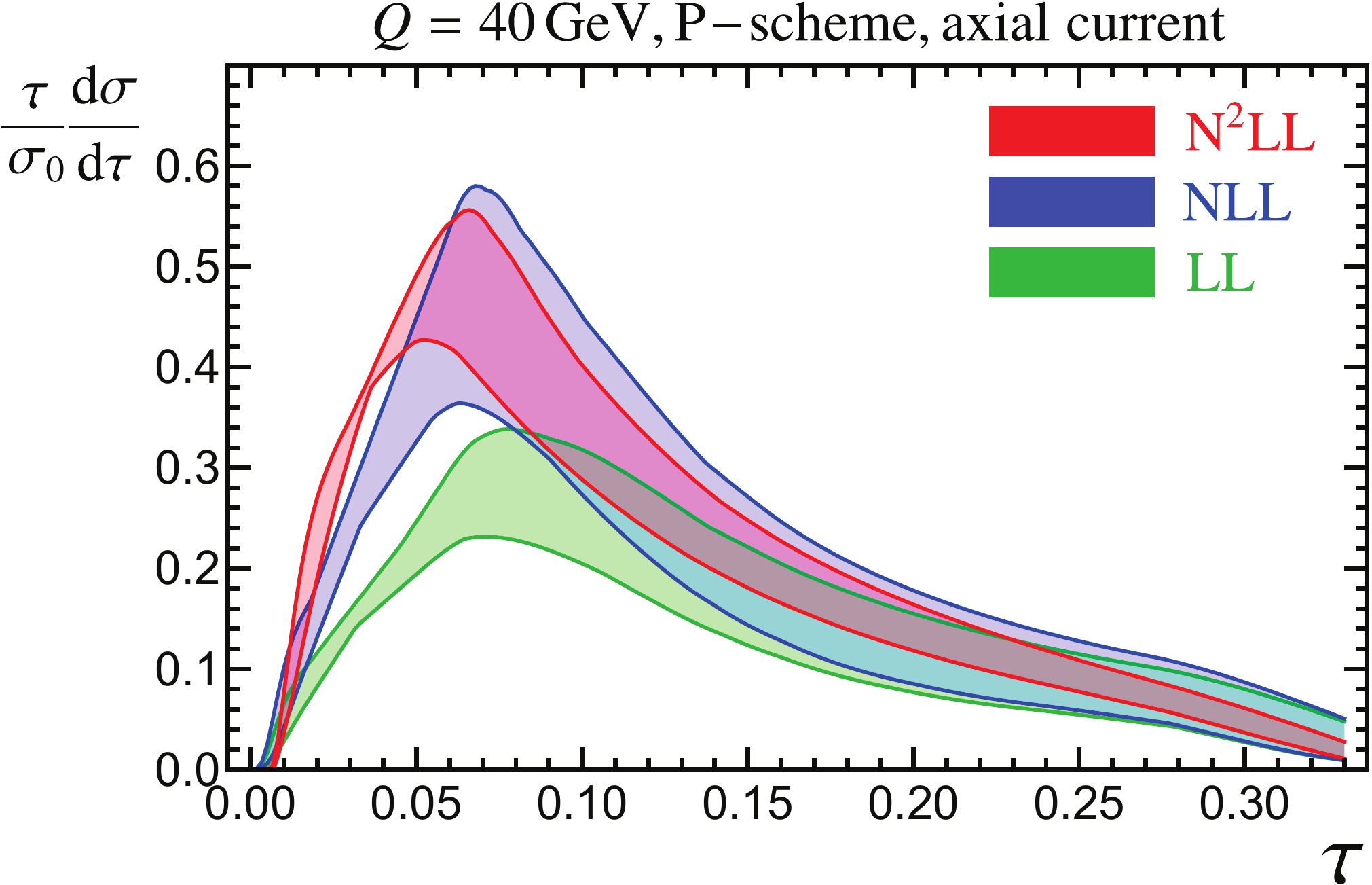}
\label{fig:40-ax}
}	\subfigure[]
{
\includegraphics[width=0.47\textwidth]{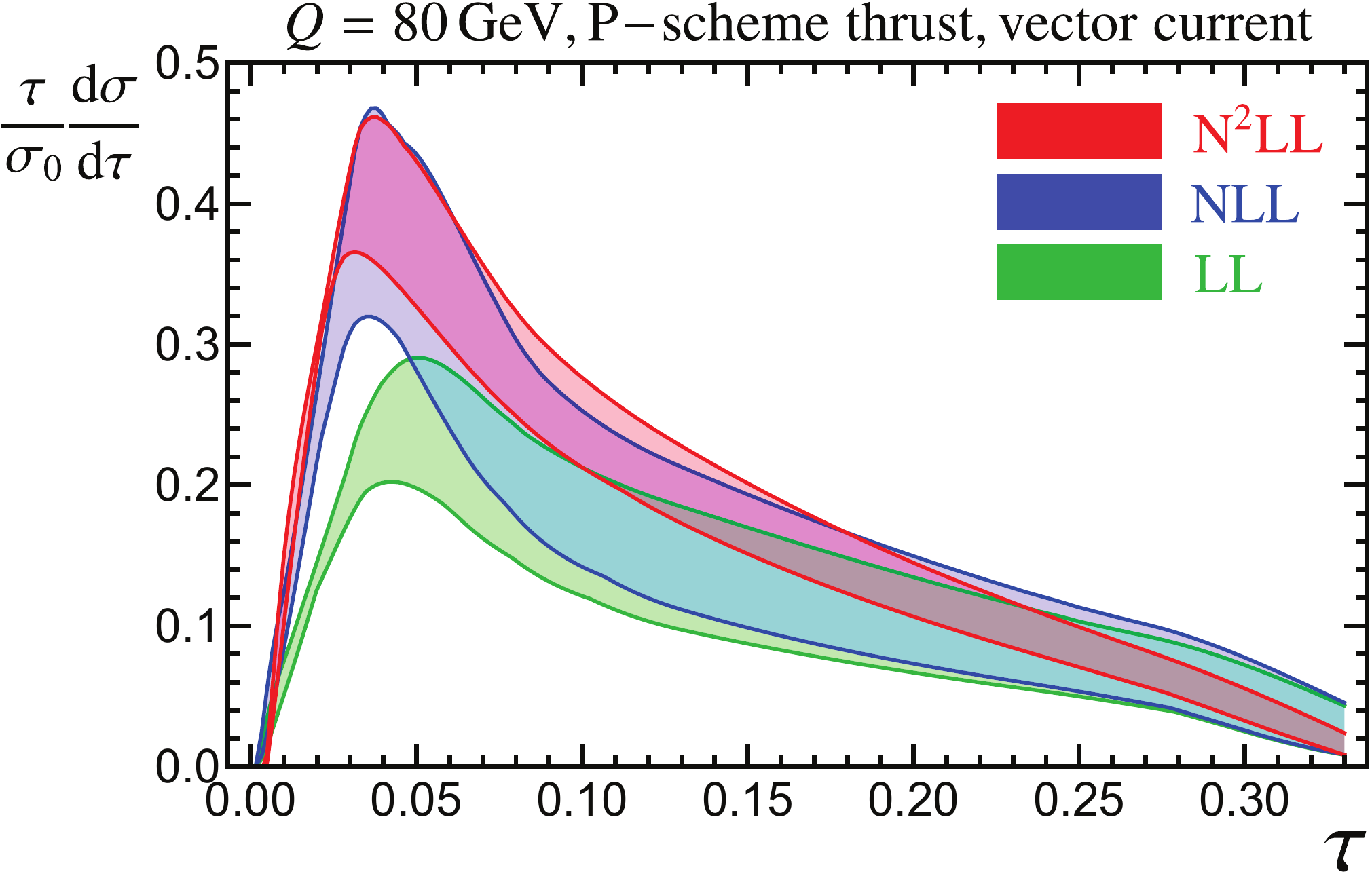}
\label{fig:80-vec}
}~~~
\subfigure[]{
\includegraphics[width=0.48\textwidth]{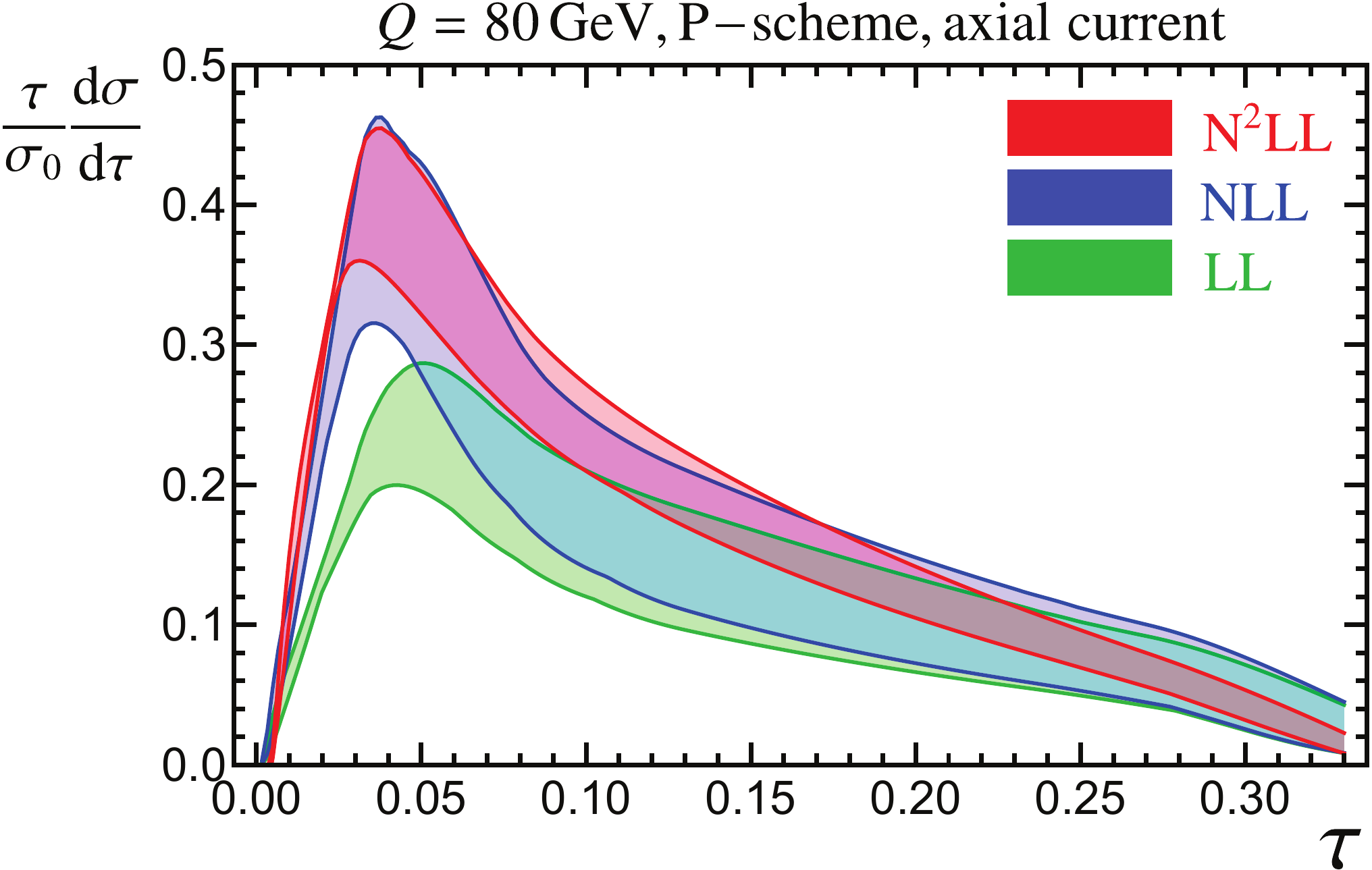}
\label{fig:80-ax}
}
\caption{Uncertainty bands for LL (green), NLL (blue) and N$^2$LL (red) for P-scheme thrust cross sections at $40$\,GeV (figures on top) and $80\,$GeV
(figures at the bottom), for vector (left figures) and axial-vector (right figures) currents. The bands are obtained with $500$ profiles generated randomly selecting
values for the parameters that define them.}
\label{fig:bands}
\end{figure*}
\!where each term contains hadronization power corrections computed as a convolution with the same shape function. In the first equality we split the full cross section (shown as
a black solid line) in SCET and non-SCET contributions, shown in magenta and green, respectively. The SCET cross section can be further divided into the sum of singular
$\dd\sigma^{\rm sing}_{m}/\dd\tau$ (shown as a red solid line) and non-distributional $\dd\sigma_{\rm nd}/\dd\tau$ (in solid blue) contributions. In our setup, the singular cross section
is defined as the contribution from terms in the SCET factorization theorem which are singular at threshold if no resummation is implemented. At N$^2$LL, these correspond to the
distributions that arise from the hard function, the 1-loop soft function, and the $J_{m=0}$, $J_m$ pieces of the one-loop jet function, with the modifications discussed in Sec.~\ref{sec:power}
to absorb the relevant mass corrections, integrated against the resummation kernels. The non-distributional terms (shown in blue) are defined as the resummed contribution from $J_{\rm nd}$
defined in Eq.~\eqref{eq:ndRun}. We observe that while the singular contribution is positive, the non-distributional is negative, and they significantly cancel each other when added
together. The singular distribution can be cast as the sum of the massless approximation $\dd\sigma^{\rm sing}_{m=0}/\dd\tau$ (shown as a dashed gray line) and singular massive corrections
$\dd\sigma^{\rm sing}_{m}/\dd\tau$ (cyan solid line). The massless approximation is quite close to the SCET cross section (specially for the vector current), as expected, since the P-scheme
decreases the sensitivity to the quark mass, and the singular massive corrections are very similar to the non-distributional term up to a global sign. We define the SCET massive corrections
$\dd\sigma^{\rm SCET}_{m}/\dd\tau$ (pink solid line) as the sum of the singular massive corrections and the non-distributional terms, which turns out to be rather small, specially
for larger values of the center-of-mass energy. The non-SCET cross section has been defined in Eq.~\eqref{eq:partonic-final} and contains non-distributional kinematic corrections coming
from the QCD fixed-order cross section. Interestingly, once we absorb all singular terms into the SCET factorization theorem, the non-SCET corrections are absolutely negligible everywhere
except in the far tail. Within the setup defined in Sec.~\ref{sec:power} only the non-distributional cross section and the massless approximation is the same for vector and axial-vector currents.
\begin{figure*}[t!]
\subfigure[]
{
\includegraphics[width=0.47\textwidth]{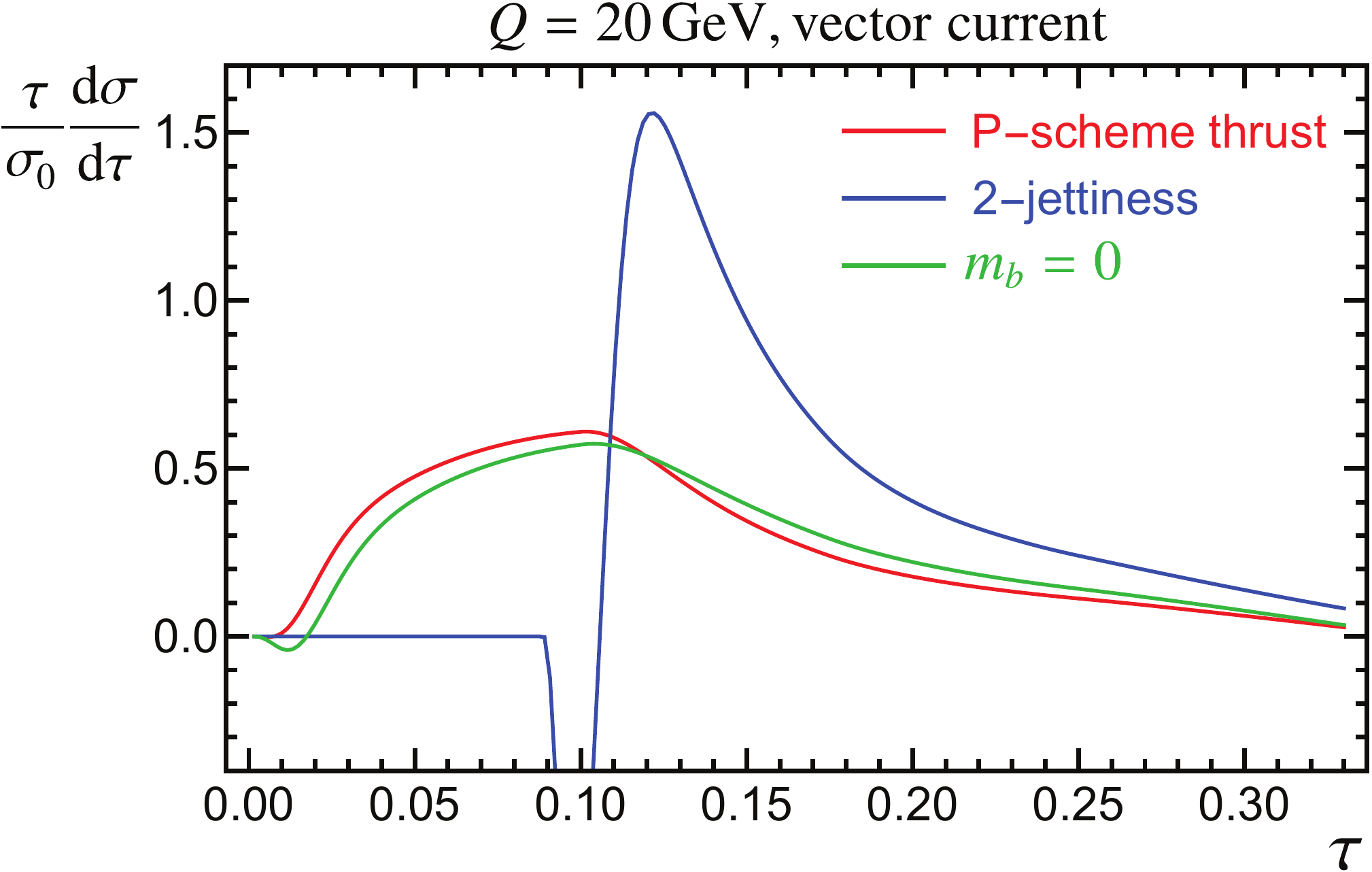}
\label{fig:schemes-20}
}~~~
\subfigure[]{
\includegraphics[width=0.48\textwidth]{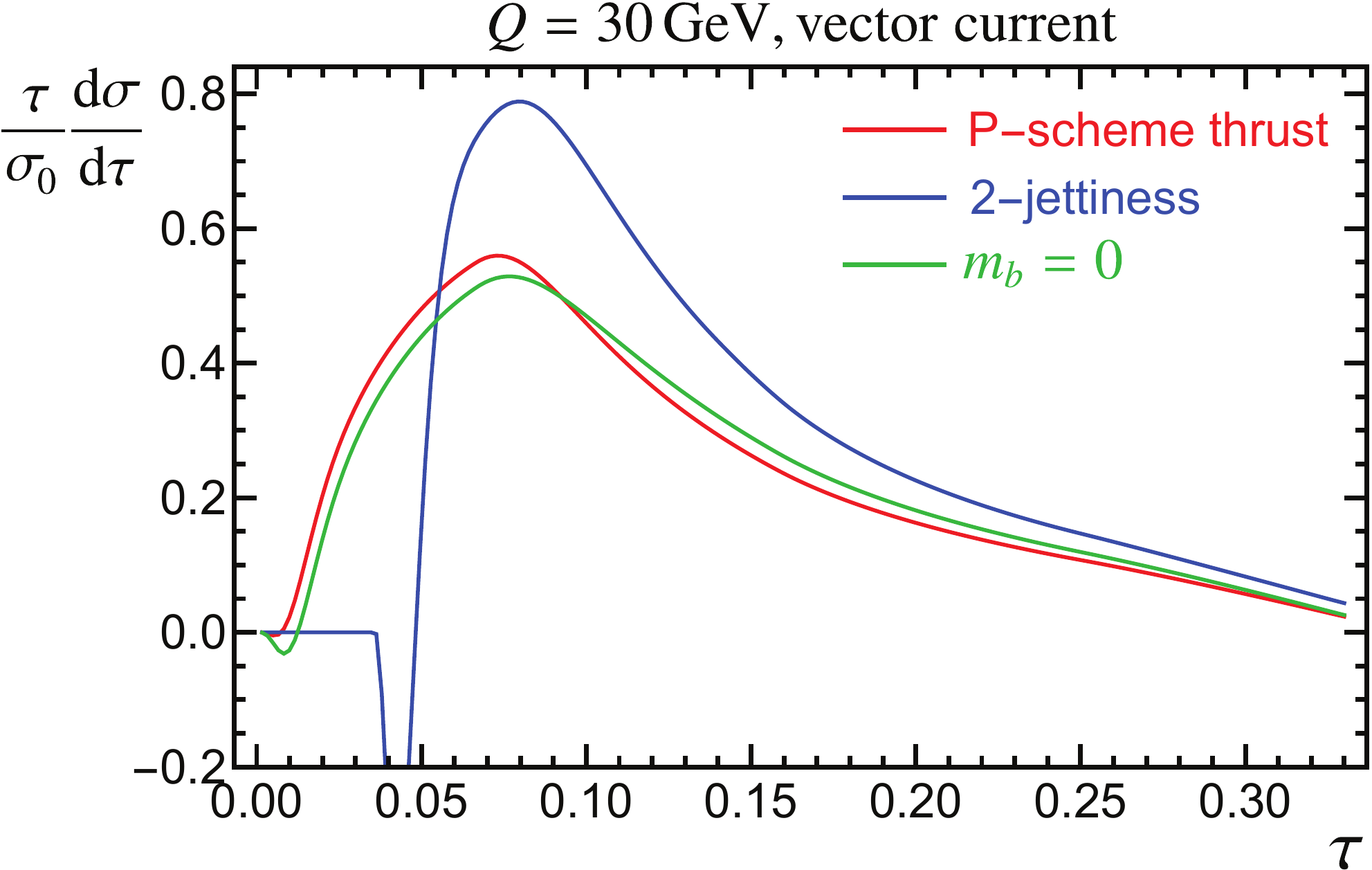}
\label{fig:schemes-30}
}	\subfigure[]
{
\includegraphics[width=0.47\textwidth]{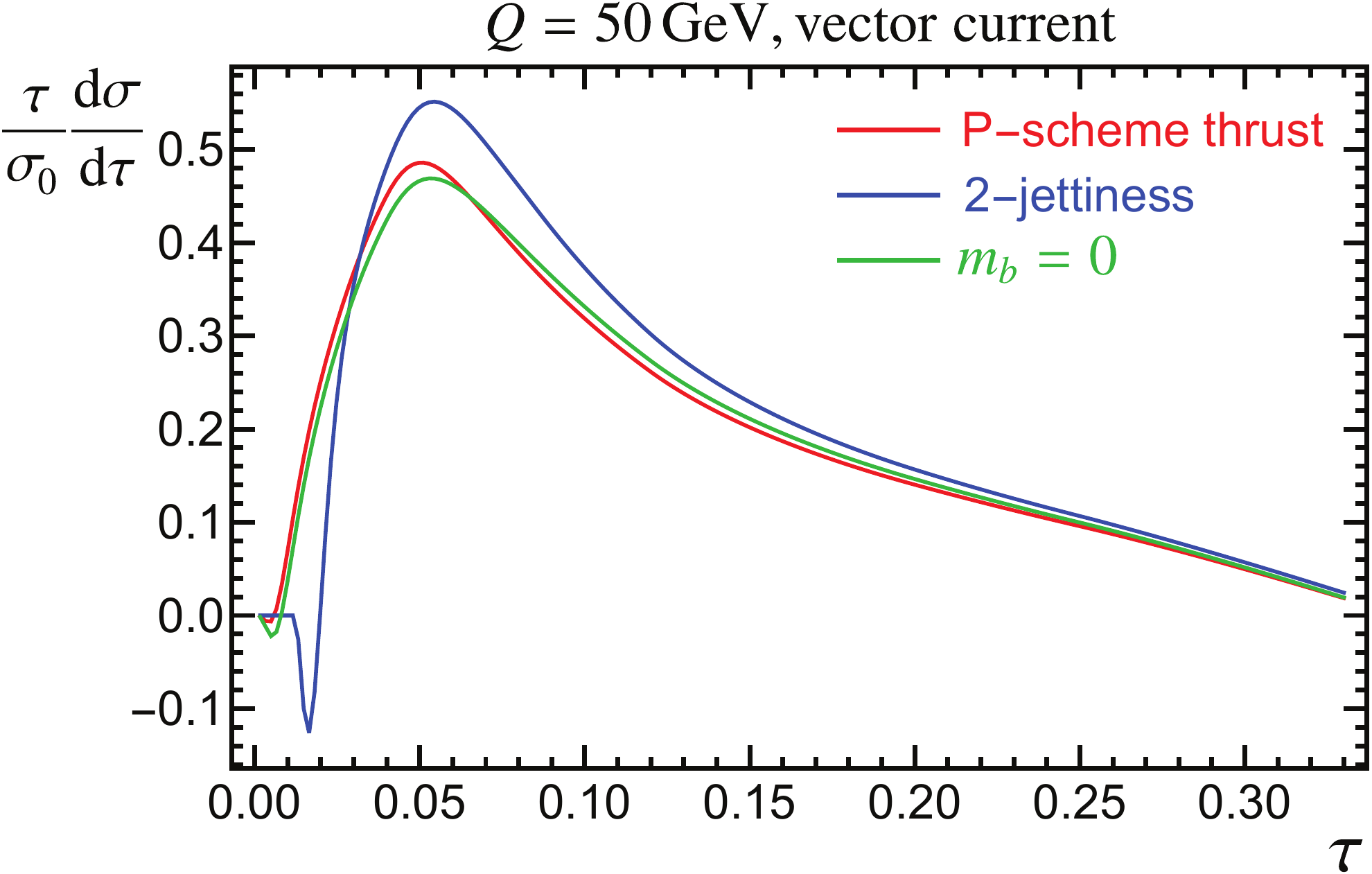}
\label{fig:schemes-50}
}~~~
\subfigure[]{
\includegraphics[width=0.48\textwidth]{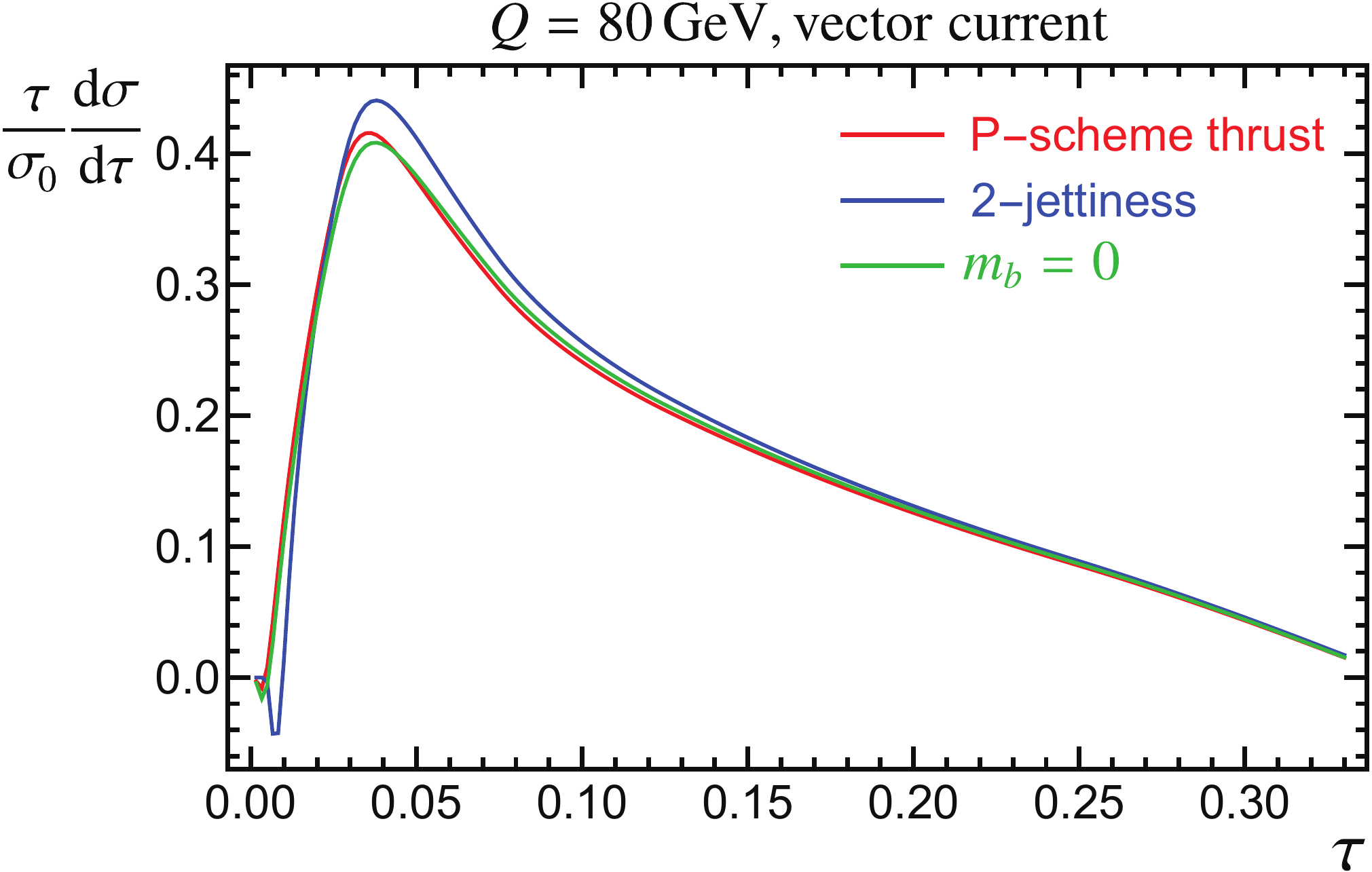}
\label{fig:schemes-80}
}
\caption{Differential cross section for massless quarks (green lines), 2-jettiness (blue lines) and P-scheme thrust (red lines) produced through the vector current. Panels (a), (b),
(c) and (d) correspond to center-of-mass energies of $20$, $30$, $50$ and $80\,$GeV, respectively.}
\label{fig:Schemes}
\end{figure*}

We study next the convergence of the (resummed) perturbative series for the differential cross section. To that end, we generate bands randomly modifying our profile functions via a flat scan on
their parameters, varying them within the ranges specified in Ref.~\cite{Hoang:2014wka}, with the exception of the non-singular scale, for which we use the following continuous variation
\begin{equation}
\mu_{\rm ns} = \frac{1}{2}[(2 + ns) \mu_H - \mu_J]\,,
\end{equation}
with $-1\leq n_s\leq 1$. In our scan we also randomly vary $\xi$ between $0$ and $1$. In Fig.~\ref{fig:bands} we show the resulting perturbative bands at LL (green), NLL (blue) and
N$^2$LL + $\mathcal{O}(\alpha_s)$ (red) for the vector and axial-vector currents, at two center-of-mass energies: $Q=40\,$GeV and $80\,$GeV. Our curves are not self-normalized, but
we nevertheless observe an excellent convergence in all cases (even at low energies) in the tail of the distribution, where higher-order bands are nicely contained in lower-order ones. 
In the peak we see a big jump between LL and the two highest orders, and the convergence is not as good as in the tail, what might indicate that the
parameters affecting mainly the peak should be varied in wider ranges. A careful inspection of the error bands reveals that the relative uncertainties for LL and NLL are nearly identical in the
whole range, and both monotonically increase as $\tau$ grows: at $Q=40\,[80]\,$GeV they change from $36\,[45]\%$ at $\tau =0.07$ to $84\,[80]\%$ at $\tau=0.28$. On the other hand,
at N$^2$LL the relative uncertainty is completely flat between $0.07\leq \tau\leq 0.3$, and smaller than the two lower orders: $36\,[30]\%$ for $Q=40\,[80]\,$GeV. We observe the same relative
uncertainties for vector and axial-vector currents.

\begin{figure}[t]\centering
\includegraphics[width=0.5\textwidth]{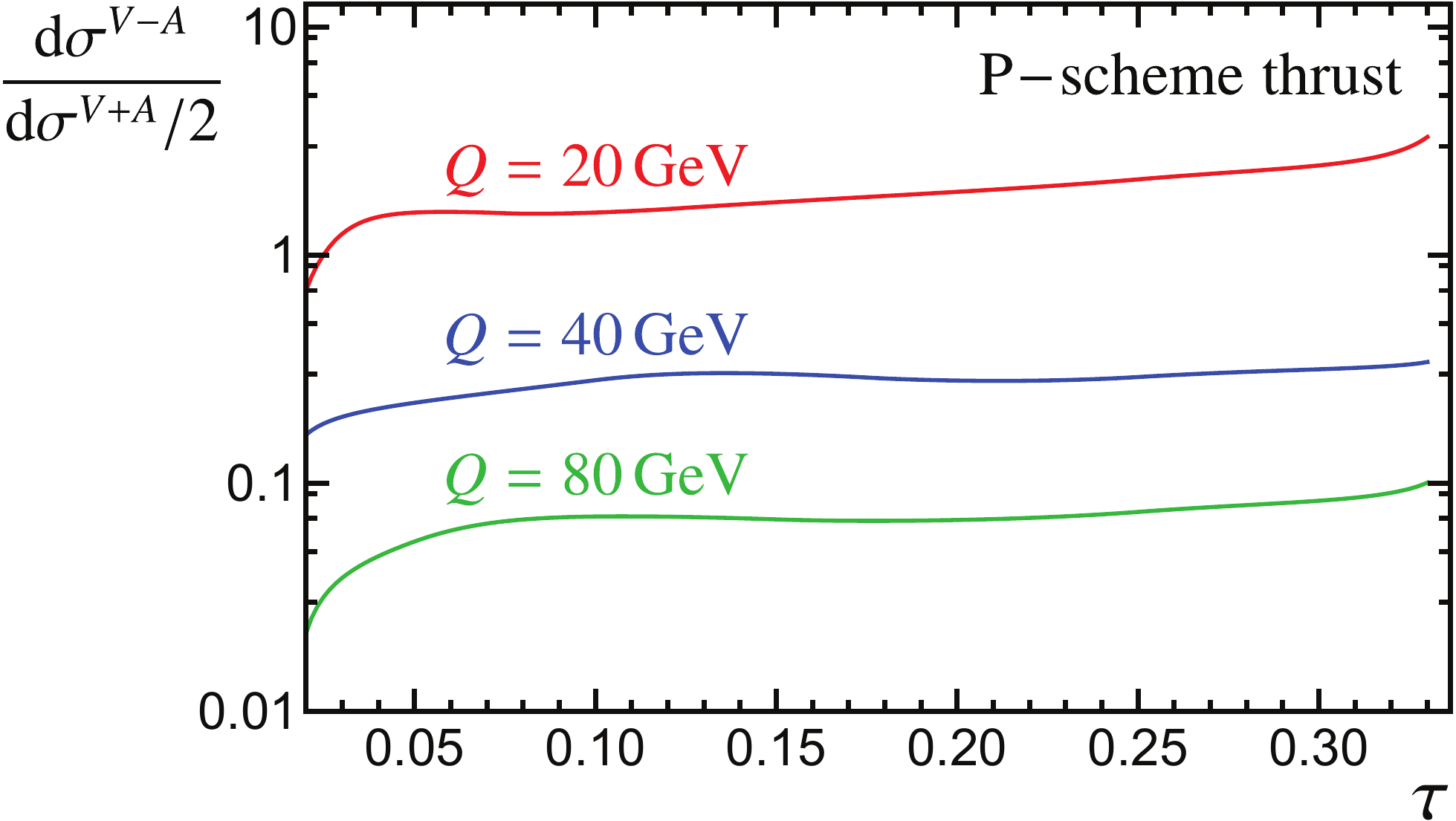}
\caption{Difference between the vector and axial-vector differential cross sections normalized to the average of the two currents. We show results for $Q=20\,$GeV (red),
$Q=40\,$GeV (blue) and $Q=80\,$GeV (green). \label{fig:V-A}}
\end{figure}

In Fig.~\ref{fig:Schemes} we compare the 2-jettiness (blue) and thrust (red) cross sections for massive quarks produced through the vector current at various center of mass energies,
as indicated in the caption of the plot. As a reference, we also show in green the massless cross section. We observe that the 2-jettiness cross section has a negative deep which becomes
more pronounced at low energies. It is produced by large logarithms which could be summed up by matching SCET to bHQET, as discussed in Sec.~\ref{sec:bHQET}. While the massless
cross section is always quite similar to massive P-scheme thrust, the 2-jettiness distribution gets quite different at low energies, with a higher peak shifted to the right. We will study this
behavior in further detail later in this section. As energies become larger, the three cross sections become similar to one another, but P-scheme thrust is always closer to the massless result.
In Fig.~\ref{fig:V-A} we plot 
the difference between the vector and axial-vector currents normalized to their average. 
To make the figure clearer, we use a logarithmic scale
on the $y$ axis. We observe, as expected, that for larger energies the difference becomes smaller, since both currents approach the (current-independent) massless result.

\begin{figure*}[t!]
\subfigure[]
{
\includegraphics[width=0.46\textwidth]{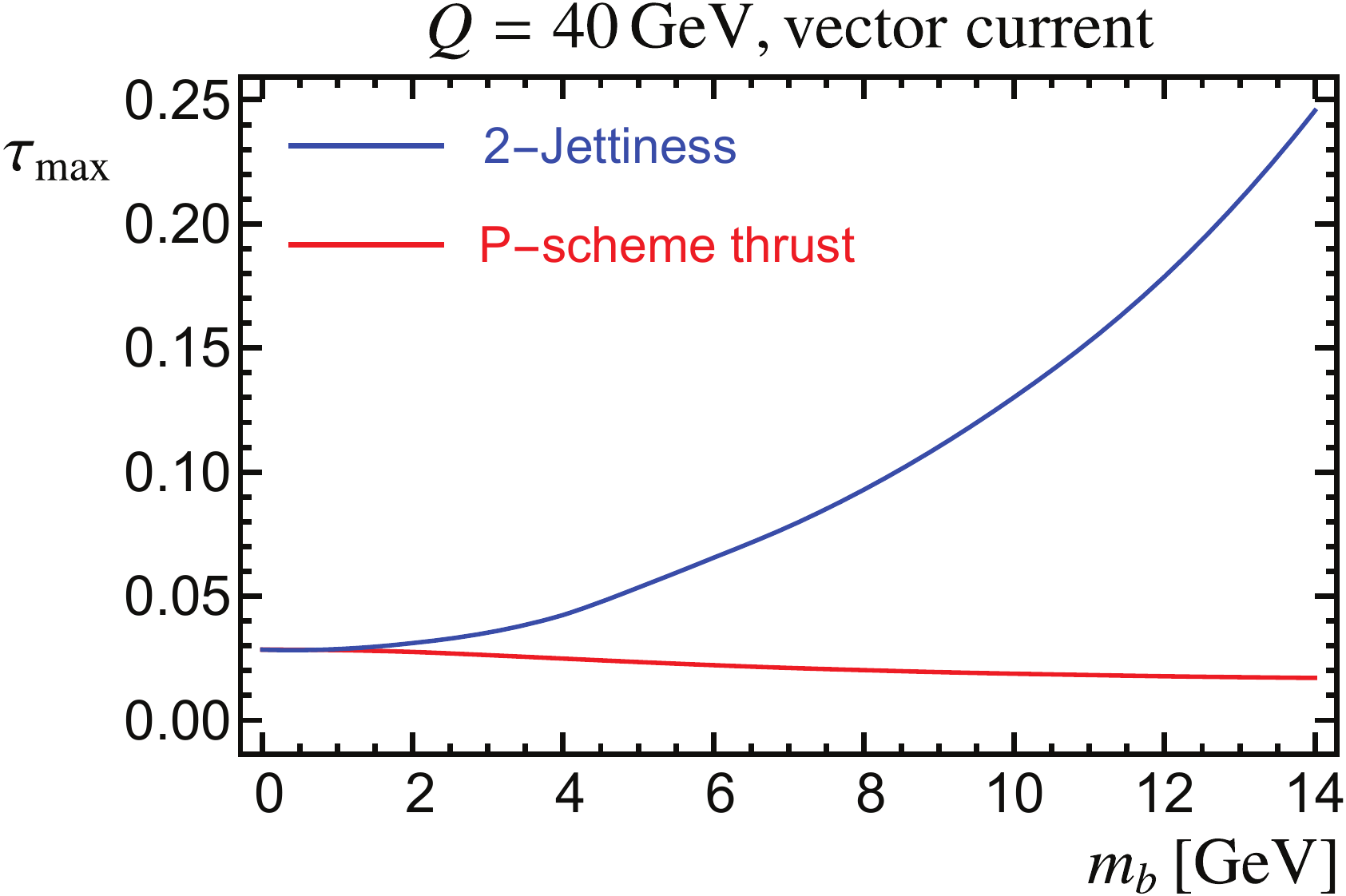}
\label{fig:Pos}
}~~~
\subfigure[]{
\includegraphics[width=0.5\textwidth]{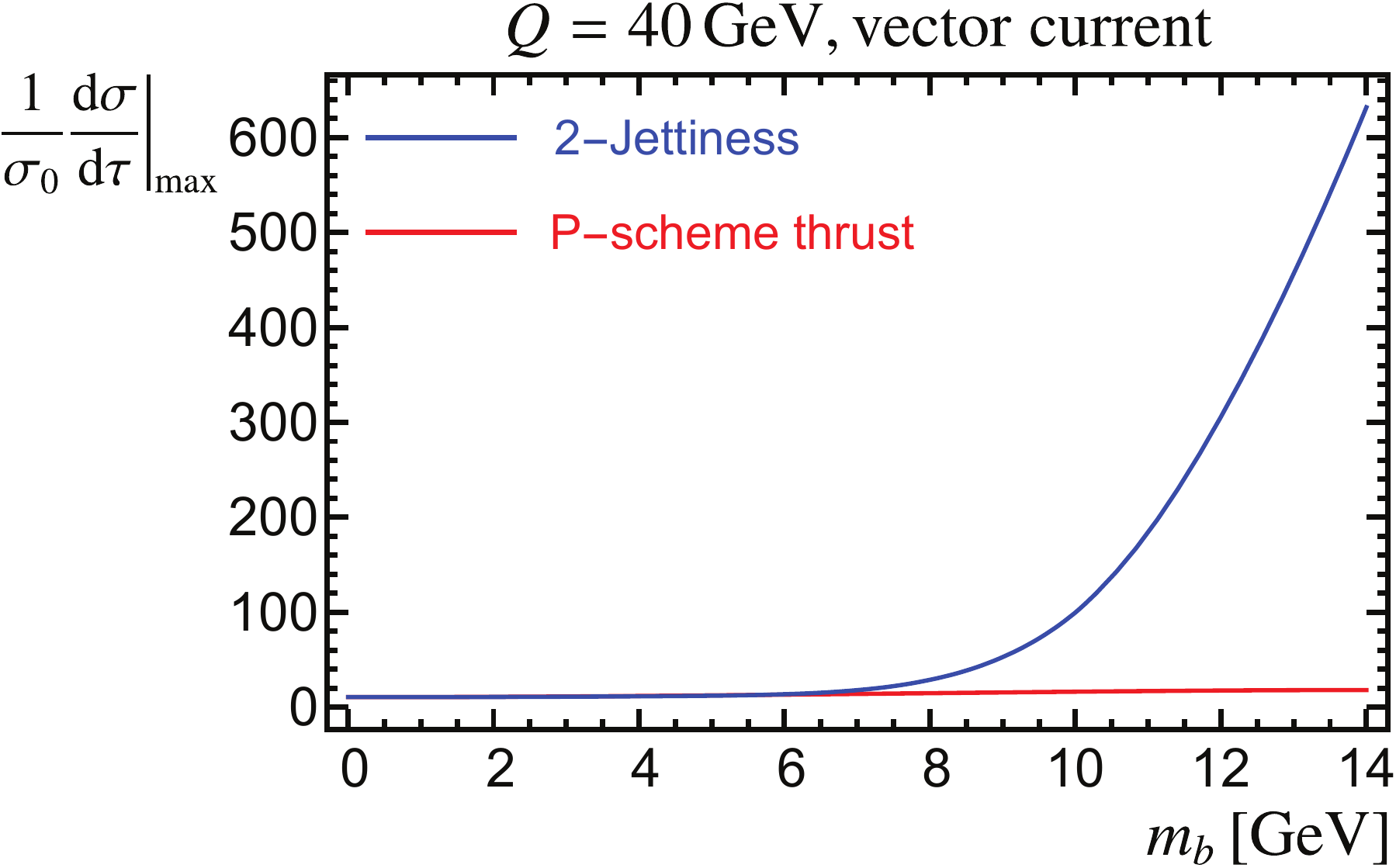}
\label{fig:Height}
}
\caption{Peak position (a) and peak height (b) for 2-jettiness (blue) and P-scheme thrust (red) massive cross section. Results correspond to default profiles, vector current and a
center-of-mass energy of $40\,$GeV, and with $m_b\equiv \mbar_b(\mbar_b)$. We vary the bottom mass between $0$ and $14\,$GeV, such that SCET still applies.}
\label{fig:Peak}
\end{figure*}

In our last analysis we study the dependence of the peak position and peak height with the heavy quark mass. Since the peak position retains some dependence on $Q$ from soft
hadronization, we fix the value of the center-of-mass energy to $40\,$GeV, such that we can make sure the peak moves only due to changes in the mass. In this case we
compare the results for thrust and 2-jettiness, since the former is relatively mass insensitive while the latter has been designed to measure the top quark mass in future linear colliders,
see e.g.\ Ref.~\cite{Fleming:2007qr}. We restrict the values of the bottom quark below $m_b=14\,$GeV to make sure we can still apply SCET and scenario~II, which should be described
using bHQET, is unimportant. The results of our study are summarized in Fig.~\ref{fig:Peak}, where one can clearly observe a flat behavior for P-scheme and an obvious quadratic dependence
for 2-jettiness. The latter is nothing but expected, since the peak position is shifted by \mbox{$\tau^J_{\min} = 1 - \sqrt{1-4\hat m^2}\simeq 2\hat m^2$}. In fact, if we perform a fit
to the 2-jettiness peak position we find $\tau_{\rm max}\simeq 0.0255 + 1.75 \hat m^2$, which follows almost exactly the blue line in Fig.~\ref{fig:Pos} and is in fair agreement with our
expectations [\,the small disagreement is expected since the peak position should be computed with $\mbar_b(\mu_J)$ and not with $\mbar_b(\mbar_b)$\,]. The dependence of the peak
height on the bottom mass is also much larger in jettiness than P-scheme thrust.

\section{Conclusions}\label{eq:conclusions}
When considering heavy quarks in the context of event shapes, depending on the scheme used in their definition the mass sensitivity of the cross section can
vary significantly. This sensitivity manifests itself already at lowest order by setting the threshold position to a non-zero value, and of course increases as the mass grows. This shifted
threshold comes solely from the jet function.

While in a recent paper we discussed how to obtain these distributions in fixed-order at NLO, in this article we have shown how to analytically compute the differential
and cumulative cross sections in the E- and P-schemes at N$^2$LL~+~$\Ord(\alpha_s)$ accuracy in SCET and bHQET. To achieve this goal, we have calculated the missing pieces,
namely the NLO jet function in those two effective field theories. We have shown that in the collinear limit the heavy quark momenta expressed in the E- and P-schemes coincide, but are different
from the original (massive) definition. This entails that for any event shape, the jet function will be identical in the former two schemes, but in the case of thrust, heavy jet mass and C-parameter
the measurement function is no longer completely inclusive, meaning that one needs to compute the jet function with cut diagrams, integrating over phase space rather than loop momenta.
We provide an optimized and compact form for the jet function definition in each EFT, written in terms of quantum and kinematic operators, that facilitates their computation. For the bHQET
jet function we explain how to rescale the integrated light-particle momenta such that the heavy quark mass drops out before any integration is carried out. In the computation
of the P-scheme SCET jet function one needs to use either sector decomposition or hypergeometric function identities to properly expand in $\varepsilon$ and extract the distributions that
appear in this limit.

Shifting its argument to relocate the threshold back to zero, the 1-loop SCET massive jet functions can be written as the sum of the massless jet function plus mass corrections. The latter can be
further divided into terms with distributions (or singular) and terms without distributions. While carrying out resummation for the former is already well known, the terms with regular functions need to
be treated in a case-by-case basis. In this article we show how to analytically RG evolve the non-distributional terms for 2-jettiness and \mbox{P-scheme} thrust, and derive rapidly-convergent
expansions that can be carried out around the threshold ($\tau - \tau_{\rm min} = 0$), ``singular'' ($Q^2\tau = m^2$)
[\,only for P- and \mbox{E-schemes}\,] and massless \mbox{($Q^2 \tau \gg m^2$)} limits, which nicely
overlap with one another such that in numerical implementations there is no need to explicitly evaluate hypergeometric functions at all (for 2-jettiness there is a small region around $Q^2\tau = m^2$
for which one cannot use expansions). Our expansions can be carried out up to any order such that the result is also arbitrarily precise. This is much faster than a direct evaluation
of $_2F_1$ and $_3F_2$ functions, which were the bottleneck of the analysis carried out in Ref.~\cite{Butenschoen:2016lpz}.

We show how to absorb into the SCET factorization theorem those mass-suppressed singular terms that appear in fixed-order corrections by a suitable redefinition of the hard and jet functions.
After this procedure is carried out, we complete our resummed expression with purely kinematic corrections (which are now entirely non-singular), which become relevant in the far tail. Hadronization
power corrections can be incorporated in the usual way by convolving with a shape function, and with this complete description we have performed some numerical investigations. We have shown
that there are strong cancellations taking place between the two types of mass corrections to the SCET factorization theorem (with or without distributions) everywhere except in the peak, and that
the remaining non-singular corrections are immaterial everywhere except in the far tail. The cancellations are stronger at larger energies, where also vector and axial-vector currents yield similar
results. We have demonstrated that the P-scheme thrust cross section is much closer to the massless prediction than for 2-jettiness by comparing cross sections as well as investigating the peak
position and height as a function of the heavy quark mass. We have also observed a nice convergence of the cross section when adding perturbative orders.

These results will be highly important for ongoing and forthcoming research in the field of event shapes with massive quarks. They will play a relevant role in the determination of $\alpha_s$ with high
precision (when the bottom quark mass cannot be neglected any longer) and in the Monte Carlo top quark mass parameter calibration. In addition, the computations we have carried out will be very valuable
for top quark mass measurement carried out at future linear colliders. Our computations can be applied to other relevant event-shapes such as angularities, groomed observables like Soft
Drop~\cite{Larkoski:2014wba} or even recoil-sensitive observables like jet-broadening. These will be presented in forthcoming publications. Extending our computations to $\Ord(\alpha_s^2)$ is
certainly challenging, but at least for the bHQET jet function, calculations of similar complexity have been carried out for the soft function e.g.\ in
Refs.~\cite{Hornig:2011iu,Kelley:2011ng,Monni:2011gb,vonManteuffel:2013vja}, and even numerical approaches have been devised in Refs.~\cite{Bell:2018vaa,Bell:2018oqa}. The massive SCET jet
function at $\Ord(\alpha_s^2)$ is definitely much more involved, and so far results only exist for 2-jettiness~\cite{Hoang:2019fze}, which is certainly simpler since it can be computed as the imaginary
part of a forward-scattering matrix element, such that the usual machinery for multi-loop computations can be applied.

\acknowledgments
This work was supported in part by the Spanish MINECO Ram\'on y Cajal program (RYC-2014-16022), the MECD grant FPA2016-78645-P, the IFT Centro de Excelencia
Severo Ochoa Program under Grant SEV-2012-0249, the EU STRONG-2020 project under the program H2020-INFRAIA-2018-1, grant agreement no.\ 824093 and the
COST Action CA16201 PARTICLEFACE. A.\,B. is supported by an FPI scholarship funded by the Spanish MICINN under grant no. BES-2017-081399. M.\,P. is partially
supported by the FWF Austrian Science Fund under the Project No. P28535-N27 and by the FWF Doctoral Program ``Particles and Interactions'' No. W1252-N27.
A.\,B. thanks the University of Salamanca for hospitality while parts of this work were completed.

\appendix
\section{Sector Decomposition}\label{sec:secDec}
The direct $\varepsilon$ expansion of $J_{a,P}^{\rm real}$ becomes much simpler if one does not have to deal with distributions, therefore we consider the cumulative jet function, and to that end
we define
\begin{equation}\label{eq:Jet-cumulant}
\Sigma_a(s_c,\mu) \equiv \!\int_0^{s_c}\!{\rm d}s\, J^{\rm real}_{a,P}(s,\mu)\,.
\end{equation}
Switching variables to $s = y s_c$ in Eq.~\eqref{eq:Jet-cumulant} and $x\to 1-x$ in Eq.~\eqref{eq:Ja-thrust} we get
\begin{align}\label{eq:SigmaI3}
\Sigma_{a}(s_c,\mu) =\,& \frac{C_F\alpha_s}{2\pi\Gamma(1-\varepsilon)}\biggl(\frac{s_c}{\mu^2}\biggr)^{\!\!-\varepsilon}I_3\biggl(\frac{m^2}{s_c}\biggr)\,,\\
I_3(t)\equiv \,& \int_0^1\!{\rm d}y\, y^{-\varepsilon}\!\! \int_0^1\!\!{\rm d}x\,\frac{(1-x)^{2-\varepsilon}x^{-1-\varepsilon}}{y (1-x)+t\,x}\,.\nonumber
\end{align}
We apply sector decomposition by splitting the $x$ integration in two segments: $(0,y)$ and $(y,1)$. In the former we switch variables to $x = z y$ and in the latter we
reverse the order of integration, which is followed by the change of variables $y = z \,x$, to find
\begin{align}
I_3(t) =& \!\int_0^1\!\!{\rm d}y\, y^{-1-2\varepsilon}\!\! \int_0^1\!\!{\rm d}z\,\frac{(1-zy)^{2-\varepsilon}z^{-1-\varepsilon}}{(1-z y)+t\,z} + \!\!
\int_0^1\!{\rm d}x\,x^{-1-2\varepsilon}(1-x)^{2-\varepsilon}\!\! \int_0^1\!\!\frac{{\rm d}z\,z^{-\varepsilon}}{(1-x)z+t}\nonumber\\
\equiv & \; I_3^\alpha(t) + I_3^\beta(t)\,.
\end{align}
Since the original singularities at $x=0,1$ have been properly separated, mapping the former at $y=0$ and the latter at $x=0$, one can expand in $\varepsilon$ before integrating.
Let us solve $I_3^\beta$ first, which has a single pole only, such that we can use Eq.~\eqref{eq:dist-expansion} on $x^{-1-2\varepsilon}$ to obtain
\begin{align}
I_3^\beta(t) =& -\!\frac{1}{2\varepsilon}\! \int_0^1\!\!\frac{{\rm d}z}{z+t}\bigl[1 - \varepsilon\log(z)\bigr] +\!\!
\int_0^1{\rm d}x \int_0^1{\rm d}z\,\frac{2t-t x-x z+z}{(t+z) (x z-t-z)}\\
=&-\!\frac{1}{2\varepsilon}\log\biggl(1+\frac{1}{t}\biggr)+\frac{1}{2}\text{Li}_2 \biggl( - \frac{1}{t}\biggr)-(1+t)\log\biggl(1+\frac{1}{t}\biggr)-\text{Li}_2 \biggl(\frac{1}{1+t}\biggr)+1\,.\nonumber
\end{align}
For $I_3^\alpha$ one must start applying Eq.~\eqref{eq:dist-expansion} to $y^{-1-2\varepsilon}$ in order to regulate the pole of the \mbox{$z$-integral}. Taking into account the plus-function
prescription and that the upper integration limit is $1$ we get
\begin{equation}
I_3^\alpha(t) = -\frac{1}{2\varepsilon}\! \int_0^1\!{\rm d}z \,\frac{z^{-1-\varepsilon}}{ 1+t\,z} -\!
\int_0^1{\rm d}y \!\int_0^1\!{\rm d}z\,\frac{1 + t z (2 - y z)-y z}{(1+t z) (1 + t z-y z)}\,,
\end{equation}
where in the second term we have already set $\varepsilon=0$. Using again Eq.~\eqref{eq:dist-expansion} to expand $z^{-1-\varepsilon}$ in $\varepsilon$
and solving the resulting integrals we arrive at
\begin{align}
I_3^\alpha(t) =& \,\frac{1}{2 \varepsilon^2}+\frac{1}{2 \varepsilon}\log (1+t)+\frac{1}{2} \,\text{Li}_2(-t) -\text{Li}_2(1-t)+\text{Li}_2\Bigl(\frac{1}{1+t}\Bigr)\\
&-\frac{t}{t-1}\log (t)+t \log (1+t)+\log (1+t)-1\,.\nonumber
\end{align}
Thus, summing $I_3^\alpha$ and $I_3^\beta $ we obtain:
\begin{equation}\label{eq:I3-expand}
I_3(t) = \,\frac{1}{2 \varepsilon^2}+\frac{1}{2 \varepsilon}\log (t)+\text{Li}_2\Bigl(\frac{1}{1-t}\Bigr)+\frac{1}{2}\log^2 (t-1) -\frac{1}{4}\log^2 (t)-\frac{1}{t-1}\log (t)+\frac{\pi^2}{12}\,.
\end{equation}
To obtain this expression, which facilitates taking the $t\to\infty$ limit (that corresponds to $s_c\to 0$), we have applied the following identities of dilogarithms:
\begin{align}
\text{Li}_2(z)=&-\!\text{Li}_2(1-z)-\log(1-z)\log(z)+\frac{\pi^2}{6}\,,\\
\text{Li}_2(z)=&-\!\text{Li}_2\Bigl(\frac{1}{z}\Bigr)-\frac{1}{2}\log^2 (-z)-\frac{\pi^2}{6}\,,\nonumber
\end{align}
where the second line holds for $z\notin(0,1)$ only. Now we insert Eq.~\eqref{eq:I3-expand} into \eqref{eq:SigmaI3} and expanding again in $\varepsilon$ becomes trivial.
To compute $J_{a,P}^{\rm real}$ we have to take the derivative of $\Sigma_{a}(s_c)$ with respect to $s_c$ taking into account that it has support only for $s_c>0$:
\begin{equation}
J_{a,P}^{\rm real}(s,\mu)=\frac{\text{d}}{\text{d} s} \Bigl[\theta (s)\,\Sigma_{a}(s,\mu)\Bigr].
\end{equation}
Using the relations in Eq.~\eqref{eq:ThetaDer}\,\footnote{To use these relations the functions multiplying $\theta(x)$ should be either $\log^n(x)$ or regular at $x=0$.
Therefore it is convenient to write $\log(t-1)$ as $\log(t)-\log(1-1/t)$.} 
and the identity given in Eq.~(3.6) of Ref.~\cite{Lepenik:2019jjk} one arrives at the result quoted in Eq.~\eqref{eq:Ja-P Result}.

\section{\boldmath Alternative Analytic Expression of $I^P_{\rm nd}$ for $s>m^2$}\label{sec:yReal}
In this appendix we present an alternative form of $I^P_{\rm nd}$ in which all terms are manifestly real for $y>1$ and where no
numerical derivatives are involved. In a first step we express ${}_2 F_1 (1, 1 + \varepsilon, 2+\tilde{\omega} , 1 - y)$ in Eq.~\eqref{eq:res-der} in terms
$_2 F_1 (1, 1 + \varepsilon, 1 - \tilde{\omega} + \varepsilon , y)$ through Eq.~\eqref{eq:id2}, and then use that for $y>1$ one has\,\footnote{To obtain this relation one simply
has to divide the integration path in Eq.~\eqref{eq:2F1} into the segments $(0,1/y)$ and $(1/y,1)$. Using
\begin{equation}
[1 - z (y \pm i \varepsilon)]^{- a} = \theta (1 - z y) (1 - y z)^{- a} +
\theta (z y - 1) (y z - 1)^{- a} [\cos (a \pi) \pm i \sin (a \pi)]\,,
\end{equation}
remapping each segment back to $(0,1)$ by a change of variables ($z\to z/y$ in the first segment and \mbox{$z\to[1-(1-1/y)x]$} in the second), and carrying out the integrals
one finds the following identity:
\begin{align}
_2 F_1 (a, b , c , y \pm i \varepsilon) =\, & \frac{\Gamma (1 - a) y^{- b}
\Gamma (c)}{\Gamma (1 - a + b) \Gamma (c - b)} \,{}_2 F_1 \!\biggl( b, 1 + b - c, 1
- a + b, \frac{1}{y} \biggr) \\
& + \frac{e^{\pm i \pi a} \Gamma (1 - a) \Gamma (c) y^{b - c} (y -
1)^{- a - b + c} }{\Gamma (b) \Gamma (1 - a - b + c)} 
\,{}_2 F_1 \!\biggl( 1 - b, c - b, 1 - a - b + c, \frac{y - 1}{y}
\biggr) . \nonumber
\end{align}
}
\begin{align}
_2 F_1 (1, 1 + \varepsilon, 1 - \tilde{\omega} + \varepsilon , y) =\, & \!-\!\frac{\tilde{\omega}\,\pi (y - 1)^{-1- \tilde{\omega} } \left[ \cot (\pi \varepsilon) + i
\, \right] y^{\tilde{\omega} - \varepsilon} \,\Gamma (1 - \tilde{\omega} + 	\varepsilon)}{\Gamma (1 - \tilde{\omega}) \Gamma (\varepsilon + 1)} \\
& \!+ \!\frac{(\tilde{\omega} - \varepsilon)}{y \varepsilon} \,{}_2 F_1 \!\biggl( 1, 1+ \tilde{\omega} - \varepsilon , 1 - \varepsilon, \frac{1}{y} \biggr) .\nonumber
\end{align}
The only involved computation left is finding an analytic expression for the derivative of \mbox{$_2 F_1 (1, 1 + \tilde{\omega} - \varepsilon , 1 - \varepsilon , 1 / y)$} with respect to
$\varepsilon$ in the $\varepsilon \rightarrow 0$ limit. The complication here arises because there are poles in $1 / \varepsilon$ such that one also needs the second
derivative, which as we shall see implies the appearance of the $_4 F_3$ function. A practical way of doing the derivative is by Taylor expanding. The first derivative
reads
\begin{equation}
\frac{{\rm d} }{{\rm d} \varepsilon} {}_2 F_1 ( 1, 1 + \tilde{\omega} - \varepsilon , 1 - \varepsilon, y ) = \frac{y\,\tilde{\omega} ( 1 - y )^{-1- \tilde{\omega}}}{(1 - \varepsilon)^2}\,
{}_3 F_2 ( 1 - \tilde{\omega}, 1 - \varepsilon, 1 - \varepsilon , 2 - \varepsilon, 2 - \varepsilon ,y )\,.
\end{equation}
For the second one needs the first derivative of the $_3 F_2$ function:
\begin{align}
& \frac{{\rm d} }{{\rm d} \varepsilon} \,{}_3 F_2 ( 1 - \tilde{\omega}, 1
- \varepsilon, 1 - \varepsilon , 2 - \varepsilon, 2 - \varepsilon , y ) = \\
& \qquad-\! \frac{2 y (1 - \tilde{\omega}) (1 - \varepsilon) }{ (2 -
\varepsilon)^3}\,{}_{4}F_3 ( 2 -
\tilde{\omega}, 2 - \varepsilon, 2 - \varepsilon, 2 - \varepsilon , 3 -
\varepsilon, 3 - \varepsilon, 3 - \varepsilon ,y )\,. \nonumber
\end{align}
With these results one can obtain the Taylor expansion of $_2 F_1 (1, 1 +\tilde{\omega} - \varepsilon, 1 - \varepsilon , 1 / y)$, which allows to compute the first derivative
of $_2 F_1 (1, 1 + \varepsilon , 1 - \tilde{\omega} + \varepsilon , y)$
\begin{align}
& \frac{\rm d}{{\rm d} \varepsilon} {}_2 F_1 (1, 1 + \varepsilon, 1 - \tilde{\omega} + \varepsilon , y) \Bigr|_{\varepsilon \rightarrow 0} \\
= \,& \frac{\tilde{\omega} (y - 1)^{- \tilde{\omega} - 1} y^{\tilde{\omega} - 1}}{12} \biggl\{ 12\, \tilde{\omega} \,{}_4 F_3 \!\biggl( 1, 1, 1, 1
- \tilde{\omega} , 2, 2, 2 , \frac{1}{y} \biggr) - 12\,{}_3 F_2 \!\biggl( 1, 1, 1 -\tilde{\omega} , 2, 2 , \frac{1}{y} \biggr) \nonumber\\
& y \Big[6 (H_{- \tilde{\omega}} - \log (y)) (2 i \pi - H_{-\tilde{\omega}} + \log (y)) - 6 \psi^{(1)} (1 - \tilde{\omega}) + 5 \pi^2\Big] \biggr\} . \nonumber
\end{align}
Using this result we arrive at the alternative expression for $I^P_{\rm nd}$
\begin{align}
&I^P_{\rm nd}(\tilde \omega, y) = \frac{\tilde{\omega} [(2 \tilde{\omega} + 5) y + \tilde{\omega} (3
\tilde{\omega} + 7) - 4 y^2]}{\Gamma(1-\tilde\omega)(1 + \tilde{\omega}) (1 - y)^2} \biggl[ H_{1 -
\tilde{\omega}} - \frac{1}{1 - \tilde{\omega}} - \log (y) \biggr] {}_2 F_1 (1, 1 , \tilde{\omega} + 2 , 1 - y) \nonumber\\
& + \frac{3 \tilde{\omega}^2 y + 3 \tilde{\omega}^2 - 5 \tilde{\omega} y
+ 14 \tilde{\omega} + y - 7}{\Gamma(2-\tilde\omega) (y - 1)^2} + (y - 1)^{-3-\tilde{\omega}} y^{\tilde{\omega} - 1} [4 y^2 - 2 (\tilde{\omega} + 5) y
- \tilde{\omega} (3 \tilde{\omega} + 7)] \nonumber\\
& \times \biggl\{ \tilde{\omega}^2 {}_4 F_3\! \biggl( 1, 1, 1, 1 - \tilde{\omega} , 2, 2, 2 , \frac{1}{y} \biggr) - \tilde{\omega} \,_3 F_2 \!
\biggl( 1, 1, 1 - \tilde{\omega} , 2, 2 , \frac{1}{y} \biggr)\\
& \times\frac{\tilde{\omega}\, y}{12} \biggl[ 5 \pi^2 - 6 \biggl(\! H_{1 -
\tilde{\omega}} - \frac{1}{1 - \tilde{\omega}} - \log (y)\! \biggr)^{\!\!2} - 6
\psi^{(1)} (1 - \tilde{\omega}) \biggr] \nonumber\\
& + y \cos (\pi \tilde{\omega}) \Gamma (1 - \tilde{\omega})
\Gamma (1 + \tilde{\omega})_2 F_1 (1, 1 , 2+\tilde{\omega} , 1 - y) \!\biggl[
H_{1 - \tilde{\omega}} - \frac{1}{1 - \tilde{\omega}} - \log (y) \biggr]\!
\biggr\}. \nonumber
\end{align}

\bibliography{thrust3}
\bibliographystyle{JHEP}

\end{document}